\documentclass[A4paper]{article}  
\usepackage{moreverb,url}

\usepackage[colorlinks,bookmarksopen,bookmarksnumbered,citecolor=black,urlcolor=red]{hyperref}
\usepackage[utf8]{inputenc}

\usepackage{graphicx}

\usepackage[margin=1in]{geometry} 

\usepackage{amsmath}
\usepackage{amsthm}
\usepackage{amssymb}

\newcommand\BibTeX{{\rmfamily B\kern-.05em \textsc{i\kern-.025em b}\kern-.08em
T\kern-.1667em\lower.7ex\hbox{E}\kern-.125emX}}

\usepackage{url}
\usepackage{ragged2e}
\usepackage{makecell}

\usepackage{multirow}
\usepackage{listings} 
\usepackage{xcolor}
\usepackage{hyperref}
\newcommand{\Cov}{\mathrm{Cov}}
\newcommand{\Corr}{\mathrm{Corr}}
\usepackage{natbib}
\usepackage{authblk}

\title{An Adaptive Phase II Trial Design for Dose Selection and Addition in Microfilarial Infections}

\author[1]{Sonja ZEHETMAYER$^\ast$}
\author[1,2]{Marta BOFILL ROIG}
\author[1]{Fabrice LOTOLA MOUGENI}
\author[3]{Sabine SPECHT}
\author[4]{Marc P. HÜBNER}
\author[1]{Martin POSCH}

\affil[1]{Center for Medical Data Science, Medical University of Vienna, Vienna, Austria}
\affil[2]{Department of Statistics and Operations Research and Institute for Research and Innovation in Health (IRIS), Universitat Politècnica de Catalunya - BarcelonaTech (UPC), Barcelona, Spain}
\affil[3]{Drugs for Neglected Diseases initiative (DNDI), Switzerland}
\affil[4]{Institute for Medical Microbiology, Immunology and Parasitology, University Hospital Bonn, Bonn, Germany\\$^5$ German Center for Infection Research (DZIF), Partner Site Bonn-Cologne, Bonn, Germany}

\begin{document}

\maketitle
\footnotetext{*Correspondence: sonja.zehetmayer@meduniwien.ac.at}
\footnotetext{On behalf of the eWHORM consortium}

\date

\begin{abstract}
{We propose a frequentist adaptive phase 2 trial design to evaluate the safety and efficacy of three treatment regimens (doses) compared to placebo for four types of helminth (worm) infections. This trial will be carried out in four Subsaharan African countries from spring 2025. Since the safety of the highest dose is not yet established, the study begins with the two lower doses and placebo. Based on safety and early efficacy results from an interim analysis, a decision will be made to either continue with the two lower doses or drop one or both and introduce the highest dose instead. This design borrows information across baskets for safety assessment, while efficacy is assessed separately for each basket.

The proposed adaptive design addresses four key challenges: 
(1) The trial must begin with only the two lower doses because reassuring safety data from these doses is required before escalating to a higher dose. (2) Due to the expected speed of recruitment,
adaptation decisions must rely on an earlier, surrogate endpoint. (3) The primary outcome is a count variable that follows a mixture distribution with an atom at 0. (4) The second stage sample size is split between the selected doses, thus the sample size per arm depends on the number of arms in the second stage.

To control the familywise error rate in the strong sense when comparing multiple doses to the control in the adaptive design, we extend the partial conditional error approach to accommodate the inclusion of new hypotheses after the interim analysis. In a comprehensive simulation study we evaluate various design options and analysis strategies, assessing the robustness of the design under different design assumptions and parameter values. We identify scenarios where the adaptive design improves the trial's ability to identify an optimal dose. Adaptive dose selection enables resource allocation to the most promising treatment arms, increasing the likelihood of selecting the optimal dose while reducing the required overall sample size and trial duration.}
{Adaptive design; Basket trial design; Clinical trial simulation, Partial conditional error rates; Surrogate endpoint;
Zero-inflated data.}
\end{abstract}

\section{Introduction}
\label{sec1}

\label{sect_intro}
For the treatment of human helminth infections, such as soil-transmitted helminths or filariae, only a limited number of drugs are currently available which have essential limitations. Single-dose treatments commonly used for soil-transmitted helminths are ineffective against the human whipworm trichuris trichiura. Additionally, the drugs used in mass drug administration (MDA) programs for filarial diseases such as river blindness (onchocerciasis) and elephantiasis (lymphatic filariasis) are not macrofilaricidal, meaning they do not kill adult worms. Moreover, individual macrofilaricidal treatments for filarial diseases require a prolonged regimen of 4-6 weeks of doxycycline, limiting their widespread use. Doxycycline is also not efficacious for treatment of the African eye worm loa loa. Infections caused by loa loa and mansonella species are not targeted by MDA programs. Furthermore, drugs used in MDA may cause life-threatening adverse events in loa loa patients and are not efficacious against mansonella perstans \citep{Risch2024}. 

The EU project "eWHORM – enabling the WHO Road Map" (Grant agreement ID: 101103053, https://ewhorm.org/) aims to conduct a clinical trial to evaluate the efficacy and safety of the drug oxfendazole (OXF) compared to placebo in adults infected with trichuriasis, mansonellosis, onchocerciasis and/or loiasis. OXF is a pan-nematode candidate that has been used in the veterinary field for decades and showed efficacy against filariae in preclinical studies. Phase I studies confirmed its safety in humans \citep{hubner2020oxfendazole,bach2020pharmacokinetics}. The trial will take place in the Democratic Republic of the Congo, the Gabonese Republic, the Republic of Cameroon, and the United Republic of Tanzania.

Instead of conducting separate trials for each disease, a basket trial will be performed using a master protocol with four substudies \citep{woodcock2017master,koenig2024current}. This approach allows the sharing of information and administrative resources. In the eWHORM trial, efficacy is assessed separately for each basket, however, for safety assessments and dose selection during interim analyses, (interim) data from all diseases may be utilized. Additionally, further exploratory secondary analyses will be considered for borrowing information across diseases with special focus on co-infections. As safety data or secondary endpoints (co-infections) are not subject of these clinical trial simulations, no borrowing between baskets is considered in the following.

When planning a clinical trial it is often necessary to perform clinical trial simulations \citep{benda2010aspects}. These simulations help to assess the influence of unknown factors in the clinical trial such as variances, treatment effects, and dropout rates on the outcome. Thus simulations are typically performed under different assumptions to evaluate whether the design is adequately powered. Additionally, various design parameters, such as sample size, futility boundaries, and statistical tests, are compared across different scenarios to inform optimal design choices. 

This manuscript describes the trial design and clinical trial simulation for the eWHORM study. 
For the substudies of onchocerciasis, mansonellosis, and loiasis adaptive two-stage designs will be performed with dose selection or adjustment in an adaptive interim analysis based on partial conditional error rates extended to the case where a hypothesis may be added at the interim analysis (\citealp{posch2005testing,bretz2009adaptive,bauer2016twenty,posch2003issues}). The conditional error function approach \citep{proschan1995designed,posch1999adaptive,muller2004general} involves recalculating the probability of rejecting the null hypothesis given the interim results. If design modifications are performed after the interim analysis, the conditional error framework ensures that the Type I error rate remains controlled. In \citet{mehta2025} an overview of adaptive multiple testing procedures based on conditional error rates is given. The procedure proposed in this manuscript where a new arm may be added to the second stage is a further extension of their so called ``nonparametric" approach based on Bonferroni adjustment, which has not been considered so far. 

If a further hypothesis is added during the course of the study, overall Type I error control must be ensured \citep{burnett2024adding,greenstreet2024multi}. Thus, in the eWHORM study, initially, a part of the total level is saved for a high dose, which, in case it is not started, is then dedicated to the initial doses. This manuscript provides details on the adaptive two-stage trial designs for onchocerciasis, mansonellosis, and loiasis including a description of the analysis methods and the adaptation strategies. Additionally, the operating characteristics of these adaptive trial designs are compared to classical fixed sample designs through comprehensive simulation studies. The simulations are performed according to the ``clinical scenario planning and evaluation" framework as proposed by \citet{benda2010aspects}. Based on the results of the simulation studies, optimized design elements were identified, such as the timing of the interim analyses, parameters of stopping rules and dose selection criteria rules at interim analysis to maximize statistical power. We explore a range of scenarios to assess the robustness of the designs with respect to deviations from the original design assumptions.

In Section \ref{sec2}, we present details on the eWHORM trial and define the adaptive test in Section \ref{Sec:adapttest}. Further, we define the hypotheses to be tested and the selection rules. Then, we define the setup of the simulation study based on the eWHORM trial in Section \ref{sect_sims} to assess operating characteristics for a large range of scenarios by means of disjunctive and marginal power, FWER and bias of estimation. Main results of the simulation study for different design elements are reported in Section \ref{sec_results}. We also compare the performance of the proposed adaptive trial with traditional multi-armed trials, illustrating the potential benefits of the adaptive trial approach. A numerical example illustrates the procedure for several selection scenarios in Section \ref{sect_example}. We close with a discussion in Section \ref{sec_discussion}. An R-package to reproduce the results is available on GitHub (\href{https://github.com/MartaBofillRoig/ewhorm_sim}{https://github.com/MartaBofillRoig/ewhorm\_sim}).

\section{Motivating example}
\label{sec2}

The eWHORM basket trial investigates the efficacy and safety of OXF in four substudies. For infection trichuriasis, the substudy is a fixed sample design and not part of the clinical trial simulations. For each of the three substudies of onchocerciasis, loiasis, and mansonellosis, we propose adaptive, two-stage, randomized, placebo-controlled, double-blind, parallel-group phase II studies to investigate the efficacy and safety of OXF. The study design for all three substudies will be identical (up to the inclusion criteria). Therefore, in the following we describe the trial design of one of the substudies and do not indicate the disease in the notation. 
In each substudy, (up to) three active OXF regimens are compared to placebo to establish efficacy: (i) five days, once daily, 400 mg; (ii) five days, once daily, 800 mg (iii) five days, twice, once daily, one month apart. In the remainder of this document we denote these doses as "low dose", "medium dose" and "high dose".

The analysis of the basket trial will follow a frequentist framework, with separate efficacy analyses conducted for each of the four diseases. There will be no borrowing of information between diseases in the efficacy analyses (\citealp{woodcock2017master,EMA2007}). Since efficacy analyses are performed separately for each disease, the Type I Error rate will be controlled for each disease individually and no multiplicity adjustment across diseases is required. However, within each substudy the familywise error rate (FWER) will be controlled at a one-sided level of $\alpha=0.025$ accounting for the comparison of multiple doses to control and the adaptive interim analysis.

The trial's primary objective is a proof of principle, demonstrating a positive dose-response relationship of OXF in the primary endpoint parasite load (in microfilariae (mf)) at Month 12 in each of the diseases. The secondary objective is to demonstrate the safety, tolerability, efficacy and pharmacokinetics characteristics of OXF in helminth infected individuals. To support the choice of doses in the second stage, the parasite load at Month 6 will be considered as an early (surrogate) endpoint, expected to be predictive for the endpoint at Month 12. 

\begin{figure}[!p]
    \centering \includegraphics{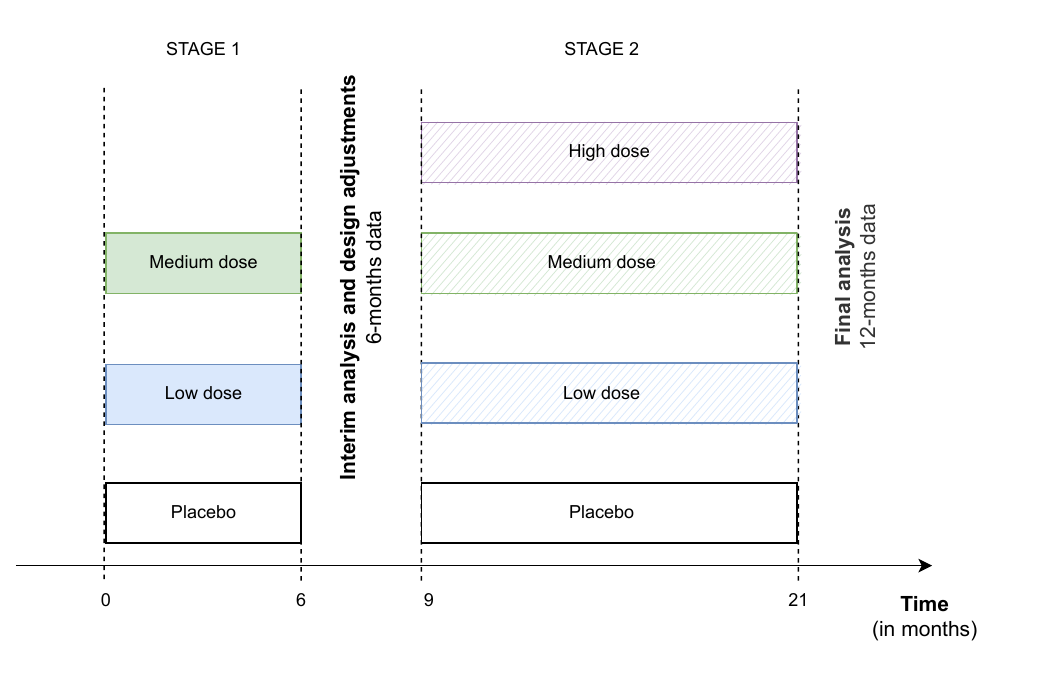}
    \caption[Scheme of the trial design]{Scheme of the trial design. The trial starts in stage 1 with placebo and medium and low doses, represented by colored bars. The trial continues in stage 2 with further investigation on the selected arms which are chosen based on interim results. The active doses that can be selected for stage 2 (low, medium and high doses) are depicted with bars with a striped pattern. The placebo group will always be continued to the second stage. Which of the active arms are selected for the second stage, depends on the efficacy and safety results in the first stage and maybe also informed by safety results obtained in the other diseases.}
    \label{fig:design}
\end{figure}

For each substudy, the clinical trial has two stages and an adaptive interim analysis (see Figure \ref{fig:design} for an illustration of the design). The fixed, overall sample size (across both stages and all doses for each substudy) is denoted by $N$. 

\begin{itemize}
    \item[] \textbf{Stage 1.}
    In the first stage, $N_1$ patients are randomized with equal allocation to three arms: placebo, low dose and medium dose. After the recruitment of the first stage patients, recruitment is paused.
    
    \item[] \textbf{Interim Analysis.}
    After the early endpoint parasite load at Month 6 (in mf) is observed for all first stage patients, an unblinded interim analysis is performed. According to the interim results, the trial may be adapted. In particular, apparently efficacious doses are selected for further evaluation in the second stage, while doses that do not show a sufficient trend at interim will be stopped for futility. If doses investigated in the first stage fail to show sufficient efficacy in the interim analysis and no safety concerns arose, the high dose arm may be selected for the second stage (which was not present in the first stage). In addition, the second stage sample size per dose may be reassessed (\citealp{muller2004general}), such that the overall second stage sample size is equal to $N_2$. The exact adaptation rule is defined further below. 
    
    \item[] \textbf{Stage 2}. 
    The trial continues to the second stage, with a placebo arm and the selected dose(s) randomising $N_2=N-N_1$ patients equally to all arms. As the number of selected doses is not fixed, the individual sample size per dose in the second stage depends on the number of doses selected in the interim analysis. After Stage 2, the final analysis is performed testing the primary endpoint parasite load at Month 12 (in mf). The adaptive multiple hypothesis testing procedure is described in the next section.
    
\end{itemize}

\section{The adaptive test}
\label{Sec:adapttest}

To obtain a hypothesis test that controls the FWER at level $\alpha$, we apply an adaptive test based on partial conditional error rates (\citealp{klinglmueller2014adaptive,posch2011Type,muller2004general}) and extend it to the case where a hypothesis may be added for stage 2. This test is based on the closure principle which guarantees strong control of the FWER (\citealp{marcus1976closed}). In the setting considered here, there are three primary individual hypotheses denoted by $H_1,H_2,H_3$. According to the closure principle, in addition to the individual hypotheses, all intersection hypotheses  need to be considered. We define an adaptive level $\alpha$ test for each of the intersection hypotheses. The closed test then rejects an individual null hypothesis $H_j$, if $H_j$ and all intersection hypotheses $J\subseteq\{1,2,3\}$ such that $j\in J$ can be rejected.

\subsection{Preplanned tests of the elementary hypotheses}

Consider the null hypotheses $H_j$, $j=\{1,2,3\}$ for the comparisons of doses $j>0$ to dose $j=0$ (note that index $j$ represents either the index of the dose or the hypothesis). For $H_j$ we define the pre-planned elementary hypothesis tests $\varphi_j^{(\gamma)}$ at level $\gamma$ as combination test setting. The stage-wise p-values for $H_j$ are denoted by $p_{j}^{(s)}$, where $s=1,2$ denotes the stage. Note that at each stage only subjects recruited in this stage are used in the test (no cumulative p-values). Furthermore, in stage 1, only p-values for $j=1,2$ are available and in stage 2 only p-values for the selected doses. Then a combined p-value is given by
$$p_j=1-\Phi[w_1\Phi^{-1}(1-p_{j}^{(1)})+ w_2\Phi^{-1}(1-p_{j}^{(2)})]$$
where $\Phi$ and $\Phi^{-1}$ denote the cumulative distribution function of the standard normal distribution and its quantile function, $w^{(1)}_j=\sqrt{N_1/N}$ and $w^{(2)}_j=\sqrt{N_2/N}$ .
	 
Dose $j=3$ and corresponding hypothesis $H_3$ is potentially added in the second stage and we set
$p_3=p_{3}^{(2)}$. 
The tests are now defined  as $\varphi_j^{(\gamma)}=\mathbf{1}_{\{p_j\leq \gamma\}}$, $j=\{1,2,3\}$, where 
 $\mathbf{1}_{\{\cdot\}}$ denotes the indicator function and the test rejects $H_j$ if  $\varphi_j^{(\gamma)}=1$ and accepts otherwise.

\subsection{Preplanned tests of the intersection hypotheses}
The level $\alpha$ intersection hypotheses are to be tested by unweighted Bonferroni tests. Thus, for the global null hypothesis, we set
$\varphi_{\{1,2,3\}}=\max(\varphi^{(\alpha/3)}_{1},\varphi^{(\alpha/3)}_{2},\varphi^{(\alpha/3)}_{3})$. For all intersection hypotheses with two elements $j,j'$ we set $\varphi_{\{j,j'\}}=\max(\varphi^{(\alpha/2)}_{j},\varphi^{(\alpha/2)}_{j'}).$

\subsection{Computation of the partial conditional error rates}

The partial conditional error rate $A_j^{(\gamma)}$ for hypotheses $j=1,2$ at level $\gamma$ is given by

$$A_j^{(\gamma)}=P(p_j\leq \gamma | p_{j}^{(1)})=1-\Phi\left(\frac{\Phi^{-1}(1-\gamma)-w^{(1)}_j \Phi^{-1}(1-p_{j}^{(1)})}{w^{(2)}_j}\right).$$
For $j=3$, $A_3^{(\gamma)}=\gamma$ because no first-stage data has been collected.

\subsection{Adapted second stage tests}

Based on the selection of doses in the interim analysis, second stage tests are performed. For doses that are not selected for the second stage, the individual null hypotheses are accepted. Also all intersection hypotheses containing only dropped hypotheses are accepted. 

Denote the set of doses selected for the second stage by  $K\subseteq\{1,2,3\}$. It is assumed that due to the specific study design, only $K=\{1,2\}$, $K=\{2,3\}$, and $K=\{3\}$ are possible sets. For all $J\subseteq\{1,2,3\}$ such that  $J\cap K=\emptyset$, the null hypothesis $H_J$ is accepted already in the interim analysis.
For all other  $J\subseteq\{1,2,3\}$ according to the partial conditional error rate principle, the adapted second stage tests $\bar \varphi_J$ rejects if either $ \sum_{k\in J\cap K}A_k^{(\alpha/|J|)}\geq 1$ or
$$\min_{j\in J\cap K}(p_{j}^{(2)}/v_{j,J})\leq \sum_{k\in J\cap K}A_k^{(\alpha/|J|)}.$$
Here, if $K=\{1,2\}$  or $K=\{3\}$ we choose 
$$v_{j,J}= \frac{A_j^{(\alpha/|J|)}}{ \sum_{k\in J\cap K}A_k^{(\alpha/|J|)}} $$ for $j\in J\cap K$.

If $K=\{2,3\}$ then for all $J\subseteq \{1,2,3\}$ such that $J\cap K=\{2,3\}$ we set
$$v_{2,J}= \frac{A_2^{(\alpha/|J|)}}{ \sum_{k\in J\cap K}A_k^{(\alpha/|J|)}},\quad v_{3,J}= \frac{A_1^{(\alpha/|J|)}+A_3^{(\alpha/|J|)}}{ \sum_{k\in J\cap K}A_k^{(\alpha/|J|)}}.$$
If $|J\cap K|=1$ we set $v_{j,J}=1$.

\section{Specification of testing framework for eWHORM trial}

In the following, we present the proposed adaptive test in the context of the eWHORM trial. First, we give several options to formulate the null hypotheses, the design including adaptation rules and the adaptive testing procedure to control the FWER in each disease. 

\subsection{Notation and hypotheses}

Let $Y_{j,t,i}$ denote the  parasite load (mf) for subject $i$ at baseline ($t=0$),  Month 6 ($t=1$) or Month 12 ($t=2$) in dose $j$, referring to placebo by $j=0$, and low, medium, and high dose by $j=\{1,2,3\}$, respectively. The distribution of $Y_{j,t,i}$ for $t>0$ can be described by  a mixture distribution with an atom at 0 (corresponding to the subjects that become free of mf, the so called "total responders") and a log-normal distribution, corresponding to subjects where positive mf values are observed (\citealp{tamarozzi2012long,wanji2016update}). We evaluate several endpoints at different follow-ups (FU), each with distinct summary measures that define the associated null hypotheses. Each null hypothesis, $H_{j,t}^{(est)}$, compares a dose $j\in \{1,2,3\}$ to placebo at FU $t$ ($t=1,2$), differing in the summary measure $est$.  

\paragraph{Geometric means at time $t$} 
Let $\delta_{j,t}^{(gm)}$ denote the geometric mean of the log-transformed mf values in dose $j$.
We consider the hypotheses 
$$H_{j,t}^{(gm)}:  \delta^{(gm)}_{0,t}-\delta^{(gm)}_{j,t}\leq 0 \  \text{versus} \ \ H'^{(gm)}_{j,t}: \delta^{(gm)}_{0,t}-\delta^{(gm)}_{j,t} >0.$$

\paragraph{Concordance at time $t$} 
Let $\delta_{j,t}^{(c)}=P(Y_{j,t,i}<Y_{0,t,i'})+\frac12 P(Y_{j,t,i}=Y_{0,t,i'})$ denote the concordance (probabilistic index, relative effect, win ratio) for dose $j$ at FU time $t$ for subject $i \neq i'$. Note that, by assumption, the concordance does not depend on $i,i'$. We consider the hypotheses
$$H_{j,t}^{(c)}:  \delta_{j,t}^{(c)}\leq \frac{1}{2} \  \text{versus} \ \ H'^{(c)}_{j,t}: \delta_{j,t}^{(c)} > \frac{1}{2}.$$

\paragraph{Concordance of the change to baseline at time $t$} 

Let $\delta_{j,t}^{(cc)}=P(Y_{j,t,i}-Y_{j,0,i}<Y_{0,t,i'}-Y_{0,0,i'})+\frac12 P(Y_{j,t,i}-Y_{j,0,i}=Y_{0,t,i'}-Y_{0,0,i'})$ denote the concordance for dose $j$ at FU time $t$. Note again, that, by assumption, the concordance does not depend on $i,i'$. We consider the hypotheses
$$H_{j,t}^{(cc)}:  \delta_{j,t}^{(cc)}\leq \frac{1}{2} \  \text{versus} \ \ H'^{(cc)}_{j,t}: \delta_{j,t}^{(cc)} > \frac{1}{2}$$

As the primary endpoint is at $t=2$ the hypotheses $H_{j,2}^{(gm)},H_{j,2}^{(c)},H_{j,2}^{(cc)}$, $j=1,2,3$, are the primary hypotheses in the respective cases.

\subsection{Analysis approaches} 
\label{sec:analysis}
For the adaptive tests described in Section \ref{Sec:adapttest}, first and second stage tests to test the above null hypotheses have to be defined. 
We consider the following analysis methods to test the null hypotheses $H_{j,t}^{(gm)}$ (approach 1), $H_{j,t}^{(c)}$ (approach 2), and $H_{j,t}^{(cc)}$ (approach 3):  
\begin{enumerate}
		\item Linear model comparing active doses against control at Month 12 adjusting for baseline parasite load (\texttt{lm}). 
	\item Wilcoxon test: nonparametric test comparing the parasite load at Month 12 between active doses and control. (\texttt{WilcoxC}). 
	\item Wilcoxon test for changes: Wilcoxon test comparing the change in parasite load at Month 12 and baseline between active doses and control (\texttt{WilcoxCC}). 
\end{enumerate}

Each of these methods gives stage-wise p-values $p_{j,t}^{(s)}$ for the comparison of active doses and placebo where $s=\{1,2\}$ denotes the stage. At each stage only subjects recruited in this stage are used in the test. 
Furthermore, in stage 1, only p-values for $j=\{1,2\}$ are available and in stage 2 only p-values for the selected doses.
For simplicity, no superscript for the type of test is included in the notation of the stage-wise p-values, as the exact procedure has no impact on the formulation of the adaptive testing procedure. The different analysis approaches will be compared in the simulation study.

\subsection{Interim analysis and dose selection rules} 
\label{sec:interim}
To adaptively select the treatment to be further investigated in stage 2, we evaluate the efficacy of low and medium doses compared to placebo based on the 6-month responses at the interim analysis. In particular, we compute the first stage p-values for the comparison of the 6-month responses $Y_{j,1,i}$ to placebo $Y_{0,1,i}$ using one of the methods described above to obtain corresponding one-sided p-values by  $p_{j,1}^{(1)}$  for dose $j$ ($j=1,2$).

To define the dose selection rule, a threshold $\alpha_1$ is pre-defined. The selection of doses for stage 2 depends on the p-values $p_{1,1}^{(1)}, p_{2,1}^{(1)}$,  the selection rules are  defined by 

\begin{enumerate}
    \item[\textbf{Case (i)} ] \textbf{$p_{1,1}^{(1)}<\alpha_1$  and $p_{2,1}^{(1)}<\alpha_1$:} Continue with doses 1 and 2, and do not start dose 3. 
    \item[\textbf{Case (ii)}] \textbf{$p_{1,1}^{(1)}<\alpha_1$ and $p_{2,1}^{(1)}>\alpha_1$:}
        Continue with doses 1 and 2, and do not start dose 3. 
    \item[\textbf{Case (iii)}] \textbf{$p_{1,1}^{(1)}>\alpha_1$ and $p_{2,1}^{(1)}<\alpha_1$:} Drop dose 1, continue with dose 2 and start dose 3. 
    \item[\textbf{Case (iv)}] \textbf{$p_{1,1}^{(1)}>\alpha_1$ and $p_{2,1}^{(1)}>\alpha_1$:} Drop doses 1 and 2, and start dose 3. 
\end{enumerate}

Note that the above rules to select the doses for the second stage are not binding. In particular, when deciding which dosing regimens should be continued, safety aspects can also be taken into account. Besides the selection of the doses, we assume that the per arm sample sizes in the second stage are reassessed to $N_2/|K|$, where $|K|$ denotes the cardinality of $K$.

\section{Simulation study} 
\label{sect_sims}

We performed a simulation study to optimise elements of the study design such as the futility threshold and the first stage sample size, as well as compare the performance of the analysis approaches. The simulation study follows the framework of clinical scenario evaluation \citep{benda2010aspects}. Hence a range of assumptions regarding the distribution of the data and design elements regarding the clinical trial are defined and compared for several operating characteristics.

We simulated data from the two-stage trial as described above, with two active doses and a placebo arm in stage 1, and one or two active dose(s) and a placebo arm in stage 2 based on the adaptive selection described in Section \ref{sec:interim}. In each stage and for each arm, we simulated logarithmic mf values at baseline, Month 6, and Month 12 based on a linear model for each patient. These values are obtained from a multivariate normal distribution with means $\mu_{j,0},\mu_{j,1}, \mu_{j,2}$ at baseline and FUs 1 and 2 for dose $j$ and covariance matrix $\Sigma$ for the multivariate normal distribution. An auto-regressive model of order 1 (AR(1)) is assumed for modeling the correlation over time and setting the correlation between FUs 0 and 1 or FUs 1 and 2 to $\rho$. Details on the distribution can be found in the Supplementary Material.

For a positive fraction of total responder, the FU values for some randomly selected patients are set to 0 resulting in a mixture distribution with an atom at zero (total responder rate, with probability $\pi$) and a component of a lognormal distribution (with probability $(1-\pi)$).

\subsection{Assumptions and design parameters}

Separate simulation studies were performed for each of the three diseases. The parameters classified as design choice or assumption of the simulations are the same for the three diseases and they are described in Table \ref{Tab:Parameters}. For some parameters, several values were considered in the simulation study; if the parameter is not varied in a simulation scenario, it is set to the standard value (highlighted in bold in Table \ref{Tab:Parameters}). The baseline parameters $\mu$ and $\sigma$ are disease specific and are shown in Table 2 in the Supplementary Material 
(baseline parameters $\mu$ and $\sigma$ and modified parameters $\mu^*$ and $\sigma^*$). For the assumed reduction rates in \% for each dose at Month 6 and Month 12 standard and modified scenarios are defined in Table \ref{Tab:scen_adaptive}.

\begin{table}[!p]
\caption[Parameters values for the simulation]{Parameters required to be specified for the simulation study classified as design choice or assumptions and selected values for the simulation studies. Standard values for simulations where the respective parameters were not varied, are given in bold.}
\begin{tabular}{p{2.4cm}p{1.8cm}p{2.2cm}p{6.2cm}}
\hline
Name&Type&Values&Description\\
\hline
number of arms & Design & 4 & Total number of arms (active and placebo doses)\\  \hline
N & Design & 200 & Total sample size per substudy \\ \hline
$N_1$ & Design & 80, 100, \textbf{120} & Total sample size of first stage, equally distributed among placebo, low and medium dose for each substudy\\ \hline
$\alpha_1$ & Design & 0.1, 0.2, \textbf{0.3}, 0.4, 0.5 & Selection threshold for the interim analysis\\ \hline
$\alpha$ & Design & 0.025 \newline (one-sided) & FWER dedicated for each disease. 
\\ \hline
test \newline (final analysis)& Design & \texttt{lm} 
\newline \texttt{WilcoxCC} \newline \textbf{\texttt{WilcoxC}} & Methods considered for the analysis of the primary endpoint: linear model adjusting for baseline data and Wilcoxon tests (for changes and FU values)\\ \hline
test \newline (selection)& Design & t-test  \newline \texttt{WilcoxCC} \newline \textbf{\texttt{WilcoxC}} & Methods considered for the interim analysis is in accordance with the final analysis; If the final analysis is based on the linear model, selection is based on t-tests.\\ \hline
$\mu$ & Assumption & see Table 2 in Supplementary Material & Mean value of baseline data based on raw microfilaria values. Specific values for each disease\\ \hline
$\sigma$ & Assumption & see Table 2 in Supplementary Material & Standard deviation of baseline data based on raw microfilaria values. Specific values for each disease\\ \hline
$\rho$ & Assumption & 0.4, \textbf{0.5}, 0.6 & Correlation between baseline and Month 6 data as well as Month 6 and Month 12 data \\ \hline
$r_{j,1}$ \newline $j=0,1,2,3$ & Assumption & see Table \ref{Tab:scen_adaptive}  
& Reduction rates in \% for each dose at Month 6 ($t=1$)\\ \hline
$r_{j,2}$ \newline $j=0,1,2,3$ & Assumption & see Table \ref{Tab:scen_adaptive}& Reduction rates in \% for each dose at Month 12 ($t=2$)\\ \hline
$\pi_j$ \newline $j=0,1,2,3$ & Assumption & see Table \ref{Tab:scen_adaptive} & Total responder rates: Proportion of patients with FU value of zeros for each dose. \\ \hline
\label{Tab:Parameters}
\end{tabular}
\end{table}

\begin{table}\centering\caption[Sims]{Illustration of simulations S1-S5. For each simulation either a design parameter and/or an assumption was varied. All simulations were performed for the analyses methods \texttt{lm}, \texttt{WilcoxCC} and \texttt{WilcoxC}.}
\begin{tabular}{cr|cccc}
\hline
&&\multicolumn{3}{c}{Design}\\ 
&& varying $\alpha_1$ & $\alpha_1=0.3$ & $\alpha_1=0.3$ \\
&& $N_1=120$ & varying $N_1$ & $N_1=120$\\
\hline Assumptions &Standard values and scenarios & S1 & S5\\
&modified reduction rates & S2 \\
&modified $\mu$ and $\sigma$ & S3\\
&varying $\rho$ &&& S4\\
\end{tabular}\label{Tab:DesignAssumption}
\end{table}

Simulations S1-S5 were performed for all 3 diseases, the varying design parameters or assumptions are described in Table \ref{Tab:DesignAssumption}. The simulations were repeated for each scenario for the three analyses methods linear model (\texttt{lm}), Wilcoxon change (\texttt{WilcoxCC}) and Wilcoxon (\texttt{WilcoxC}) (a total of 50000 runs for each scenario).

\begin{table}\centering\caption[Standard scenarios of reduction rates]{Scenarios of reduction rates in \% (low dose/medium dose/high dose). The proportion of total responders at Month 6 and 12 in each setting is assumed to be 20 percentage points less than the reduction rate at Month 12. Throughout we assumed that also in the control group there is a proportion of 10\% total responders at Months 6 and  12 but no change in the lognormal component (reduction rate = 0). The modified and the standard scenario only differ in the assumptions for the rates at Month 6 (marked by bold numbers).}
\begin{tabular}{cl|ll|lll}
\hline
Scenario&&\multicolumn{3}{c}{Reduction rates}\\ 
&&\multicolumn{2}{c}{Standard scenario} &\multicolumn{2}{c}{Modified scenario}\\ 
Number &&Month 6 & Month 12 &Month 6 & Month 12\\ \hline
1&No effect & 0/0/0& 0/0/0 &0/0/0&0/0/0\\ 2&Efficacy only in high dose & 0/0/50 &0/0/60 &0/0/\textbf{40} &0/0/60\\
3& Trend (a) &0/30/50 &0/40/60 & 0/\textbf{20/40}&0/40/60\\
4& Trend (b) &0/40/50 &0/50/60 & 0/\textbf{30/40}&0/50/60\\
5&All doses effective  &40/40/40 &50/50/50 & \textbf{30/30/30}&50/50/50\\
\hline
\end{tabular}\label{Tab:scen_adaptive}
\end{table}

\subsection{Single-stage multi-armed trials MA1 and MA2}
\paragraph{Multi-armed trial 1 (MA1): } For comparison with the adaptive design, multi-armed fixed sample trials were considered in the simulations. In MA1, two multi-armed trials are imitated, the low and medium dose are investigated independently from the high dose (separate placebo group each). 
The high dose is only tested if neither the low nor the medium dose is significant. A detailed description can be found in the Supplementary Material, Section 4. 
This design is anticipated to be more efficient, as it does not rely on an early endpoint. However, it will require more time as the high dose can only be started if the low dose does not show an effect.

\paragraph{MA2:} A second multi-armed trial with a fixed sample design is considered with equal sample sizes for each of the three active doses and a shared placebo arm. In this design, all comparisons are conducted simultaneously. This approach is expected to have greater power than MA1 due to larger group-wise sample sizes. However, this trial design is included solely as a benchmark, as it is deemed ethically unsuitable (see Supplementary Material, Section 4, for more details).

\subsection{Operating characteristics}

The different trial designs are evaluated with regard to a range of operating characteristics. To take into account the special features of the trial design, we consider operating characteristics based on the pairwise comparisons and trial-level operating characteristics, e.g., as marginal, disjunctive or conditional power (see Table 3 in the Supplementary Material).

For simulations analyzed with the \texttt{WilcoxC} method we assessed the summary measure concordance (\citealp{newcombe2006confidence1}, \citealp{konietschke2012rank}). In adaptive clinical trials, point estimates are often biased because they do not fully account for potential and realized trial modifications - such as changes in sample sizes, early stopping, or dropping/adding treatment arms - based on interim data \citep{bauer2010selection,brannath2006estimation}. Bias in the simulations was quantified as the difference between the true value of the single stage design and the mean value derived from the simulations. Since the true value is unknown, it was approximated using a Monte Carlo simulation of multi-arm trial MA2 under the same parameters and scenarios as in simulation study with sample sizes scaled by a factor of 5000. 

The overall concordance and confidence intervals (CIs) from the simulated data across both stages for each dose were estimated using the unconditional, conditional and inverse normal method. The approaches are defined in the Supplementary Material.

\section{Results}
\label{sec_results}
The resulting power values for simulations S1, S3, and S4 for mansonellosis are displayed in Figs. (\ref{fig:MansPowSim1})-(\ref{fig:MansPowSim5}). Power values for S2, S5 and error rates, selection and conditional probabilities for mansonellosis as well as results for onchocerciasis and loiasis are displayed in the Supplementary Material. Below we give a general summary of the findings, which apply to a large extent across the disease-specific simulation studies.

\subsection{Impact of design parameters on operating characteristics}
\paragraph{Futility boundary $\alpha_1$:} A high value of $\alpha_1$ increases the probability of selecting the low or medium dose in the interim analysis, thereby reducing the likelihood of starting the high dose in stage 2. Consequently, for larger $\alpha_1$, higher power values are observed for the low and medium doses, while the power for the high dose decreases (see, e.g, Figs. \ref{fig:MansPowSim1}-\ref{fig:MansPowSim3}, black lines). In scenarios 2 and 3 ("Efficacy only in high dose" and "Trend a"), the disjunctive power decreases substantially as $\alpha_1$ increases. For instance, in scenario 2, the reduction in disjunctive power exceeds 0.4 across all analysis methods while in scenario 3, it is approximately 0.1 for $\alpha_1=0.1$ versus $\alpha_1=0.5$.
\paragraph{Total first stage sample size $N_1$:} With a fixed total sample size $N$ for the trial, an increase in the first stage sample size $N_1$ results in a smaller $N_2$. Consequently, higher power is observed for the low and medium doses in stage 1 while power for the high dose decreases (Fig. \ref{fig:MansPowSim5}). The disjunctive power of the adaptive design decreases as $N_1$ increases, although the effect of $N_1$ on power is less pronounced for the proposed values than that of $\alpha_1$. E.g. for simulation S5 in Fig. \ref{fig:MansPowSim5}, scenarios 2 and 3 ("Efficacy only in high dose" and "Trend a") exhibit a reduction in disjunctive power of approximately 0.1 as $N_1$ increases.

\paragraph{Analysis method:} Comparing analysis methods \texttt{lm}, \texttt{WilcoxC} and \texttt{WilcoxCC} for the adaptive design reveals no uniformly best test across the considered scenarios. In many cases, \texttt{WilcoxC} has the highest power while \texttt{lm} has the lowest (e.g., Simulation S1 in Fig. \ref{fig:MansPowSim1}) with power differences reaching up to 20\% points in scenario 3 for the medium dose for $\alpha_1=0.5$. These differences are even more pronounced for loiasis, while for onchocerciasis (Figs. 33 and 18 in Supplementary Material), the power differences are negligible. However, in some scenarios \texttt{lm} achieves higher power than both \texttt{WilcoxC} and the \texttt{WilcoxCC} (e.g, simulation S3 in Fig. \ref{fig:MansPowSim3} with modified baseline parameters $\mu$ and $\sigma$ where differences reach up to 0.05\% points for the two lower doses). Across all simulations \texttt{WilcoxCC} did not emerge as the most powerful analysis method in any scenario.

\paragraph{Control of the error rate:} For the scenarios under the global null hypothesis (scenario "No effect") the FWER is controlled at level $\alpha$ (disjunctive power) with only occasional excesses. Only for \texttt{lm} the FWER is inflated but still lower than 0.03 (see Fig. 4 in the Supplementary Material). 

\subsection{Influence of assumptions on operating characteristics}
\begin{figure}[!p]
    \centering     \includegraphics[width=1\linewidth,page=1]{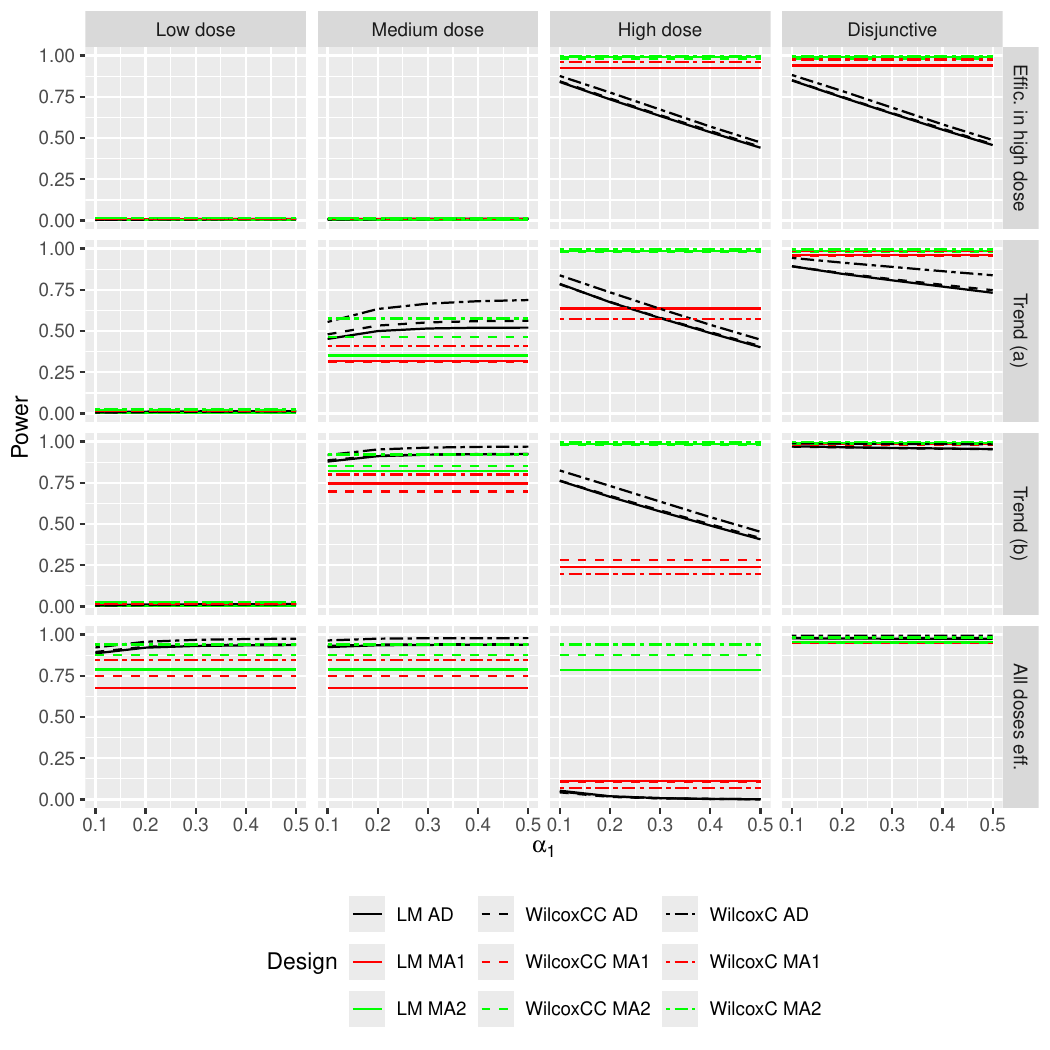}
    \caption[Plot 1]{Mansonellosis S1: Simulated power values for the low, medium, and high doses and the disjunctive power are depicted (in the columns) for simulation S1 (standard reduction rates) as a function of $\alpha_1=\{0.1,0.2,0.3,0.4,0.5\}$ for several scenarios of reduction rates (in the rows).}
    \label{fig:MansPowSim1}
\end{figure}

\begin{figure}[!p]
    \centering     \includegraphics[width=1\linewidth,page=3]{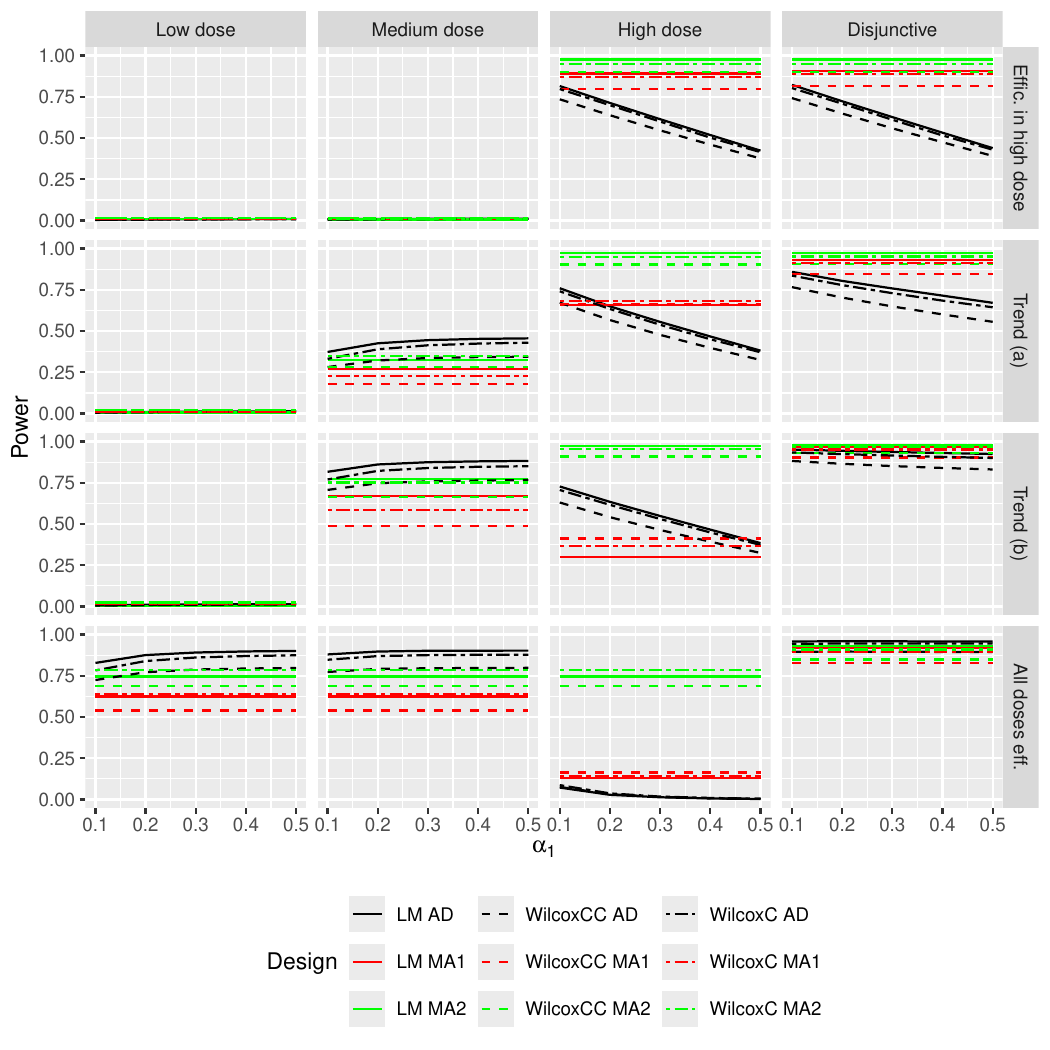}
    \caption[Plot 1]{Mansonellosis S3: Simulated power values for the low, medium, and high doses and the disjunctive power are depicted (in the columns) for simulation S3 (modified baseline parameters) as a function of $\alpha_1=\{0.1,0.2,0.3,0.4,0.5\}$ for several scenarios of reduction rates (in the rows).}
    \label{fig:MansPowSim3}
\end{figure}

\begin{figure}[!p]
    \centering     \includegraphics[width=1\linewidth,page=5]{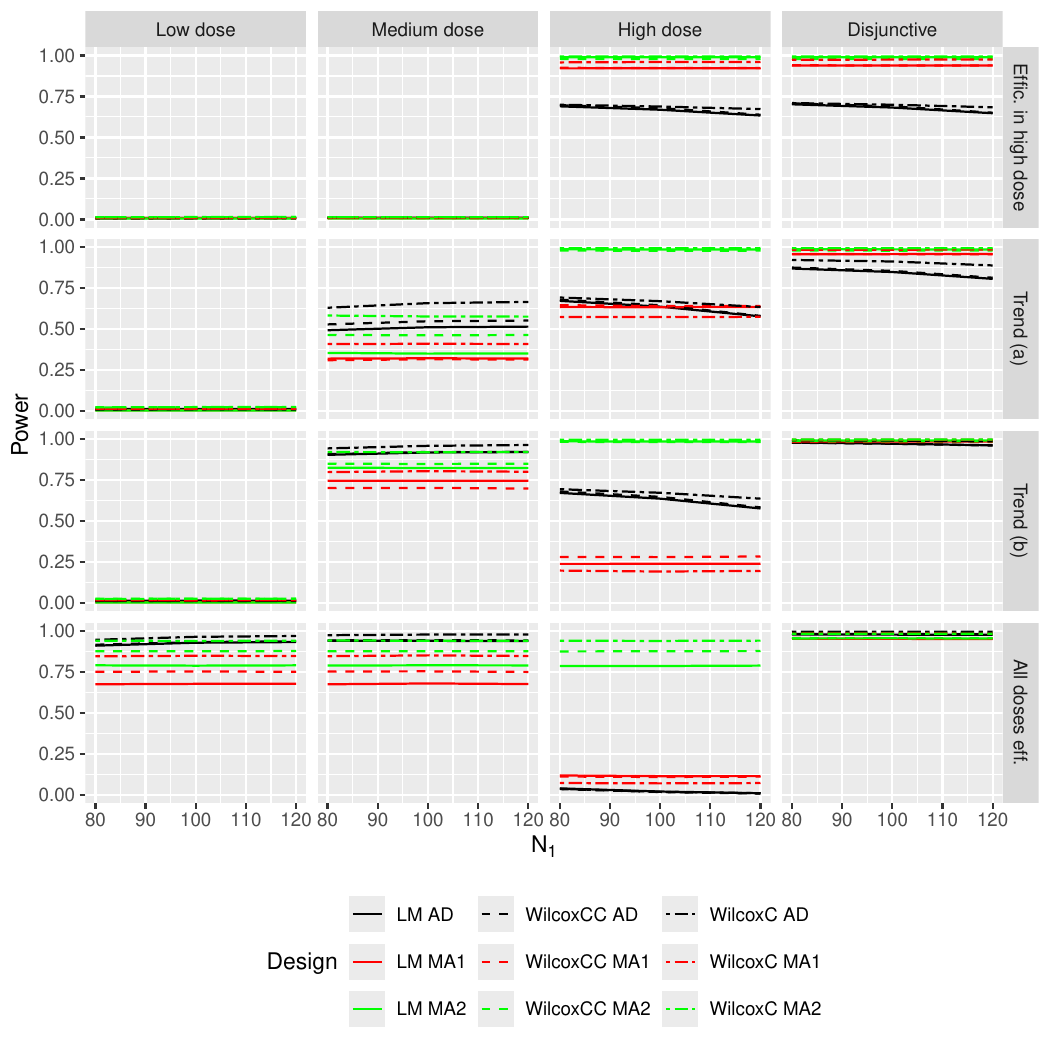}
    \caption[Plot 1]{Mansonellosis S5: Simulated power values for the low, medium, and high doses and the disjunctive power are depicted (in the columns) for simulation S1 as a function of $N_1=\{80,100,120\}$ for several scenarios of reduction rates (in the rows).}
    \label{fig:MansPowSim5}
\end{figure}

\paragraph{Reduction rates:} The effect sizes of the active doses compared to placebo are defined as a relative reduction of the baseline values. The influence of the reduction rates is shown twice: First as outlined in Table \ref{Tab:scen_adaptive}, each simulation S1-S5 consists of 5 standard scenarios defined by different reduction rates of the doses. As shown, for example in Fig. \ref{fig:MansPowSim1}, higher reduction rates lead to higher power: A difference in the reduction rate of 10\% points between scenarios 3 and 4 (Trend (a) and Trend (b)) for the medium dose has a high impact on marginal power of the medium dose and the disjunctive power. The latter jumps from 0.81 to 0.99 for \texttt{WilcoxC} and $\alpha_1=0.5$. 
Second, the 5 standard scenarios are slightly modified for simulation S2 (see Table \ref{Tab:scen_adaptive}): The reduction rates for Month 6 are decreased by 10 percentage points, whereas the reduction rates for Month 12 are not changed. The comparison of Simulations S1 and S2 shows that smaller reduction rates for Month 6 lead to slightly reduced power values for low and medium doses for small $\alpha_1$. For $\alpha_1=0.5$ there is hardly any difference.

\paragraph{Correlation $\rho$ between follow-ups:} For increasing $\rho$, slightly higher power values for the low and medium doses and lower power values for the high dose can be observed (see Fig. 3 in the Supplementary Material).

\paragraph{Baseline values $\mu$ and $\sigma$} In simulation S3 modified baseline values with higher variances and lower mean values are considered to mimic a worst case scenario (see Fig. \ref{fig:MansPowSim3}). Slightly different results are found compared to the standard baseline values (see Fig. \ref{fig:MansPowSim1}): The resulting power is smaller and for fmost scenarios \texttt{lm} is slightly more powerful than \texttt{WilcoxC} for the adaptive design.

\subsection{Evaluation of adaptive study design}

\paragraph{Comparison with single stage designs} In all scenarios MA2 has greater or equal power compared to MA1 attributed to the shared placebo arm. The comparison of MA2 with the adaptive design depends on the parameters: For the high dose, MA2 is always more powerful. For the low and medium doses, the adaptive design is always more powerful than MA2 for the \texttt{WilcoxC} method when $\alpha_1 \geq 0.3$. However, for lower $\alpha_1$, this advantage is not guaranteed. For example in Fig. \ref{fig:MansPowSim1}, scenario 3 ("Trend a"), the medium dose shows slightly lower power in the adaptive design compared to MA2 when $\alpha_1=0.1$. This difference is more pronounced in the same scenario in  Fig. 6 in the Supplemental material for simulation S2. The disjunctive power is higher for MA2 in scenarios "Efficacy only in high dose" and "Trend a", for the other scenarios, power approaches the value of 1 with only minor deviations in the adaptive design.

\subsection{Bias of concordance in adaptive design}
Table \ref{Tab:Bias} summarizes the average true concordance, estimation bias, CI limit, and coverage of the CI for the conditional inverse normal method for Simulation S1 and $\alpha_1=0.3$ (detailed results for the unconditional and conditional estimation can be found in the Supplementary Material). For the conditional and the inverse normal method, a positive bias was observed for the low and the medium doses when these doses were continued to the second stage, suggesting that concordance was overestimated in the simulations. The coverage of the CI for the inverse normal method consistently equaled or exceeded 97.5\% for all scenarios considered in Table \ref{Tab:Bias}. For the unconditional method, a small but negative bias was observed for the low and the medium dose across all scenarios except when all doses are effective, indicating that the true concordance slightly exceeds the simulated mean concordance.

\begin{table}
\footnotesize
\centering\caption[Concordance, bias standard scenario]{The simulated true concordance values (C) of each scenario for the three doses are reported with  bias (mean C - true C), lower CI limit (mean value over simulation runs), and coverage of the CI (in \%) for inverse normal method for Simulation S1 (standard scenario) for $\alpha_1=0.3$ and analysis method \texttt{WilcoxC}. A total of 50000 simulation runs was performed for each scenario.  
}
\begin{tabular}{l|ccc}
\hline
Scenario &Low dose& Medium dose& High dose\\ \hline
& True concordance  & True concordance  & True concordance \\
& Bias/CI/Coverage& Bias/CI/Coverage &Bias/CI/Coverage \\
\hline
No effect                   &0.5  &0.5    &0.50\\ &0.028, 0.43, 98.1 & 0.016, 0.42, 97.9  & 0, 0.35, 98.9\\
Efficacy only in high dose  &0.5  &0.5   &0.77 \\ & 0.028, 0.43, 98.1&0.016, 0.42, 97.9& 0.001, 0.61, 99.4\\
Trend (a)                   &0.5  &0.64  &0.77 \\ &0.028, 0.43, 98.1&0.004, 0.55, 97.5&0.001, 0.61, 98.6\\
Trend (b)                   &0.5  &0.71  &0.77 \\ &0.028, 0.43, 98.1&0.000, 0.61, 97.6&0.001, 0.61, 98.5\\
All doses effective         &0.71  &0.71   &0.71 \\ &0.002, 0.61, 97.6&0.001, 0.61, 97.5&-0.005, 0.54, 98.7\\
\hline
\end{tabular}\label{Tab:Bias}
\end{table}

\subsection{Summary and recommendations}
Based on the presented simulation study and external considerations, the trial was planned to be conducted and started in May 2025 with a first-stage total sample size of $N_1 = 120$ and a second-stage total sample size of $N_2 = 80$. A selection threshold $\alpha_1 = 0.3$ applied to the p-values of the Month 6 mf load comparisons in the interim analysis for dose selection was chosen for the three diseases. Even though a higher disjunctive power might have been observed for a larger $\alpha_1$ and $N_1$, the decision was made against them as the principal investigators favored to find effects in lower doses, high power values for the low and medium dose have thus higher priority than for the high dose. No clearly uniformly best analysis method was determined, however, assumptions from the simulation study might not be met in the trial (e.g., normal distribution of logarithmic data) and thus the trial was planned with the more robust Wilcoxon tests (\texttt{WilxcoxC}) for parasite load at Month 12.

\section{Data example} 
\label{sect_example}

Since no real data from the eWHORM trial is available yet, we present a hypothetical examples to illustrate the procedure. Based on simulation S1, one simulation run for the scenarios "Trend" and "All doses effective" is presented in Table \ref{Tab:Scen}, histograms of the data can be found in the Supplemental Material (Figs. 2-5). 
For the scenario "All doses effective", $\alpha_1=0.3$ leads to continuation of both the low and the medium dose  whereas for the simulation scenario "Trend", only the medium dose is continued to the second stage and the high dose is started. The complete closed test with all corresponding intersection and elementary hypotheses for both scenarios with their specific partial conditional error rates can be found in the Supplemental Material. For both scenarios, both doses in the second stage are rejected after stage 2. Table \ref{Tab:Scen} shows the concordance values and the lower limit of the 97.5\% confidence limit of the doses started in stage 2.

\begin{table}
\small \centering
\caption{Data example from one simulation run for scenarios "Trend" and "All doses effective" showing surrogate, first and second stage p-values for low, medium and high dose for $\alpha_1=0.3$. The observed concordance values $\Delta_{j,2}$ for doses j=1,2,3 at Month 12 are reported with the lower limit of the one-sided 97.5\% CI.\label{tab1}
}
\begin{tabular}{p{2.75cm} p{0.5cm} p{0.5cm}p{0.5cm}p{0.5cm}p{0.5cm} p{0.6cm}p{0.6cm}p{1.5cm} p{1.5cm}p{1.5cm}}
\hline
Simulation scenario &$p_{1,1}^{(1)}$ & $p_{2,1}^{(1)}$ &$p_{1,2}^{(1)}$ & $p_{2,2}^{(1)}$ &$p_{1,2}^{(2)}$ & $p_{2,2}^{(2)}$ &$p_{3,2}^{(2)}$ & $\Delta_{1,2}$ (CI) & $\Delta_{2,2} $(CI) & $\Delta_{3,2}$ (CI) \\\hline
Trend & 0.391 &0.047 &0.692 &0.057 &- &0.015 &0.000&    - &0.64 (0.54) & 0.87  (0.73)\\
All doses effective &
0.002 &0.005& 0.009& 0.001& 0.045& 0.014& -& 0.65 (0.55) & 0.70 (0.60)   & - \\
\hline \label{Tab:Scen}
\end{tabular}
\end{table}

\section{Discussion}
\label{sec_discussion}

For the implementation of an adaptive basket trial design for infections in the eWHORM project (”eWHORM – enabling the WHO Road Map”) extensive simulations were performed to support decisions for an optimal trial design. To control the FWER, adjusting for the comparison of multiple doses to placebo in each disease, the partial conditional error approach was chosen and extended to allow for the inclusion of new hypotheses after an interim analysis. If the high dose is not started, the initially foreseen local level is not lost for the final test decision of the observed doses as it is reallocated by use of adaptive partial conditional error rates. 

A large number of simulations were performed to assess operating characteristics. The simulation of the observation was based on the assumption of a normal distribution of the logarithmic observations and a proportion of total responder with a FU value of zero. 
The simulations showed that a non-parametric Wilcoxon test (\texttt{WilcoxC}) comparing parasite load at Month 12 resulted in high power values to identify effective doses. In many scenarios, the linear model showed similar or even higher power values, however, as the assumptions in the presented data simulation might deviate from real data as a result the linear model might not be appropriate. These assumptions are, e.g., same variances for each group and each FU or no normal distribution of logarithmic values of patients who are not total responders. The impact of a scenario with different variances for each group and each FU on the analysis strategy or skewed distribution of the logarithmic data on the operating characteristics has not been investigated.


To account for the total responders at FU the use of alternative analysis strategies is subject to further research. Possible strategies are two-part models for zero inflated data as described in \citet{gleiss2015two} or the use of zero-inflated and two-part mixed effects model with R-package GLMMadaptive \citep{rizopoulos2022package}. The latter one has the advantage that the analysis of longitudinal data considers early FUs and thus missing values have less impact.

This study is designed as basket trial with four substudies. One of the major advantages of a basket trial is to share information. In the eWHORM trial safety data of different doses between the 4 diseases mansonelliosis, loiasis, onchocerciasis, and trichuriasis is shared, no borrowing of data for the analyses between substudies will be considered for the final primary analysis. However, for safety assessments and dose selection during interim analyses, (interim) data from all diseases may be utilized. In further exploratory analysis a Bayesian analysis for borrowing of information across diseases with special focus on co-infections is subject to further research. 

Clinical trial simulations are powerful tools, but they come with several limitations due to unknown parameters and probably simplified assumptions. For this specific trial, assumptions, e.g, for baseline parameters, correlation between surrogate and primary endpoint and assumed effect sizes of surrogate and primary endpoint, and the total responder rates were made. Thus in addition to standard scenarios, a large number of modified parameters were observed and their impact on operating characteristics was investigated. These simulation results, however, depend a lot on the correlation between surrogate endpoint and Month 6 and true endpoint at Month 12. In the simulations, we assume a correlation between 0.4 and 0.6, if, however, the true correlation is very low, the Month 6 observations are no appropriate surrogate endpoints. The error can be in two directions: Either the effects of the low and the medium dose at Month 6  are higher than for Month 12, then the trial might end without significant results; Or the effects of the low and the medium dose are lower then there is a higher chance for the high dose to start in the stage 2.

\section{Software}
\label{sec5}
R programs implementing the methods and code to reproduce the results of this paper are available through the repository: \url{https://github.com/MartaBofillRoig/ewhorm_sim/}

\section{Supplementary Material}
\label{sec6}

Supplementary Material is available below. 
It includes a detailed summary of the trial and additional material for the adapted second stage test, disease specific baseline parameters for the simulation study, for the estimation of baseline mf values in the different diseases, specification of the multi-armed trials. In addition the operating characteristics are defined and more results on the bias are presented. Numerical examples are shown. Finally Figures for simulations S1-S5 are depicted for all diseases.

\section*{Acknowledgments}

The eWHORM project (No 101103053) supported by the Global Health EDCTP3 Joint Undertaking and its members as well as the Swiss Confederation. Views and opinions expressed are those of the author(s) only and do not necessarily reflect those of the European Union or the Global Health European and Developing Countries Clinical Trials Partnership (EDCTP3). Neither the European Union nor the granting authority can be held responsible for them

M. Bofill Roig is a Serra Húnter Fellow and was additionally supported by Grant PID2023-148033OB-C21 funded by MICIU/AEI/10.13039/501100011033 and by FEDER/UE.

The authors are grateful to the eWHORM investigators not included as authors for their work: Karen Dequatre Cheeseman, Jennifer Keiser, Ghyslain Momba Ngoma, Dieudonne Mumba Ngoyi, Michael Ramharter, Samuel Wanji, Achim Hoerauf, Benjamin Lenz, Annina Schnoz.

{\it Conflict of Interest}: None declared.

\bibliographystyle{apalike}
\bibliography{refs}

\newpage

\appendix
\renewcommand\thefigure{\thesection.\arabic{figure}} 
\setcounter{table}{0}
\renewcommand{\thetable}{A\arabic{table}}

\section{Supplemental Material}




\section*{Abbreviations} 
\addcontentsline{toc}{section}{\protect\numberline{}Abbreviations}

\begin{table}[!h] 
\begin{tabular}{p{3cm}p{12cm}}
c & Wilcoxon test of follow-up values (abbreviation in Figures)\\
cc& Wilcoxon test of differences (abbreviation in Figures)\\FWER & familywise error rate\\
lm & linear model (abbreviation in Figures)\\
MA & Multi-armed\\
mf & microfilaria\\
OXF & Oxfendazole\\
\end{tabular} \label{Tab:Abbrev}
\end{table}

\newpage

\pagenumbering{arabic}

\section{Adapted second stage test}
\begin{figure}[!p]
\centering
\begin{tabular}{c|c}
\includegraphics[width=0.48\textwidth,page=1]{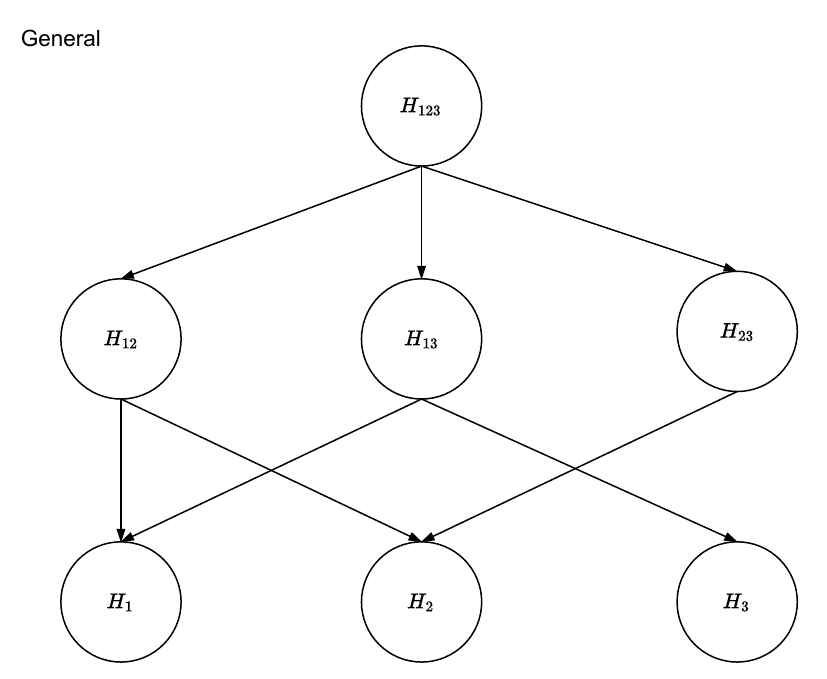}&
\includegraphics[width=0.48\textwidth,page=2]{./fig_closedtesting2}\\ \hline
\includegraphics[width=0.48\textwidth,page=3]{./fig_closedtesting2}&
\includegraphics[width=0.48\textwidth,page=4]{./fig_closedtesting2}
\end{tabular}
\caption[Diagram of the hypotheses tested]{Diagram of the hypotheses tested in each scenario depending on the selection at the interim. The diagram on the top-left shows the full closure of hypotheses. Hypotheses in grey nodes refer to those tested in the final analysis; dashed nodes refer to hypotheses which are accepted at the interim analysis. Arrows illustrate the intersection hypotheses contained in the subsequent hypotheses.}
\label{fig:closedtesting}
\end{figure}

Fig. \ref{fig:closedtesting} describes the hypotheses to be tested depending on the selection in the interim analysis.
\begin{table}[h]
    \centering
    \caption[Cases based on the selection rule and hypotheses to consider]{The intersection hypotheses to be tested in the final analysis depending on the doses selected in the different cases described in Section 4.3 in the main manuscript. All other (intersection) hypotheses are accepted already in the interim analysis.}
    \begin{tabular}{p{1cm}p{5cm}p{7.6cm}}
     \hline
        Case & Selection rule  & Selection scenario \\ \hline
        (i) & Continue with doses 1 and 2, and do not start dose 3   & Selection scenario 1 in Figure \ref{fig:closedtesting}.\\ \hline
        (ii)
        & Continue with doses 1 and 2, and do not start dose 3.  &  Selection scenario 1 in Figure \ref{fig:closedtesting}.  \\ 
        \hline
        (iii) & Drop dose 1, continue with dose 2 and start dose 3.   & Selection scenario 3 in Figure \ref{fig:closedtesting}.  \\ \hline
        (iv) & Drop doses 1 and 2, and start dose 3.   &   Selection scenario 2 in Figure \ref{fig:closedtesting}.  \\ \hline
    \end{tabular} 
\label{tab:closedtesting}
\end{table}

\newpage

\section{Disease specific baseline parameters for simulation study}

For the simulation study assumptions on the mean value $\mu$ and standard deviation $\sigma$ of the baseline data of microfilaria values are needed. These parameters should be based on real data sets and can either be calculated from the arithmetic means or the geometric means. 

Table \ref{Tab:Baselineparameters} shows the considered baseline parameters $\mu$ and $\sigma$ for the simulation study for each disease and the source where the values have been collected (literature citation or internal data). Note that for the literature sources of loiasis the inclusion criteria for the primary endpoint are not completely in agreement with our study. In eWHORM for loiasis the inclusion criterium is baseline mf between 100 and 8000 compared to 1 and 15000 in \citet{campillo2022safety}, thus for the simulations the value of $\mu$ from \citet{campillo2022safety} was set to 4000 (instead of 5000) and for $\sigma$ the published was value chosen, even though a smaller value can be expected for the given inclusion criteria. This implies that the simulation study gives a lower bound for the statistical power. To account for deviations in $\mu$ and $\sigma$ in a further simulation we considered modified parameters $\mu^*$ and $\sigma^*$ (see Table \ref{Tab:Baselineparameters}).

\begin{table}[htb]
\caption[Considered baseline parameters]{Considered baseline parameters for the arithmetic mean $\mu$ and standard deviation $\sigma$ of the diseases on the original scale for each disease for Simulations S1, S2, S4, and S5 are derived from internal data or the literature. In Simulation S3 the parameters are modified to $\mu^*$ and $\sigma^*$.
\newline $^a$ inclusion criteria: 1-15000}
\begin{tabular}{p{2.5cm}p{1.2cm}p{1.2cm}p{4.5cm}|p{1.2cm}p{1.2cm}}
\hline
Disease&$\mu$&$\sigma$& Source & $\mu^*$&$\sigma^*$\\
\hline
mansonellosis& 1838 & 2565 &  Internal data & 1000 & 3500
\\
onchocerciasis &19 & 30 & Internal data  
&15& 40\\
loiasis &5000&4000&\citet{campillo2022safety}$^a$ & 4000 & 5000\\
\hline
\end{tabular} \label{Tab:Baselineparameters}
\end{table}

\newpage
\section{Estimating the baseline microfilaria distribution in the different diseases}

\subsection{Estimation of the geometric mean from the arithmetic mean}
\label{sec:estimategeom}
The distribution of microfilaria values of patients who are not total responder tends to be skewed (see, e.g.\cite{debrah2019efficacy}), thus for the simulations we consider a logarithmic transformation of the data as in \cite{debrah2019efficacy}. In the following we derive the calculation of the parameters of the normal distribution for the simulation studies based on reported (raw) mf values in the literature.

\paragraph{Definition of values on "original" scale (log-normal distribution)}

\noindent Let $X_{j, t,i}$ denote the parasite load mf value for dose group $j$ ($j=0,1,2,3$) at time $t$ ($t=0,1,2$) for subject $i$ with arithmetic mean $\bar X_{j, t}$ and  standard deviation  $s_{j, t}$. It is assumed that the original  observations follow a log-normal distribution. At baseline, we assume equal mean and standard deviations for all groups, $\bar x_0=\bar x_{j,0}$ and $s_0=s_{j, 0}$ for all $j=0,1,2,3$.

\paragraph{Definition of values on log scale}

\noindent We assume that $Y_{j,t,i}=\log(X_{i,t,j})$ follows a normal distribution with mean $\mu_{j,t}$ and standard deviation $\sigma_{j,t}$. The parameters can be estimated as a function of $\bar x_{j, t}$ and $s^2_{j, t}$: 
\begin{equation}
    \hat \mu_{j, t} = \log(\frac{\bar x_{j, t}^2}{ \sqrt{\bar x_{j, t}^2 + s_{j, t}^2} }) \label{Eq:log_mu}
\end{equation}

\begin{equation}
    \hat \sigma_{j, t} = \sqrt{\log(1 + \frac{s_{j, t}^2}{\bar x_{j, t}^2} )} \label{Eq:log_sigma}
\end{equation}  

We assume that the true $\sigma_{j,t}=\sigma_{0,0}$ is constant for each $j$ and each $t$ and thus also $\hat \sigma_{j,t}=\hat \sigma_{0,0}$ is constant for each $j$ and $t$. For the calculation of $\hat \mu_{j,t}$, $t=1,2$ in Eq. (\ref{Eq:log_mu}), we first recalculate $s^2_{j,t}=\bar x_{j,t} \sqrt{(\exp(\hat \sigma_{j,t}^2) - 1)}$.

\paragraph{Effect size}
The reduction rates at follow-up time 1 or 2 (Month 6 or 12) are given by $r_{j,t}$. Thus $\bar x_{j,t}=(1-r_{j,t})\bar x_{j,0}$ for $t=1,2$.
The mean change of dose $j$ between baseline and follow-up time $t$ is given by $\delta_{j, t} = \log(\bar x_{j, t}) - \log(\bar x_{j, 0}) = \log(x_{j, t}/x_{j, 0})=\log(1-r_{j,t})$. 

The effect size is given by $\Delta_{j,t}=\delta_{j, t}-\delta_{0, t}=\log((1-r_{j,t})/(1-r_{0,t})$

\subsection{Distribution of observations}
The distribution of the logarithmic microfilaria values is thus given by: 
\[\begin{pmatrix}
Y_{j,0,i} \\
Y_{j,1,i}\\
Y_{j,2,i}
\end{pmatrix}\sim N\left(\begin{pmatrix}
\mu_{j,0}\\
\mu_{j,1} \\
\mu_{j,2} 

\end{pmatrix},\Sigma \right).
\]

\noindent $\Sigma$ is the variance-covariance matrix with elements $\sigma_{j,t}^2=\sigma^2_{0,0}$ on the main diagonal and $\Cov(y_{j, t},y_{j,t1})$, $t\neq t_1$ on the off-diagonals. Assuming an auto-regressive model of order 1 (AR(1)) for modeling the correlation over time and setting the correlation between time points 0 and 1 or between time points 1 and 2 to $\rho$, respectively, $\Corr(y_{j,0}, y_{j,2}) = \rho^2$ and $\Cov(y_{j, t_{k}}, y_{j, t_{k+l}}) = \rho^l \sigma_{0,0}\sigma_{0,0}$ with $t=0,1,2$, $k=1,2$, $l=1,2$ and $k+l\leq 3$ . Thus, 
$$\Sigma = \begin{pmatrix}
 \sigma^2_{0,0}  & \rho  \sigma^2_{0,0}, & \rho^2  \sigma^2_{0,0}\\

\rho \sigma^2_{0,0}  & \sigma^2_{0,0} & \rho  \sigma^2_{0,0}\\
\rho^2  \sigma^2_{0,0}  & \rho  \sigma^2_{0,0} & \sigma^2_{0,0}
\end{pmatrix}$$

To define subjects who are total responder, which means the parasite load for the FUs is set to 0, random values from the binomial distribution with probability $\pi_j$ for the doses $j=0,1,2,3$ are generated.

\newpage
\section{Specification of multi-arm trials: } 
\paragraph{Multi-arm trial 1 - MA1: } For comparison with the adaptive design, multi-armed fixed sample trials were considered in the simulations. In MA1, two multi-armed trials are imitated, the low and medium dose are investigated independently from the high dose (separate placebo group each). 
The high dose is only tested if neither the low nor the medium dose is significant. The two trials are specified as follows: 

\begin{itemize}
    \item Study 1: Low and medium doses compared to placebo: 
        \begin{itemize}
            \item 3/5 of the total sample size $N$ is equally divided among the 3 groups placebo, low and medium dose. 
            \item Each dose is compared to placebo with the chosen test at significance level $\alpha^*=2\alpha/3$ and Bonferroni-Holm adjusted p-values are calculated.
        \end{itemize}
    \item Study 2: High dose compared to placebo (only started if neither the low nor the medium dose is significant):
        \begin{itemize}
            \item 2/5 of $N$ is equally divided among the groups placebo and high dose. 
            \item The high dose is compared to placebo at significance level $\alpha^*=\alpha/3$.
        \end{itemize}
\end{itemize}

\paragraph{Multi-arm trial 2 - MA2}
In MA2 a trial with a fixed sample design is considered with equal sample sizes for each of the three active doses and a shared placebo arm. In this design all comparisons are conducted simultaneously. This approach is expected to have greater power than MA1 due to larger group-wise sample sizes. However, this trial design is included solely as a benchmark, as it deemed ethically unsuitable. Initiating the high-dose group without adequate safety data from the lower doses is not considered appropriate. The trial is specified as follows:

\begin{itemize}
    \item The total sample size $N$ is equally divided among the 4 doses. 
    \item Pairwise comparisons of the active doses to placebo are performed at significance level $\alpha$ and Bonferroni-Holm adjusted p-values are derived.
\end{itemize}

\newpage

\section{Operating characteristics}
Table \ref{Tab:Operatingch} shows the operating characteristics for the simulation study from the main manuscript

\begin{table}[htb]
\caption{Operating characteristics for simulation study}
\begin{tabular}{p{5cm}p{10cm}}
\hline
Name&Description\\
\hline
Error control & - Per-sub-study FWER (within disease)\\
& - Type I error (individual treatment-control comparison) \\
\hline
Power &- Disjunctive power: probability of finding at least one true dose effect (within disease) \\
& - (Marginal) power for individual treatment-control comparison: probability to reject a dose\\ 
& - Conditional power for individual treatment-control comparison: probability to reject a dose conditional on selecting the dose for the second stage \\ \hline
Selection probabilities& - frequency (\%) that the dose is selected for/started in stage 2 \\ \hline
Properties on the estimates & - Bias \\
& - 95\% Confidence interval \\
\hline
\end{tabular}\label{Tab:Operatingch}
\end{table}

\subsection{Estimation and simulation of bias}
In adaptive clinical trials point estimates are often biased because they do not fully account for potential and realized trial modifications - such as changes in sample sizes, early stopping, or dropping/adding treatment arms - based on interim data \citep{bauer2010selection,robertson2023point,brannath2006estimation}.
For simulations analyzed with the \texttt{WilcoxC} method we assessed the summary measure concordance \citep{newcombe2006confidence1,konietschke2012rank,brunner2018rank}. Bias in the simulations was quantified as the difference between the true value of the single stage design and the mean value derived from the simulations. Since the true value is unknown, it was approximated using a Monte Carlo simulation of multi-arm trial MA2 under the same parameters and scenarios as in simulation study with sample sizes scaled by a factor of 5000. 

The overall concordance from the simulated data across both stages for each dose was estimated using three approaches: \begin{itemize}
    \item Unconditional estimation: The overall concordance for $H_1$ and $H_2$ was calculated as the weighted mean of concordance values from stages 1 and 2 with the proportions of actual stage-wise sample sizes as weights. If a dose was observed only in one stage, the overall concordance was set to the corresponding value from that stage. 
    \item Conditional estimation: Similar to the unconditional estimation, but overall concordance was calculated only if the dose was continued after the interim analysis.
Note that for the high dose the unconditional and conditional estimates are always equivalent.

\item Conditional inverse normal method: A median-unbiased estimator of the concordance for each dose was derived using the inverse normal method \citep{lehmacher1999adaptive} with weights $w_j^{(s)}$ for $H_j$ and stage $s$ as defined for the computation of the conditional errors: 
    $$P(\delta^{(cc)}_{j,t})=1-\Phi(w_j^{(1)} Z^{(1)}(\delta^{(cc)}_{j,t})+w_j^{(2)} Z^{(2)}(\delta^{(cc)}_{j,t}))$$ where solving $P(\delta^{(cc)}_{j,t})=0.5$ yields a median-unbiased estimator with $Z^{(s)}(\delta^{(cc)})=(\hat \delta^{(s),(cc)}_{j,t}-\delta^{(s),(cc)}_{j,t})/\sigma_\delta^{(cc)})$, for stages $s=1,2$, dose $j$ and FU $t$. The standard deviation $\sigma_\delta^{(cc)}$ was estimated with the modified Hanley–McNeil approach as proposed by \citet{newcombe2006confidence2} (Method 5). If a dose was observed only in stage 1, no overall concordance was determined.
\end{itemize}

Asymptotic 97.5\% confidence intervals (one-sided) were computed for the three approaches applying the modified Hanley–McNeil approach for variance estimation \citep{newcombe2006confidence2}. Confidence bounds were derived iteratively by solving the resulting equations, e.g., for the inverse normal method this involved solving $P(\delta^{(cc)}_{j,t})=1-\alpha$.  

\newpage

\newpage
\section{Numerical example}
\begin{figure}[!p]
    \centering     \includegraphics[width=1\linewidth,page=1]{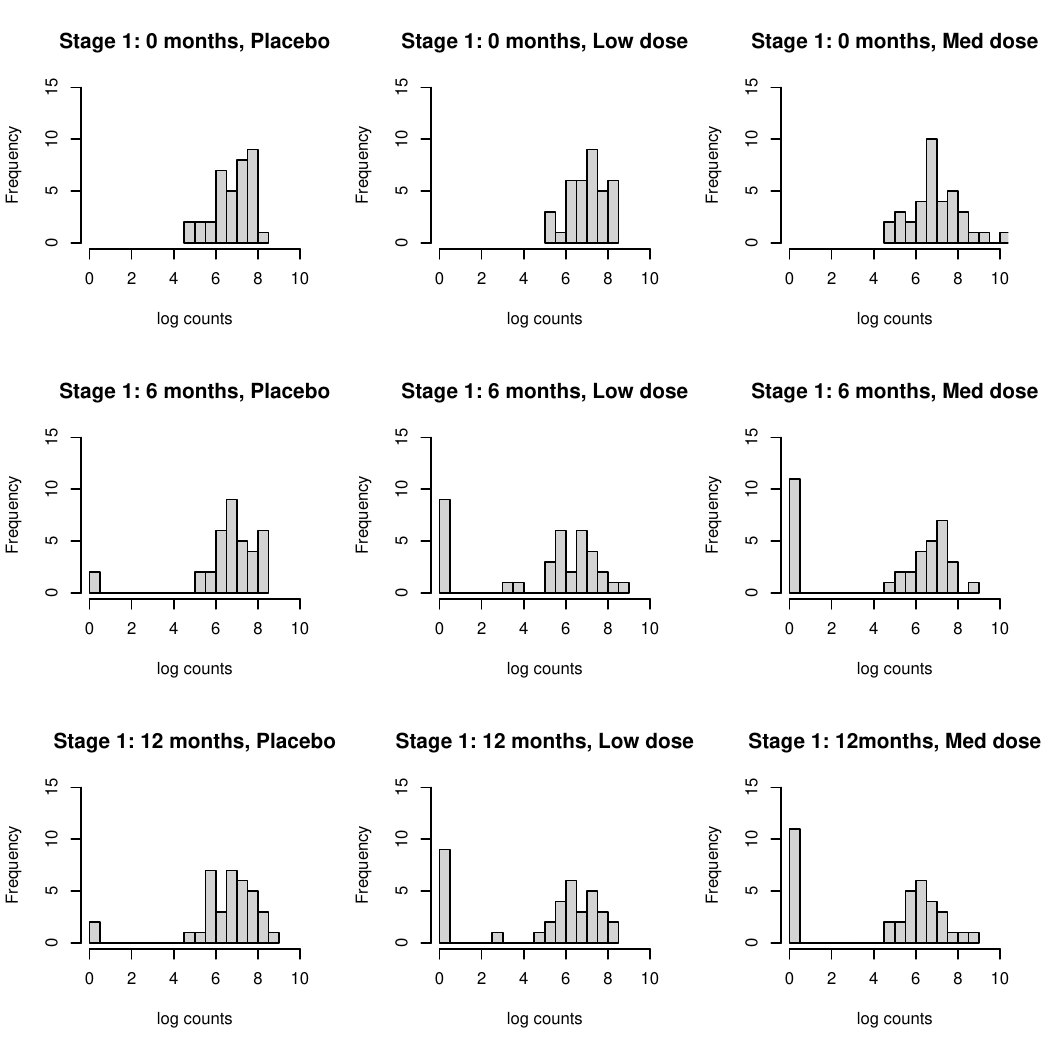}
    \caption[Numerical example - first stage data scenario "all doses effective"]{First stage data of numerical example for scenario "all doses effective"}
    \label{fig:H1}
\end{figure}

\begin{figure}[!p]
    \centering     \includegraphics[width=1\linewidth,page=2]{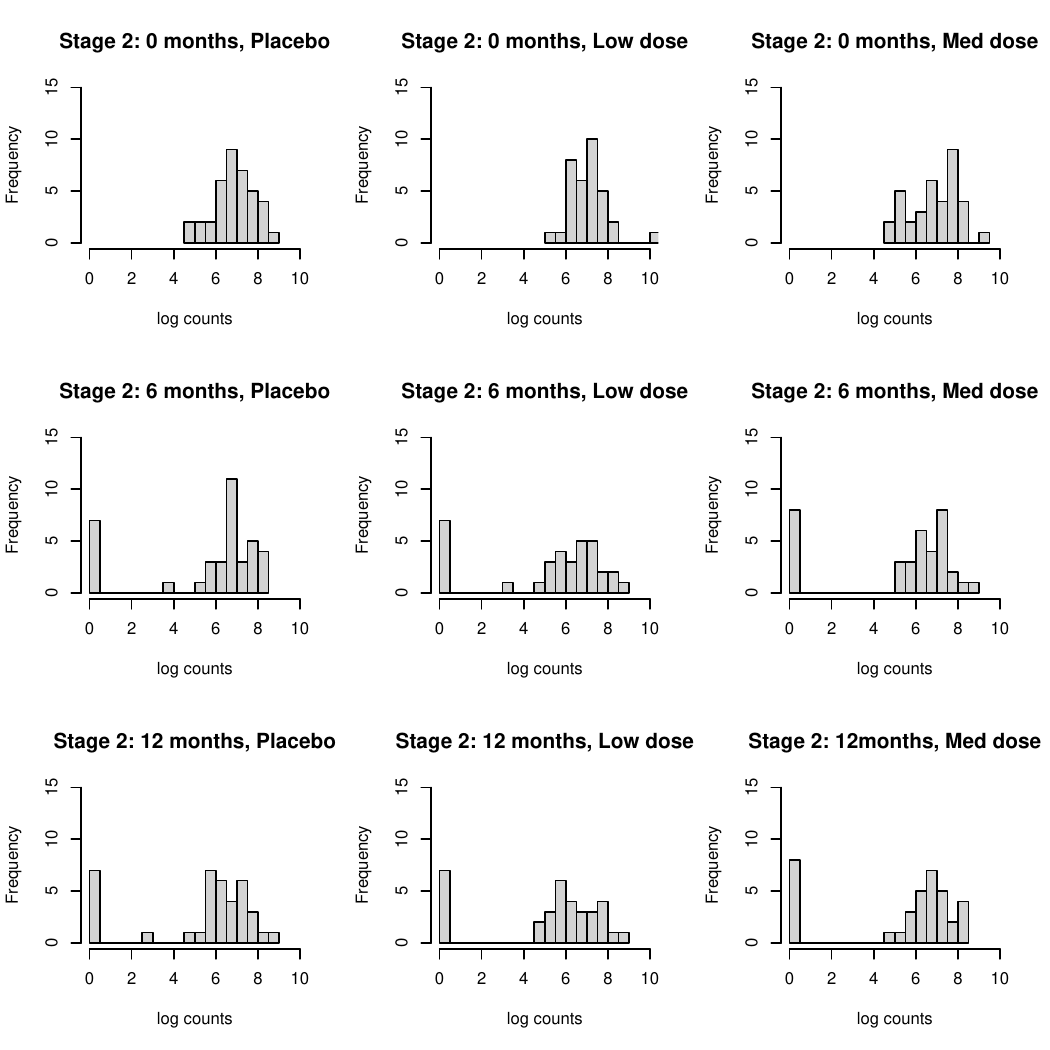}
    \caption[Numerical example - second stage data scenario "all doses effective"]{Second stage data of numerical example for scenario "all doses effective"}
    \label{fig:H2}
\end{figure}

\begin{figure}[!p]
    \centering     \includegraphics[width=1\linewidth,page=3]{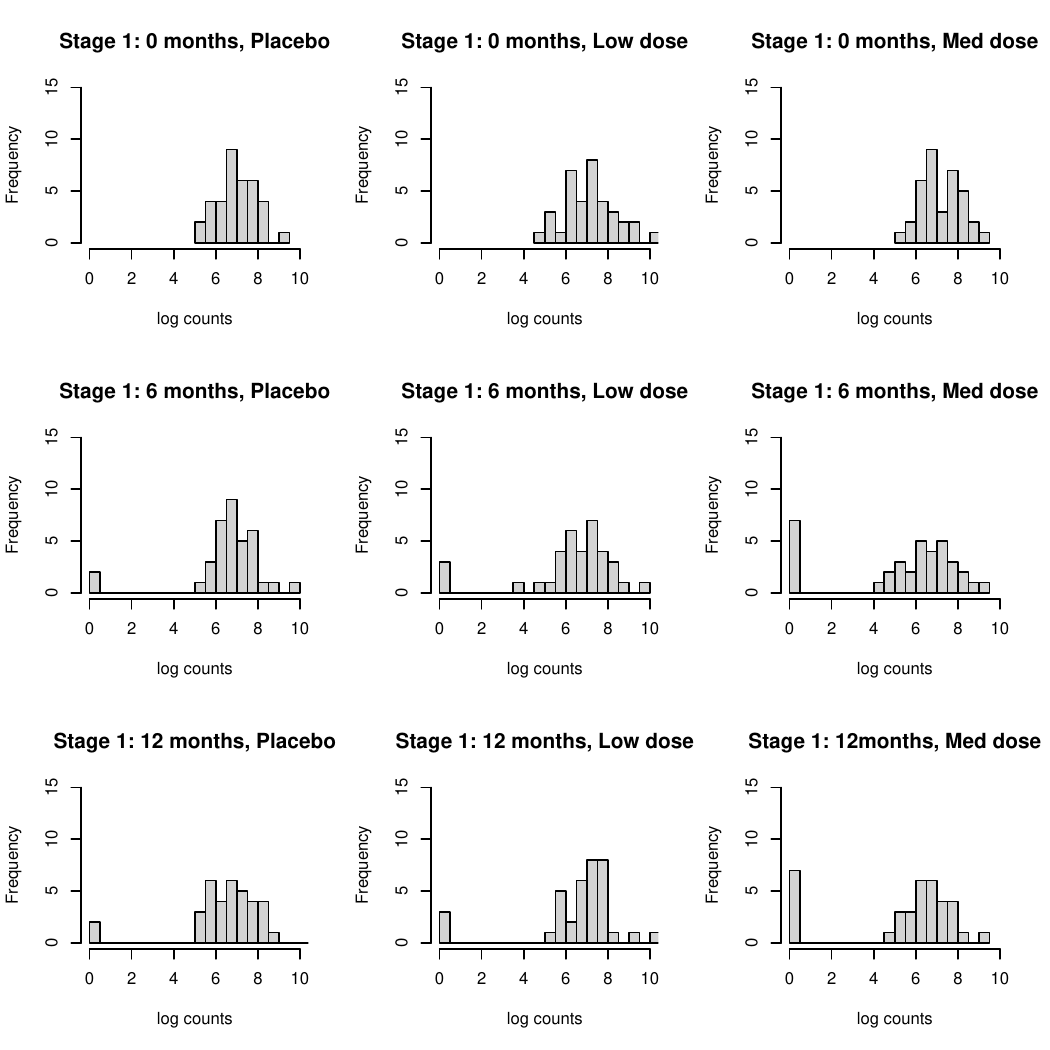}
    \caption[Numerical example - first stage data scenario "Trend"]{First stage data of numerical example for scenario "Trend"}
    \label{fig:H3}
\end{figure}

\begin{figure}[!p]
    \centering     \includegraphics[width=1\linewidth,page=4]{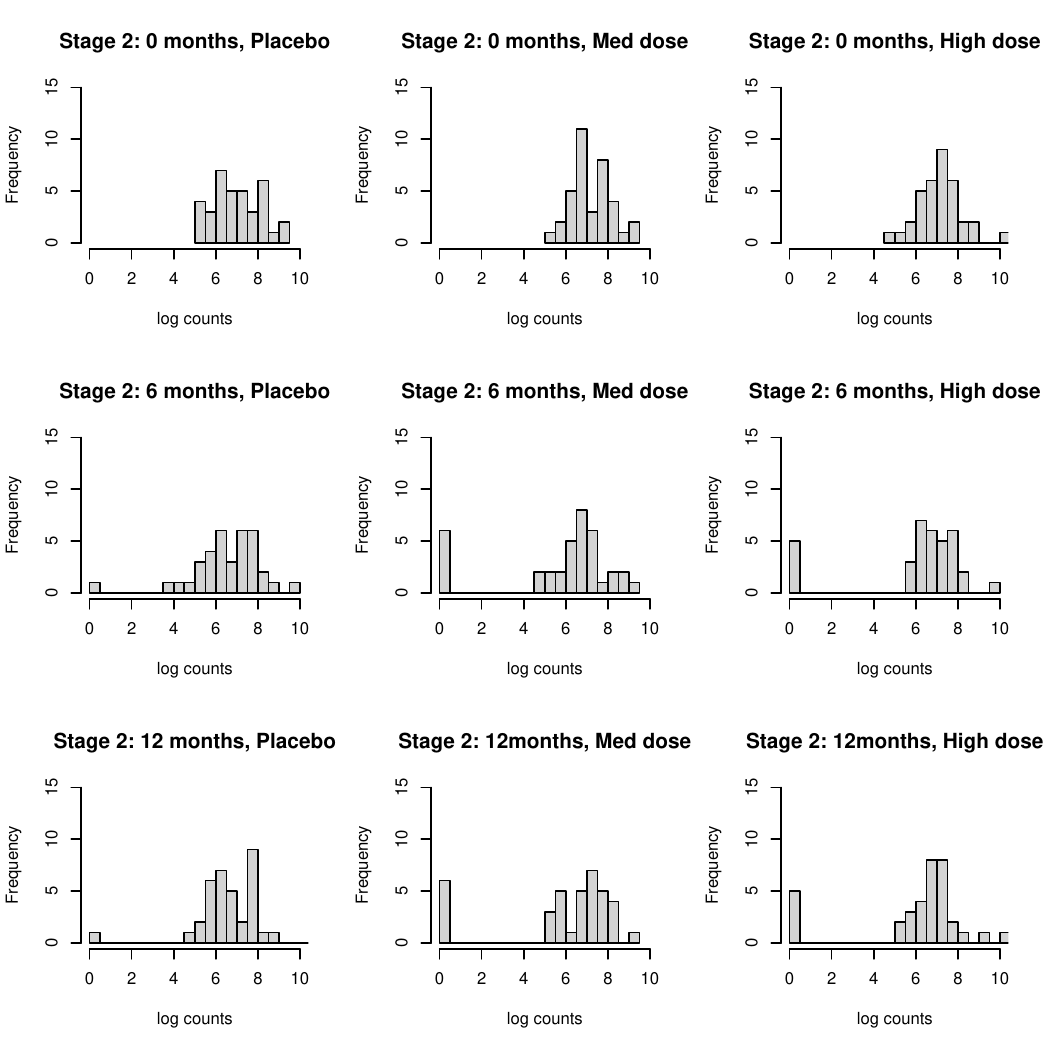}
    \caption[Numerical example - second stage data scenario "Trend"]{First stage data of numerical example for scenario "Trend"}
    \label{fig:H4}
\end{figure}

In the following the partial conditional error rates for the examples discussed in Section 7 in the manuscript are specified. Histograms of the simulated raw data are shown in Figs. (\ref{fig:H1}) - (\ref{fig:H4}).

\paragraph{Selection scenario "All doses effective":} For surrogate p-values $p_{1,1}^{(1)}$ and $p_{2,1}^{(1)}$ for the low and the medium doses, both doses are continued to the second stage, the high dose is not started. According to the closed test all following intersection hypotheses and the elementary hypotheses are tested at their specific partial conditional error rates (marked in red if condition is TRUE):

\begin{table*}[h!]
\centering 
 \begin{tabular}{rlllllll} 
$H_{123}$: & $p_{1,2}^{(2)}\leq 0.198$ & OR & \textcolor{red}{$p_{2,2}^{(2)}\leq 0.597$} &OR& $0.795\geq 1$\\
$H_{12}$: & \textcolor{red}{$p_{1,2}^{(2)}\leq 0.274$} & OR& \textcolor{red}{$p_{2,2}^{(2)}\leq 0.685$} &OR& $0.96\geq 1$\\
$H_{13}$: & \textcolor{red}{$p_{1,2}^{(2)}\leq 0.286$ } \\
$H_{23}$: & &&\textcolor{red}{$p_{2,2}^{(2)}\leq 0.697$ }\\
$H_{1}$: & \textcolor{red}{$p_{1,2}^{(2)}\leq 0.446$} \\
$H_{2}$: &&& \textcolor{red}{$p_{2,2}^{(2)}\leq 0.829$} \\
\end{tabular}
\end{table*}

Both the low and the medium dose are rejected in the end.

\paragraph{Selection scenario "Trend":  }
For surrogate p-values $p_{1,1}^{(1)}$ and $p_{2,1}^{(1)}$ for the low and the medium dose, the medium dose is continued to the second stage and the high dose is started. Closed test with all corresponding intersection and elementary hypotheses with their specific partial conditional error rates:
\begin{table*}[h!]
\centering 
 \begin{tabular}{rlllll} 
$H_{123}$:  & \textcolor{red}{$p_{2,2}^{(2)}\leq 0.59$} & OR & \textcolor{red}{$p_{3,2}^{(2)}\leq 0.204$} & OR & $0.795\geq 1$\\
$H_{12}$:   & \textcolor{red}{$p_{2,2}^{(2)}\leq 0.959$} \\
$H_{23}$:   & \textcolor{red}{$p_{2,2}^{(2)}\leq 0.685$} & OR & \textcolor{red}{$p_{3,2}^{(2)}\leq 0.013$}\\
$H_{13}$: & & &\textcolor{red}{$p_{3,2}^{(2)}\leq 0.286$} \\
$H_{2}$: & \textcolor{red}{$p_{2,2}^{(2)}\leq 0.829$} \\
$H_{3}$: &&& \textcolor{red}{$p_{3,2}^{(2)}\leq 0.025$ }\\
\end{tabular}
\end{table*}

Both the medium and the high dose are rejected in the end, the low dose is retained.

\newpage

\newpage

\newpage

\newpage

\newpage

\section{Unconditional and conditional bias}
Tables \ref{Tab:UncondBias} and \ref{Tab:CondBias} show the unconditional and the conditional bias, lower CI limit and coverage of the confidence intervals related to Table 4 in the main manuscript.

\begin{table}[htb]
\footnotesize
\centering\caption[Concordance, bias, and CI for the standard scenario]{The simulated true concordance values (C) of each scenario for the three doses are reported with  bias (mean C - true C), lower CI limit (mean value over simulation runs), and coverage of the confidence intervals (in \%) for \textbf{unconditional} method for Simulation 1 (standard scenario) for $\alpha_1=0.3$. A total of 50000 simulation runs was performed for each scenario.  }
\begin{tabular}{l|ccc}
\hline
Scenario &Low dose& Medium dose& High dose\\ \hline
& True concordance  & True concordance  & True concordance \\
& Bias/CI/Coverage& Bias/CI/Coverage &Bias/CI/Coverage \\
\hline
No effect                   &0.5  &0.5    &0.50\\ &-0.006, 0.35, 99.9 & -0.007, 0.35, 99.9  & 0.000, 0.34, 98.9 \\
Efficacy only in high dose  &0.5  &0.5   &0.77 \\ & -0.006, 0.35, 99.9&-0.007, 0.35, 99.9& 0.001, 0.61, 99.4\\
Trend (a)                   &0.5  &0.64  &0.77 \\ &-0.006, 0.35, 99.9&-0.005, 0.47, 99.9&0.001, 0.61, 98.6\\
Trend (b)                   &0.5  &0.71  &0.77 \\ &-0.006, 0.37, 99.9&0.001, 0.53, 1&0.001, 0.61, 98.5\\
All doses effective         &0.71  &0.71   &0.71 \\ &0.000, 0.53, 1&0.001, 0.53, 1&-0.005, 0.54, 98.7\\
\hline
\end{tabular}\label{Tab:UncondBias}
\end{table}

\begin{table}[htb]
\footnotesize
\centering\caption[Concordance, bias, and CI for the standard scenario]{The simulated true concordance values (C) of each scenario for the three doses are reported with  bias (mean C - true C), lower CI limit (mean value over simulation runs), and coverage of the confidence intervals (in \%) for \textbf{conditional} method for Simulation 1 (standard scenario) for $\alpha_1=0.3$. A total of 50000 simulation runs was performed for each scenario.  
}
\begin{tabular}{l|ccc}
\hline
Scenario &Low dose& Medium dose& High dose\\ \hline
& True concordance  & True concordance  & True concordance \\
& Bias/CI/Coverage& Bias/CI/Coverage &Bias/CI/Coverage \\
\hline
No effect                   &0.5  &0.5    &0.50\\ &0.029, 0.36, 99.9 & 0.017, 0.35, 99.9  & 0, 0.35, 98.9\\
Efficacy only in high dose  &0.5  &0.5   &0.77 \\ & 0.029, 0.36, 99.9&0.017, 0.35, 99.9& 0.001, 0.61, 99.4\\
Trend (a)                   &0.5  &0.64  &0.77 \\ &0.029, 0.36, 99.9&0.005, 0.47, 99.9&0.001, 0.61, 98.6\\
Trend (b)                   &0.5  &0.71  &0.77 \\ &0.029, 0.36, 99.9&0.000, 0.53, 99.9&0.001, 0.61, 98.5\\
All doses effective         &0.71  &0.71   &0.71 \\ &0.002, 0.53, 1&0.001, 0.53, 99.9&-0.005, 0.54, 98.7\\
\hline
\end{tabular}\label{Tab:CondBias}
\end{table}

\newpage

\section*{Acknowledgments}

The eWHORM project (No 101103053) supported by the Global Health EDCTP3 Joint Undertaking and its members as well as the Swiss Confederation. Views and opinions expressed are those of the author(s) only and do not necessarily reflect those of the European Union or the Global Health European and Developing Countries Clinical Trials Partnership (EDCTP3). Neither the European Union nor the granting authority can be held responsible for them

M. Bofill Roig is a Serra Húnter Fellow and was additionally supported by Grant PID2023-148033OB-C21 funded by MICIU/AEI/10.13039/501100011033 and by FEDER/UE.

The authors are grateful to the eWHORM investigators not included as authors for their work: Karen Dequatre Cheeseman, Jennifer Keiser, Ghyslain Momba Ngoma, Dieudonne Mumba Ngoyi, Michael Ramharter, Samuel Wanji, Achim Hoerauf, Benjamin Lenz, Annina Schnoz.
\newpage

\begin{figure}[!p]
    \centering     \includegraphics[width=1\linewidth,page=2]{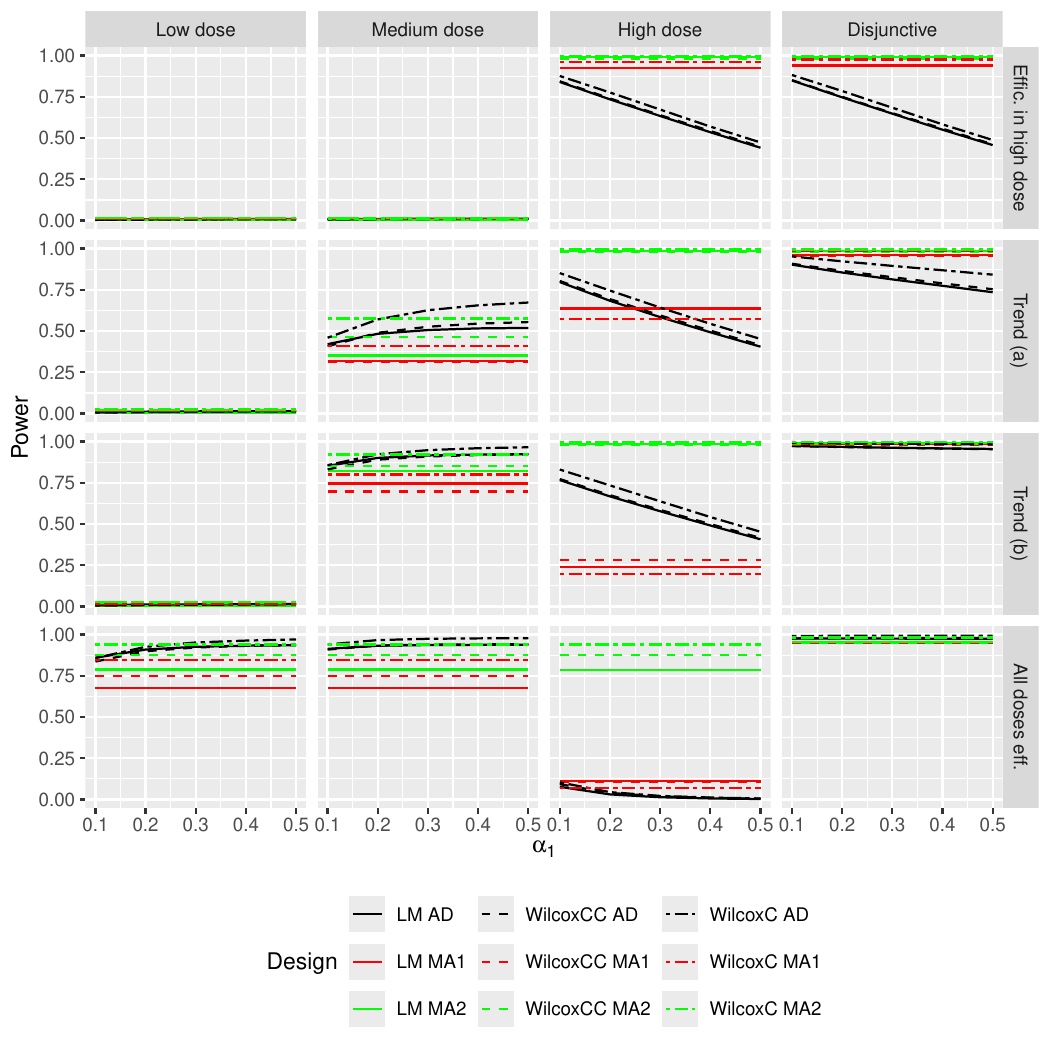}
    \caption[Mansonellosis S2]{Mansonellosis S2: Simulated power values for the low, medium, and high doses and the disjunctive power are depicted (in the columns) for simulation S2 (modified reduction rates) as a function of $\alpha_1$ for several scenarios of reduction rates (in the rows).}
    \label{fig:MansPowSim2}
\end{figure}

\begin{figure}[!p]
    \centering     \includegraphics[width=1\linewidth,page=4]{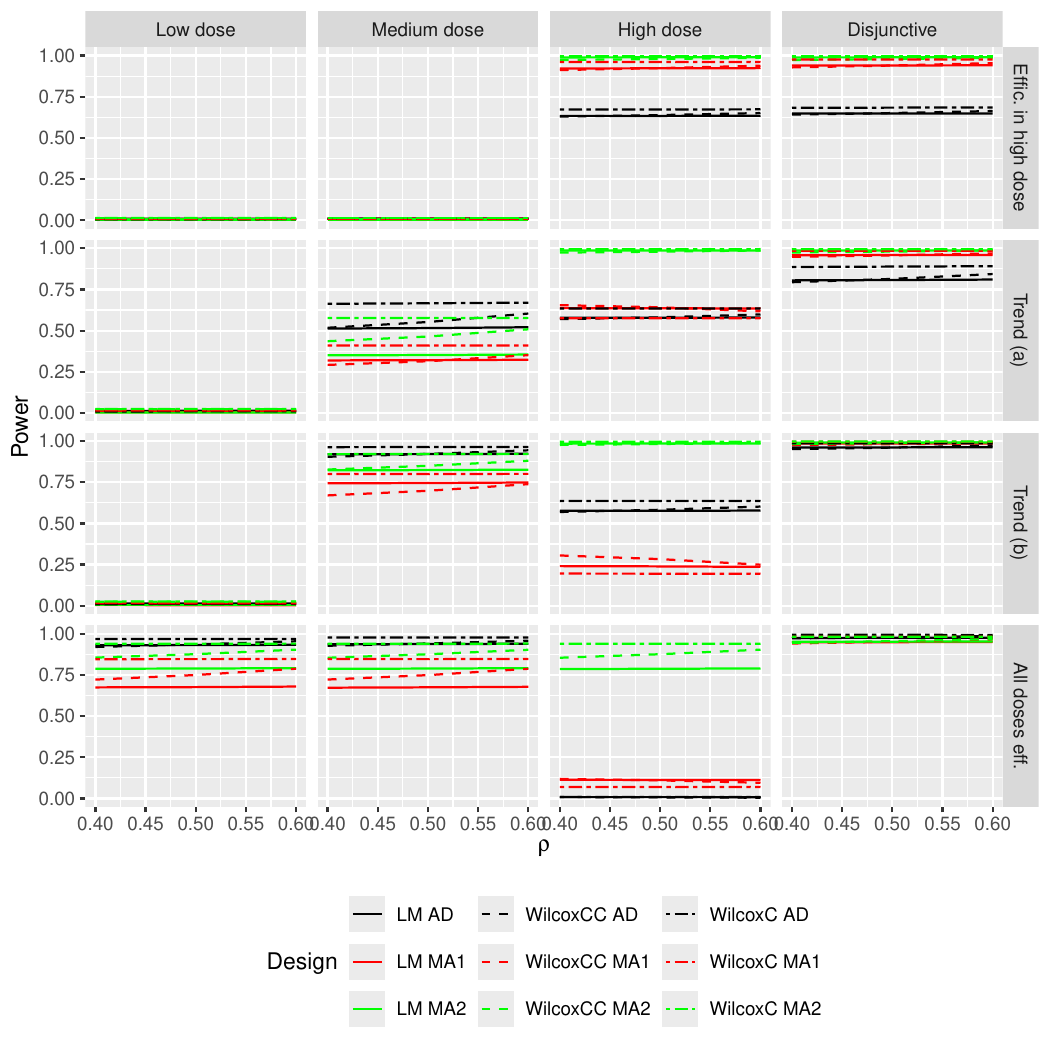}
    \caption[Mansonellosis S4]{Mansonellosis S4: Simulated power values for the low, medium, and high doses and the disjunctive power are depicted (in the columns) for simulation S4 as a function of $\rho$ for several scenarios of reduction rates (in the rows).}
    \label{fig:MansPowSim4}
\end{figure}

\begin{figure}[!p]
    \centering     \includegraphics[width=1\linewidth,page=6]{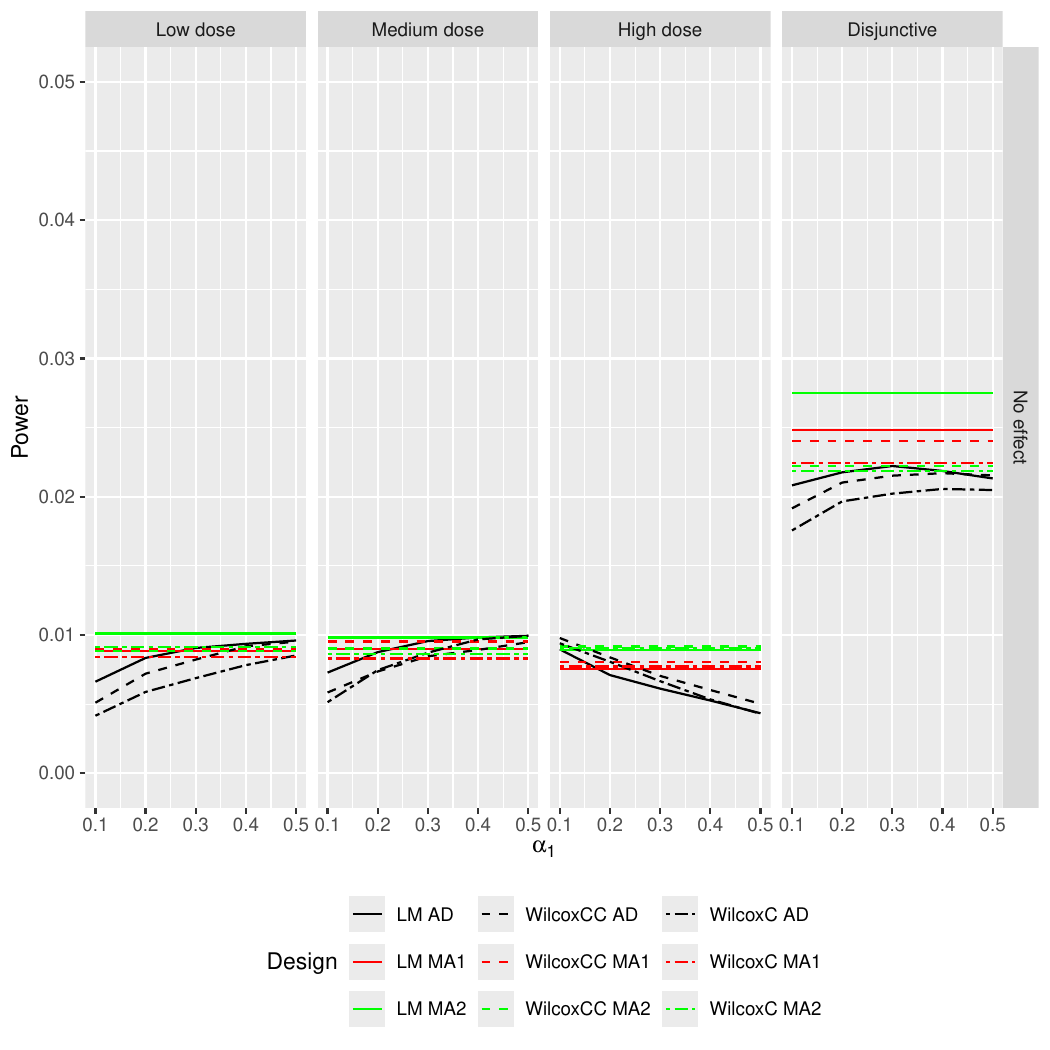}
    \caption[Mansonellosis S1 - Error rates]{Mansonellosis S1: Type I error rates for the low, medium, and high doses and the overall type I error rate are shown (in the columns) for simulation S1 as a function of $\alpha_1$ for scenarios of reduction rates 1 ("No effect").}
    \label{fig:MansErrorSim1}
\end{figure}

\begin{figure}[!p]
    \centering     \includegraphics[width=1\linewidth,page=7]{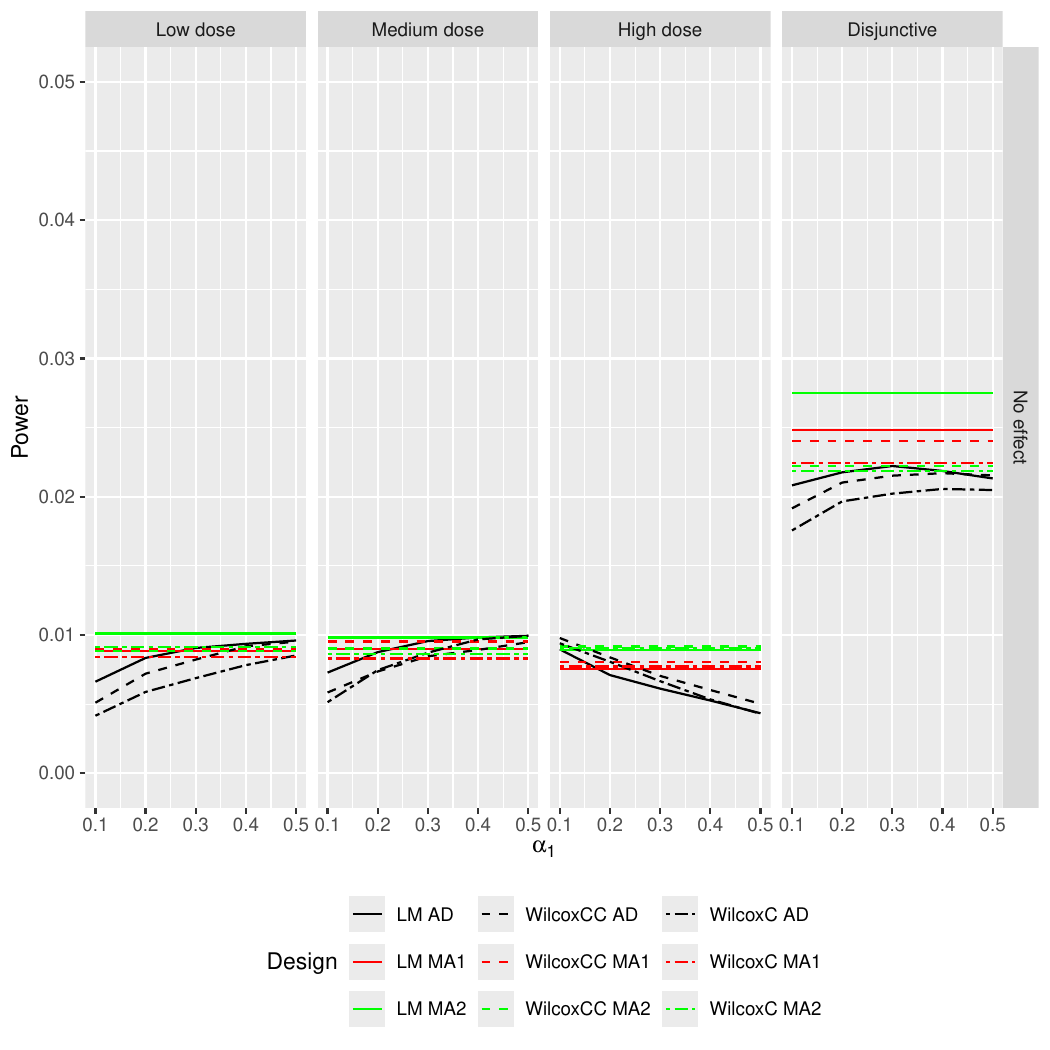}
    \caption[Mansonellosis S2 - Error rate]{Mansonellosis S2: Type I error rates for the low, medium, and high doses and the overall type I error rate are shown (in the columns) for simulation S2 (modified reduction rates) as a function of $\alpha_1$ for scenarios of reduction rates 1 ("No effect").}
    \label{fig:ScenariosWorst}
\end{figure}

\begin{figure}[!p]
    \centering     \includegraphics[width=1\linewidth,page=8]{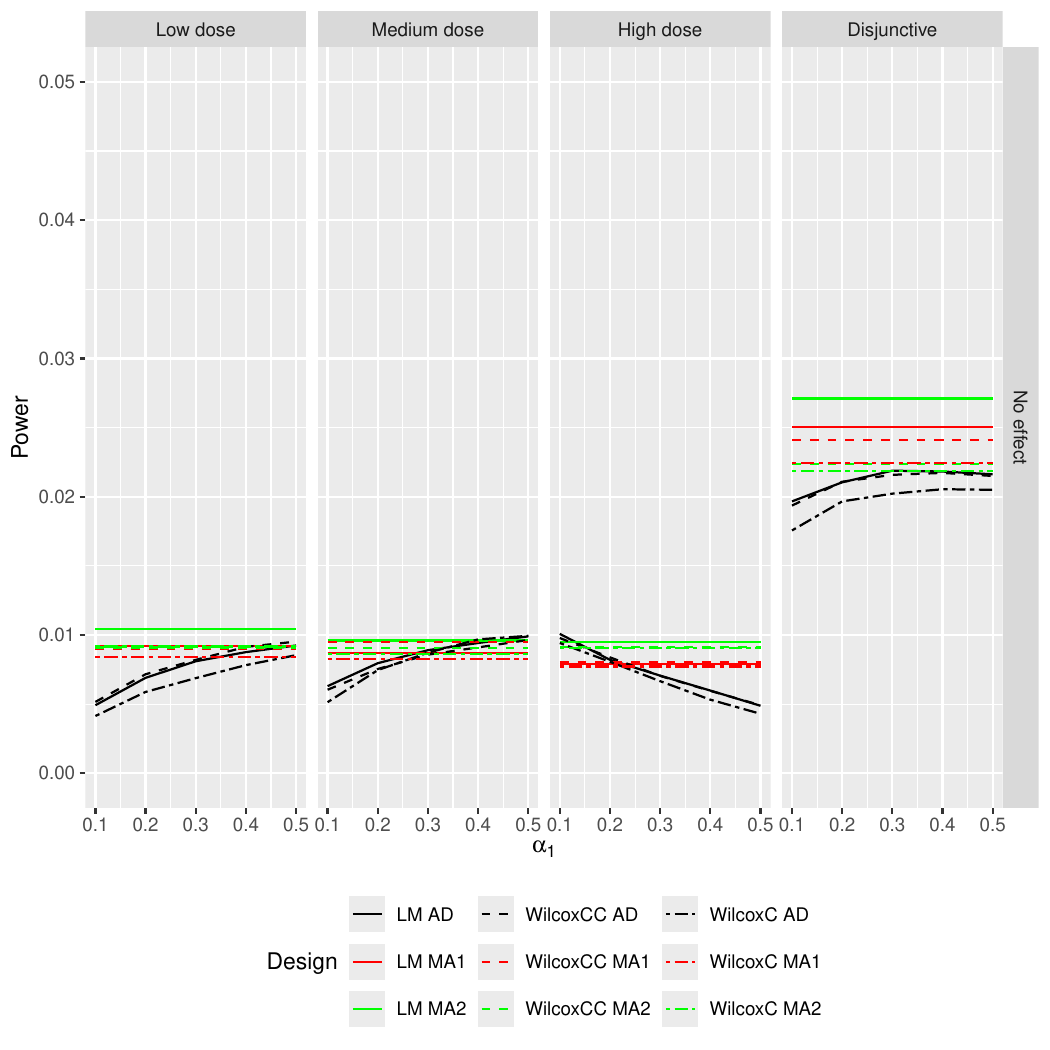}
    \caption[Mansonellosis S3 - Error rate]{Mansonellosis S3: Type I error rates for the low, medium, and high doses and the overall type I error rate are shown (in the columns) for simulation S3 (modified baseline parameters) as a function of $\alpha_1$ for scenarios of reduction rates 1 ("No effect").}
    \label{fig:ScenariosWorst}
\end{figure}

\begin{figure}[!p]
    \centering     \includegraphics[width=1\linewidth,page=9]{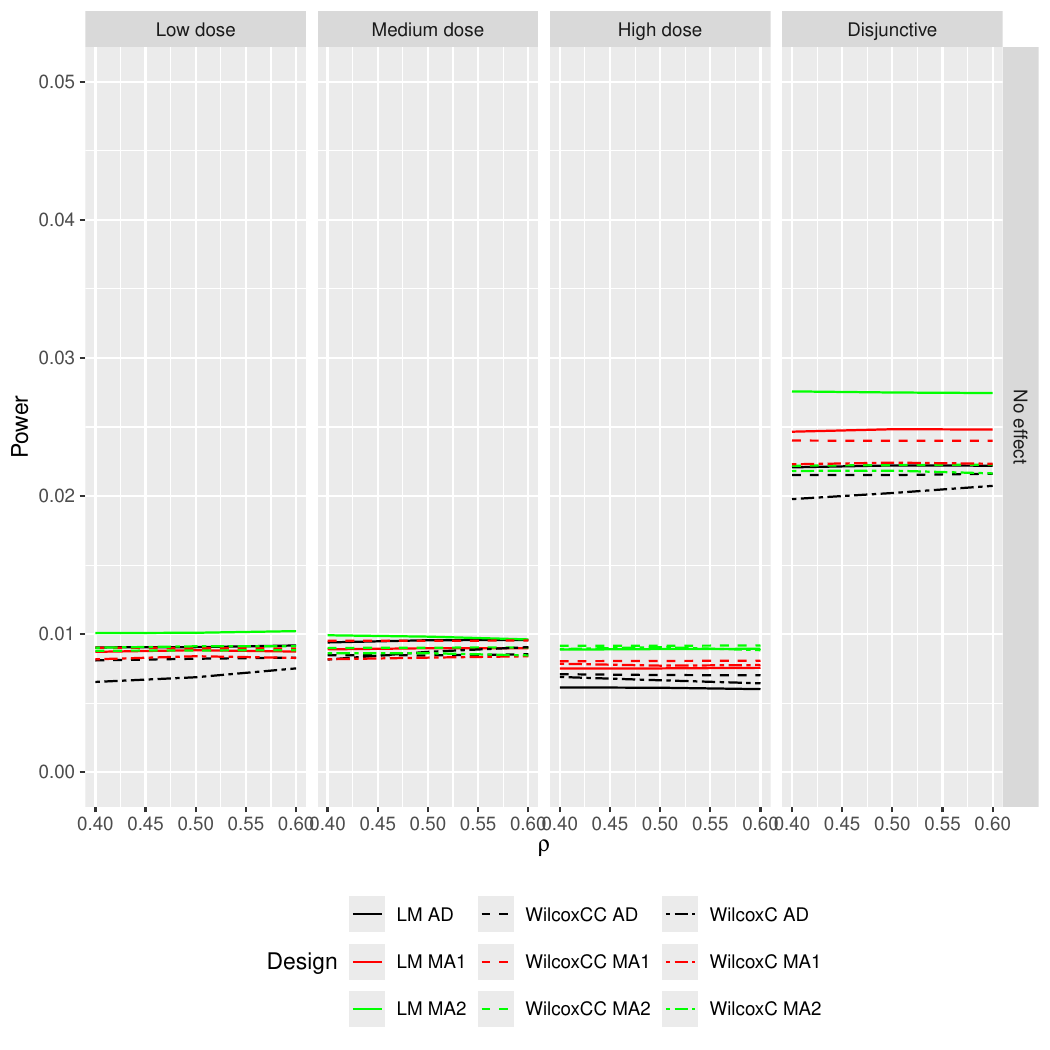}
    \caption[Mansonellosis S4 - Error rate]{Mansonellosis S4: Type I error rates for the low, medium, and high doses and the overall type I error rate are shown (in the columns) for simulation S4 as a function of $\rho$ for scenarios of reduction rates 1 ("No effect").}
    \label{fig:ScenariosWorst}
\end{figure}

\begin{figure}[!p]
    \centering     \includegraphics[width=1\linewidth,page=10]{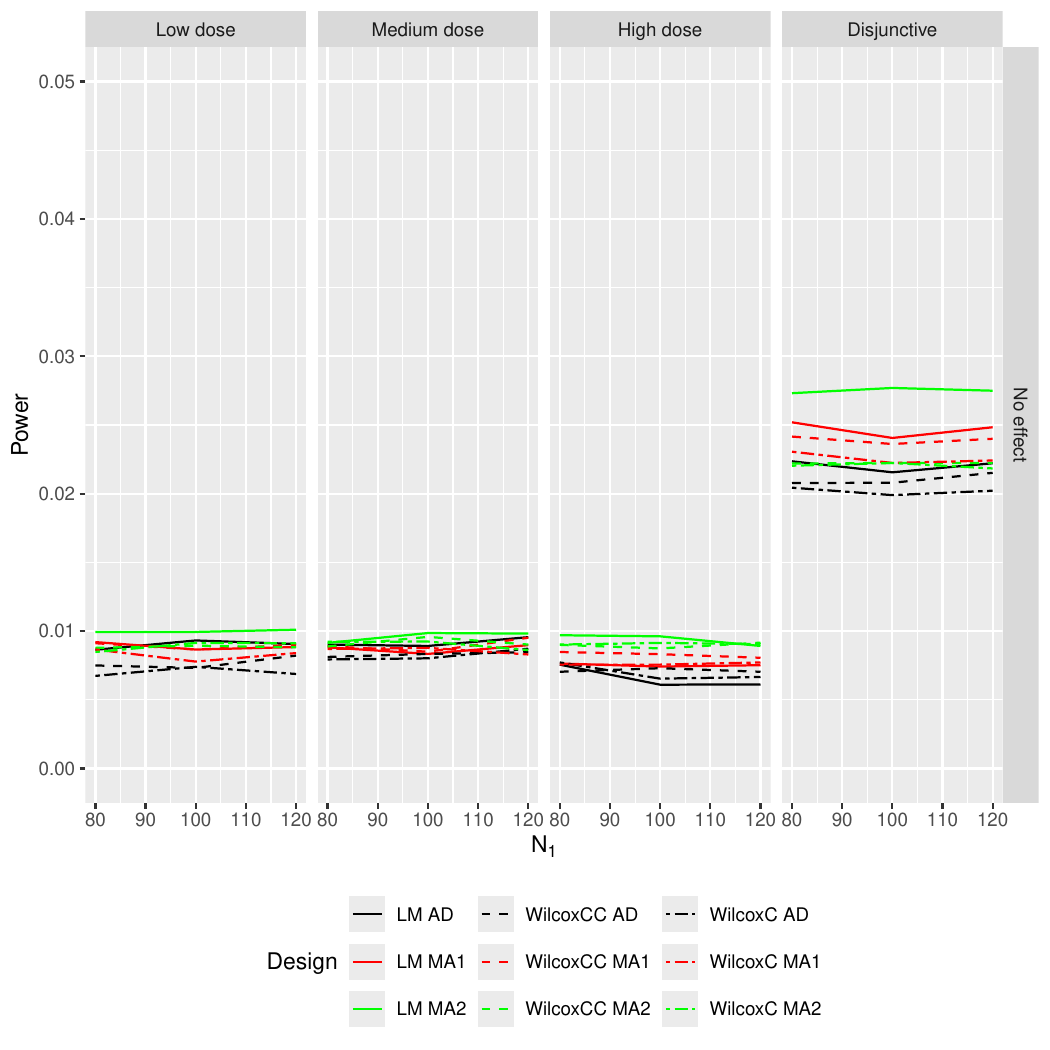}
    \caption[Mansonellosis S5 - Error rate]{Mansonellosis S5: Type I error rates for the low, medium, and high doses and the overall type I error rate are shown (in the columns) for simulation S5 as a function of $N_1$ for scenarios of reduction rates 1 ("No effect").}
    \label{fig:ScenariosWorst}
\end{figure}

\begin{figure}[!p]
    \centering     \includegraphics[width=1\linewidth,page=11]{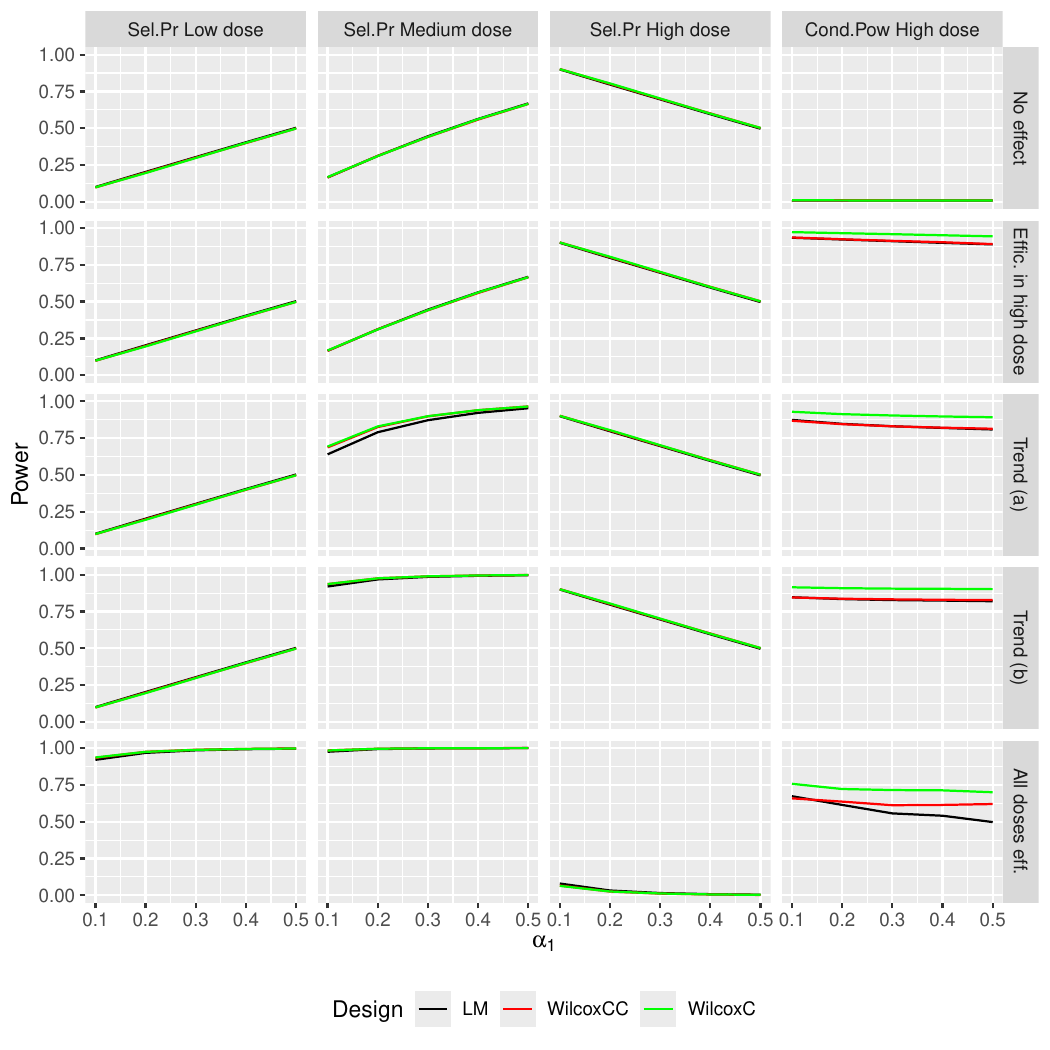}
    \caption[Mansonellosis S1 - Selection probabilities and conditional probability]{Mansonellosis S1: Selection probabilities (Sel.Prob) for all doses and conditional probability of high dose are depicted (in the columns) for simulation S1 (standard reduction rates) as a function of $\alpha_1 = \{0.1, 0.2, 0.3, 0.4, 0.5\} $ for scenarios of reduction rates 1.}
    \label{fig:ScenariosWorst}
\end{figure}

\begin{figure}[!p]
    \centering     \includegraphics[width=1\linewidth,page=12]{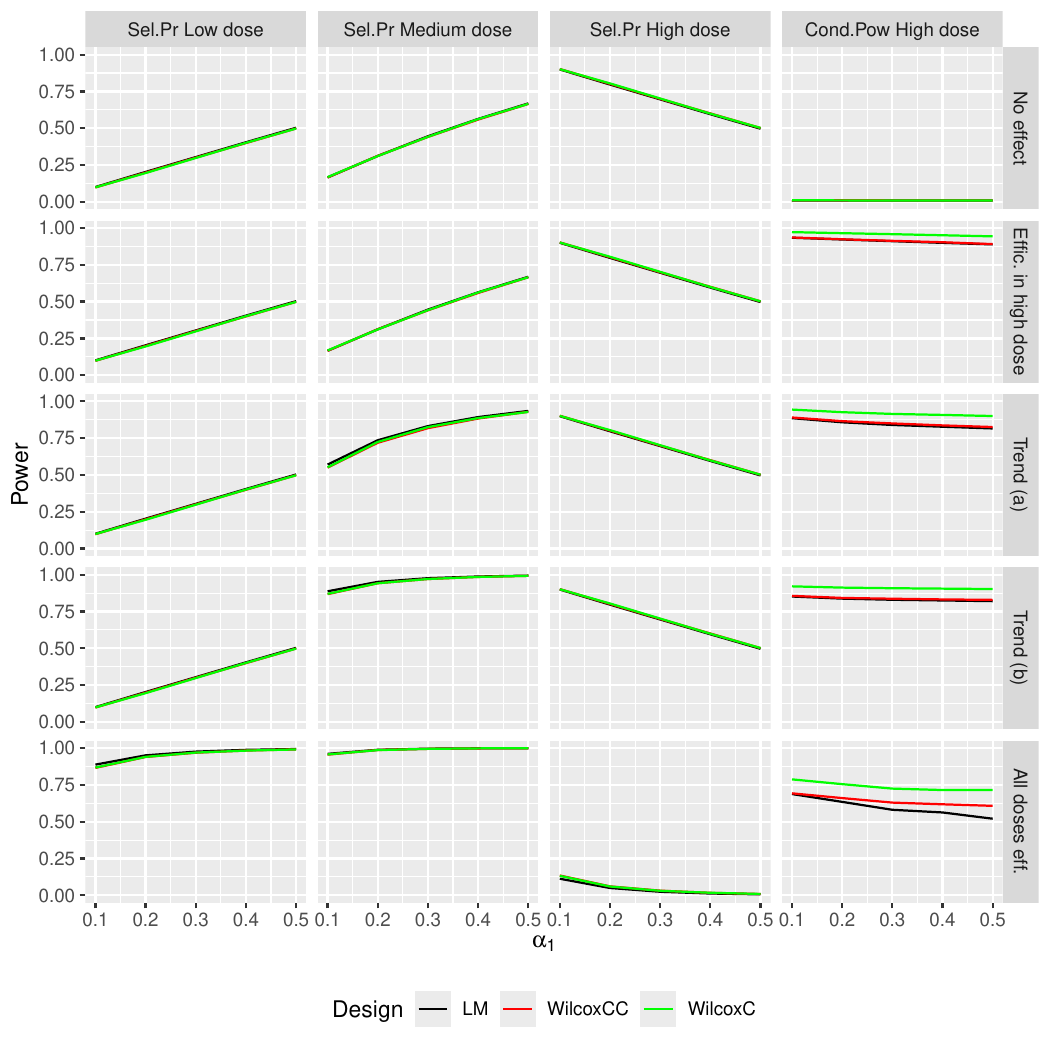}
    \caption[Mansonellosis S2 - Selection probabilities and conditional probability]{Mansonellosis S2: Selection probabilities (Sel.Pr) for all doses and conditional probability (Cond.Pow) for the high dose are depicted (in the columns) for simulation S2 (modified reduction rates) as a function of $\alpha_1$ for several scenarios of reduction rates (in the rows).}
    \label{fig:ScenariosWorst}
\end{figure}

\begin{figure}[!p]
    \centering     \includegraphics[width=1\linewidth,page=13]{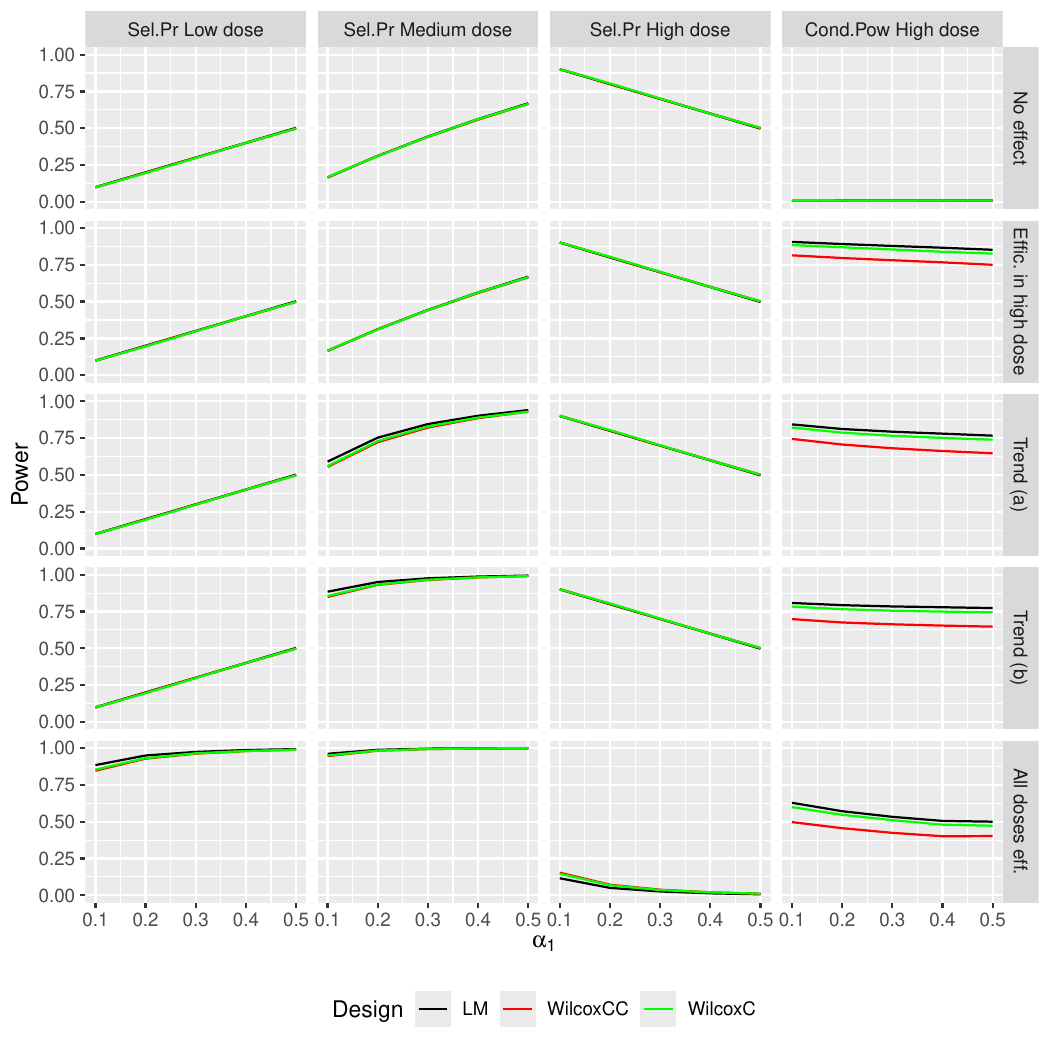}
    \caption[Mansonellosis S3 - Selection probabilities and conditional probability]{Mansonellosis S3: Selection probabilities (Sel.Pr) for all doses and conditional probability (Cond.Pow) for the high dose are depicted (in the columns) for simulation S3 (modified baseline parameters) as a function of $\alpha_1 = \{0.1, 0.2, 0.3, 0.4, 0.5\}$ for several scenarios of reduction rates (in the rows).}
    \label{fig:ScenariosWorst}
\end{figure}

\begin{figure}[!p]
    \centering     \includegraphics[width=1\linewidth,page=14]{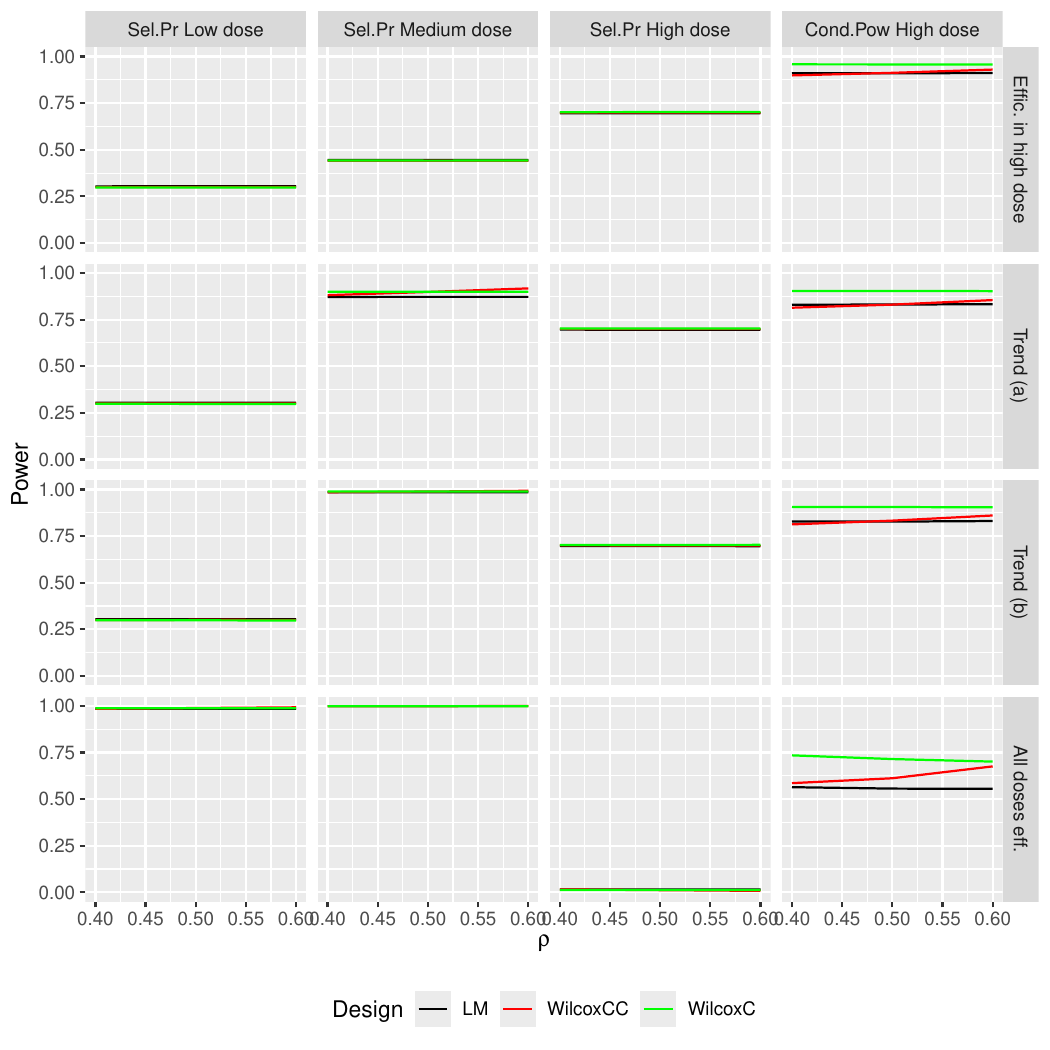}
    \caption[Mansonellosis S4 - Selection probabilities and conditional probability]{Mansonellosis S4: Selection probabilities (Sel.Pr) for all doses and conditional probability (Cond.Pow) for the high dose are depicted (in the columns) for simulation S4 (modified baseline parameters) as a function of $\rho$ for several scenarios of reduction rates (in the rows).}
    \label{fig:ScenariosWorst}
\end{figure}

\begin{figure}[!p]
    \centering     \includegraphics[width=1\linewidth,page=15]{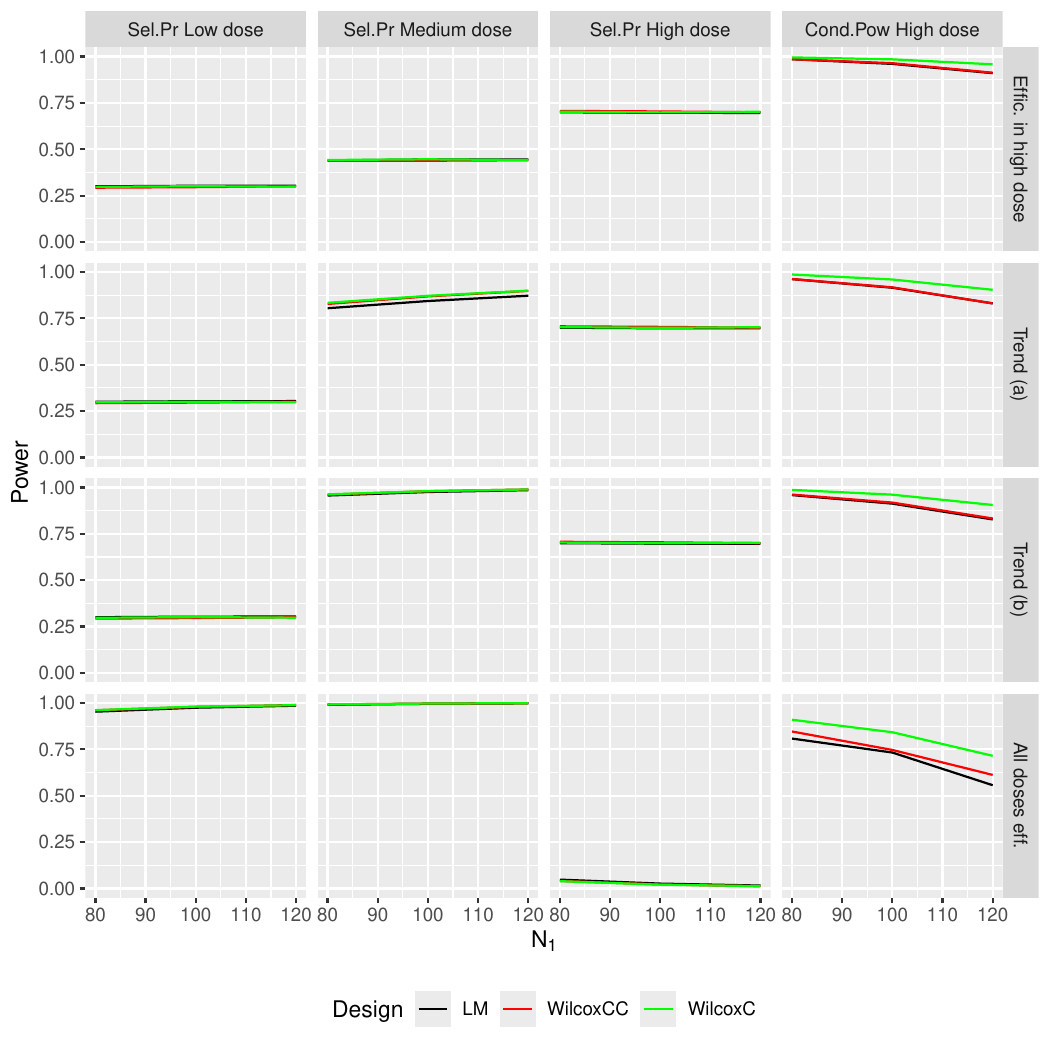}
    \caption[Mansonellosis S5 - Selection probabilities and conditional probability] {Mansonellosis S5 - Selection probabilities (Sel.Pr) for all doses and conditional probability (Cond.Pow) for the high dose are depicted (in the columns) for simulation S5 (modified baseline parameters) as a function of $N_1$ for several scenarios of reduction rates (in the rows).}
    \label{fig:ScenariosWorst}
\end{figure}

\begin{figure}[!p]
    \centering     \includegraphics[width=1\linewidth,page=1]{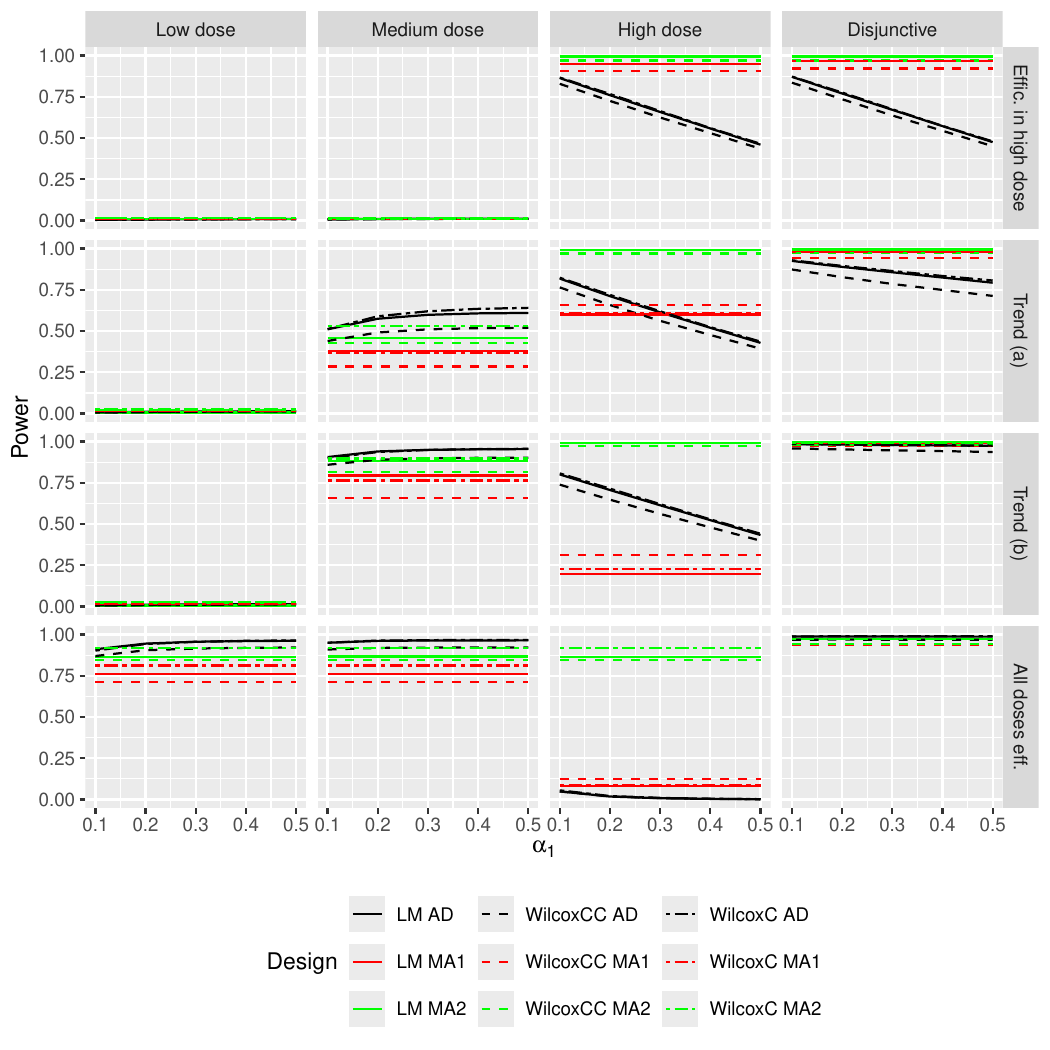}
    \caption[Onchocerciasis S1]{Onchocerciasis S1: Simulated power values for the low, medium, and high doses and the disjunctive
    power are depicted (in the columns) for simulation S1 (standard reduction rates) as a function of $\alpha_1 = \{0.1, 0.2, 0.3, 0.4, 0.5\}$ for several scenarios of reduction rates (in the rows)}
    \label{fig:ScenariosWorst}
\end{figure}

\begin{figure}[!p]
    \centering     \includegraphics[width=1\linewidth,page=2]{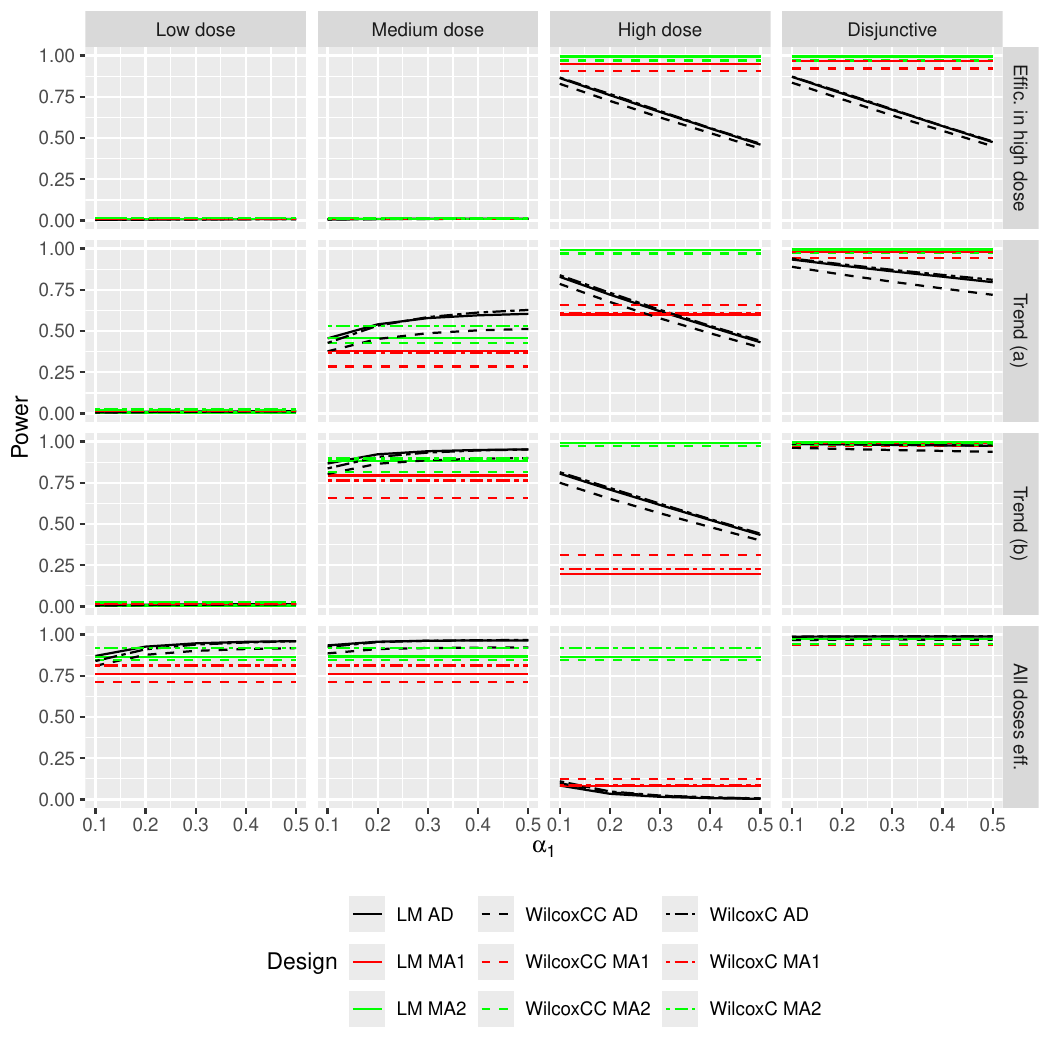}
    \caption[Onchocerciasis S2]{Onchocerciasis S2: Simulated power values for the low, medium, and high doses and the disjunctive power are depicted (in the columns) for simulation S2 (modified reduction rates) as a function of $\alpha_1$ for several scenarios of reduction rates (in the rows).}
    \label{fig:ScenariosWorst}
\end{figure}

\begin{figure}[!p]
    \centering     \includegraphics[width=1\linewidth,page=3]{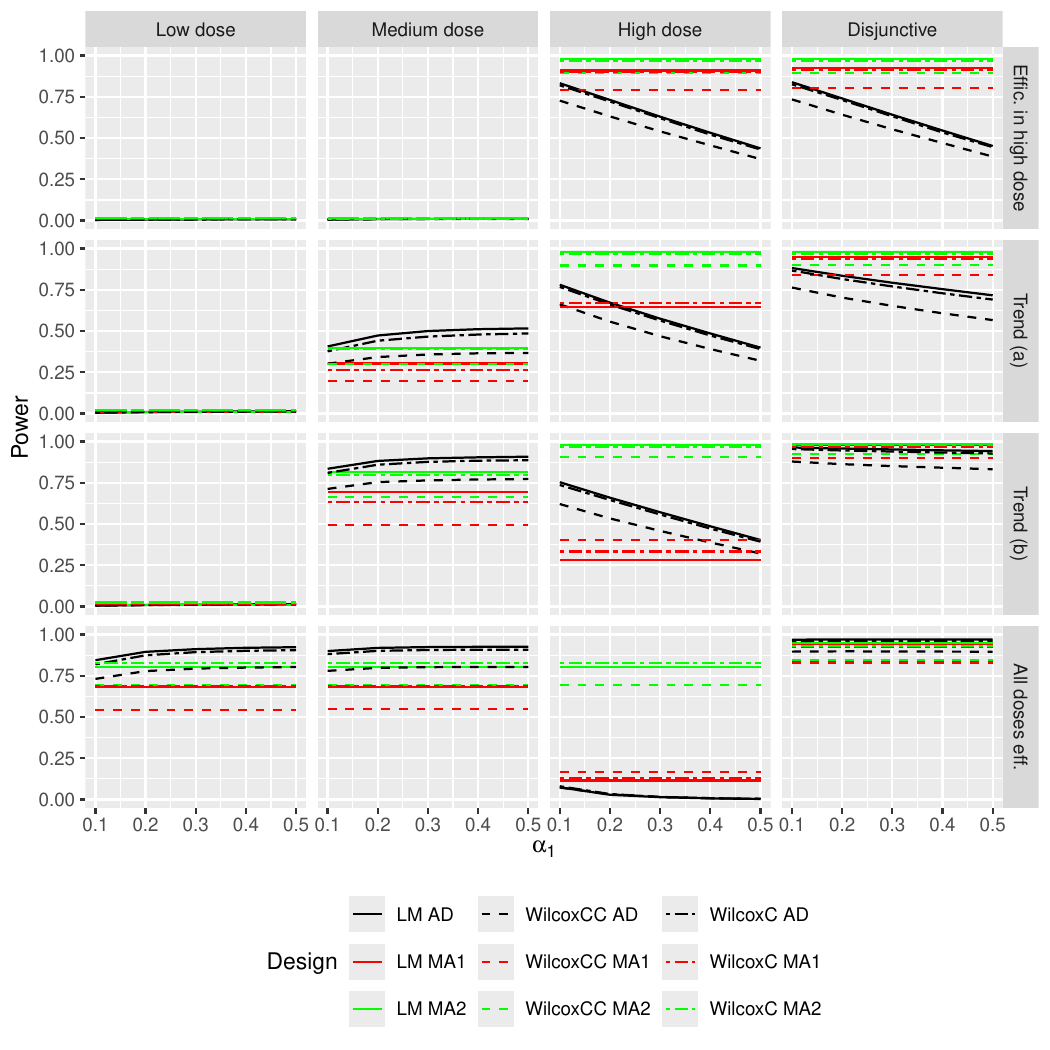}
    \caption[Onchocerciasis S3]{Onchocerciasis S3: Simulated power values for the low, medium, and high doses and the disjunctive power are depicted (in the columns) for simulation S3 (modified baseline parameters) as a function of $\alpha_1$ for several scenarios of reduction rates (in the rows). }
    \label{fig:ScenariosWorst}
\end{figure}

\begin{figure}[!p]
    \centering     \includegraphics[width=1\linewidth,page=4]{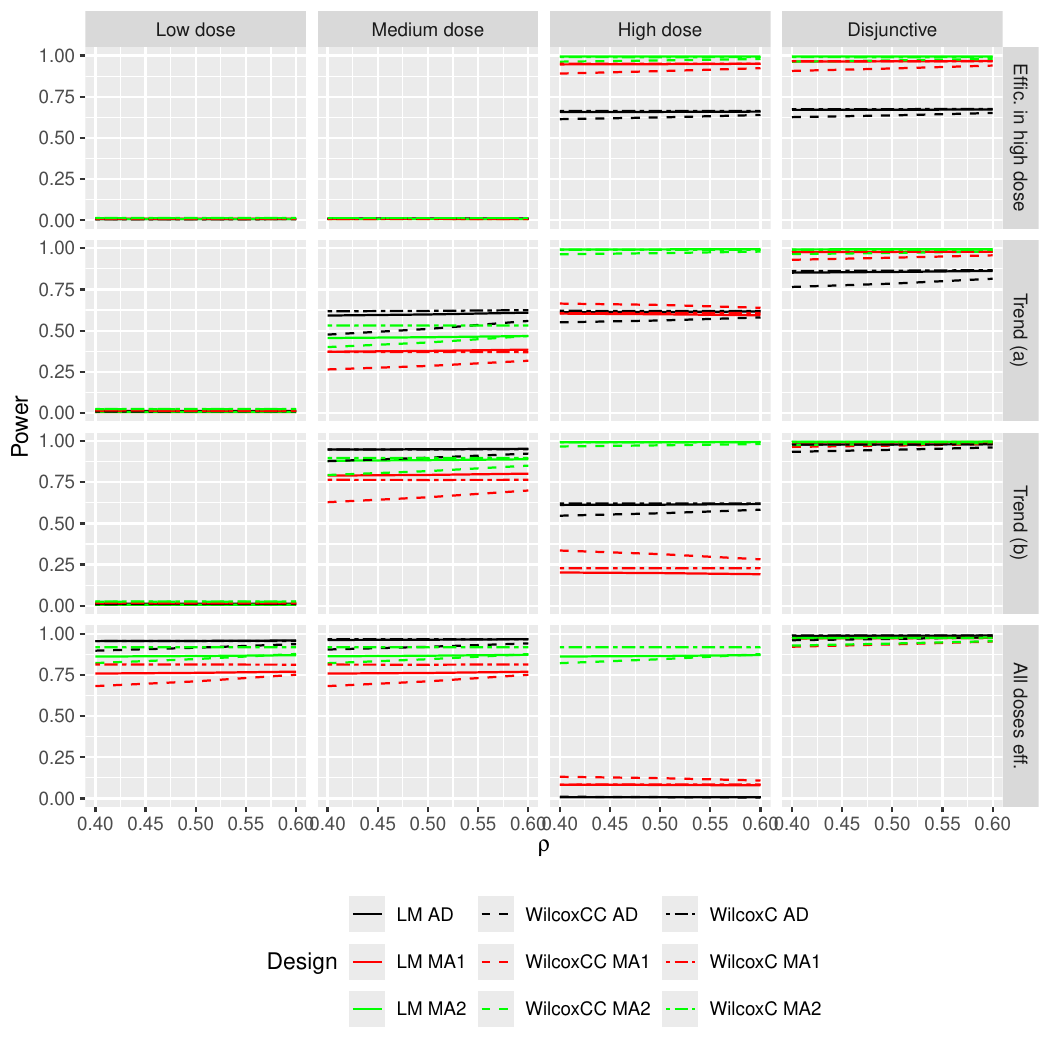}
    \caption[Onchocerciasis S4]{Onchocerciasis S4: Simulated power values for the low, medium, and high doses and the disjunctive power are depicted (in the columns) for simulation S4 (modified baseline parameters) as a function of $\rho$ for several scenarios of reduction rates (in the rows). }
    \label{fig:ScenariosWorst}
\end{figure}

\begin{figure}[!p]
    \centering     \includegraphics[width=1\linewidth,page=5]{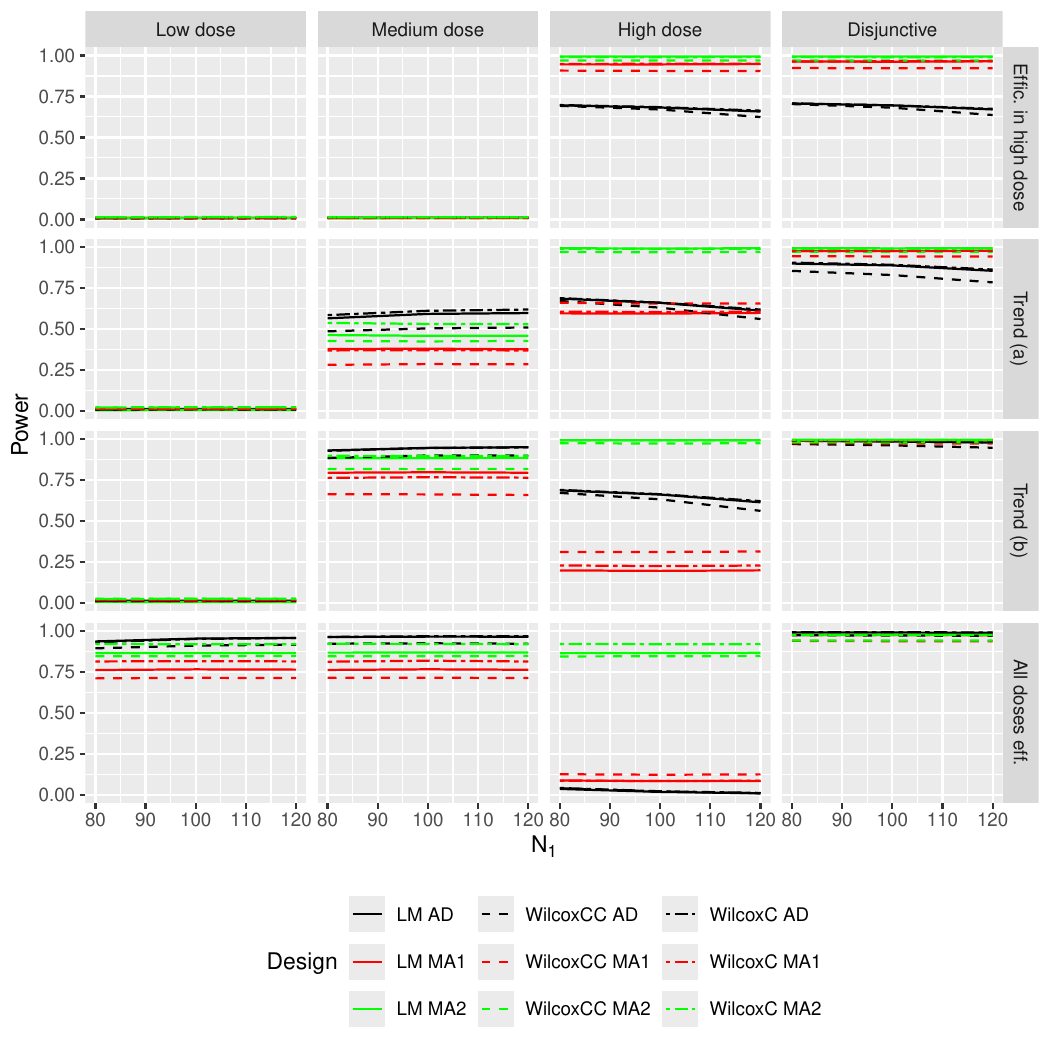}
    \caption[Onchocerciasis S5]{Onchocerciasis S5: Simulated power values for the low, medium, and high doses and the disjunctive power are depicted (in the columns) for simulation S5 (modified baseline parameters) as a function of $N_1$ for several scenarios of reduction rates (in the rows). }
    \label{fig:ScenariosWorst}
\end{figure}

\begin{figure}[!p]
    \centering     \includegraphics[width=1\linewidth,page=6]{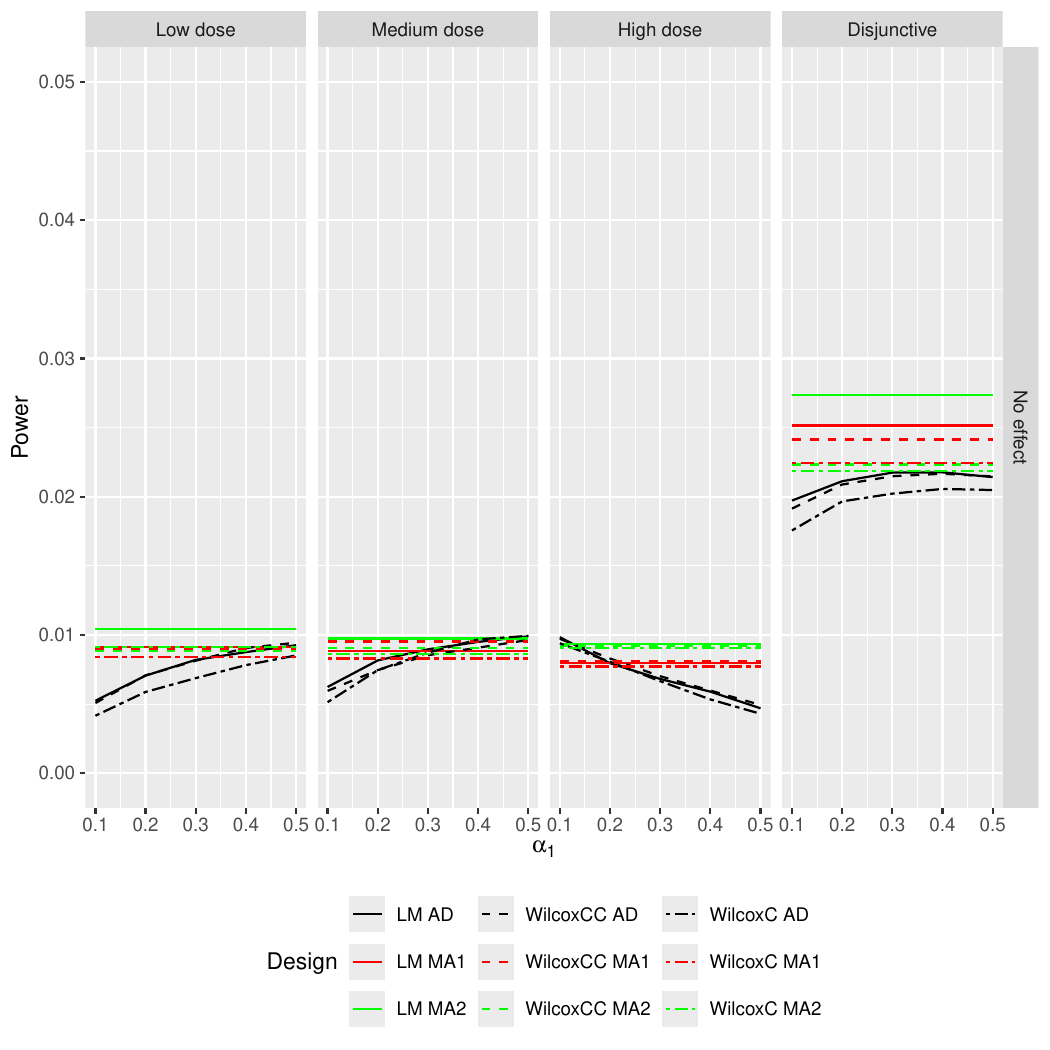}
    \caption[Onchocerciasis S1 - Error rates]{Onchocerciasis S1: Type I error rates for the low, medium, and high doses and the overall type I error rate are shown (in the columns) for simulation S1 as a function of $\alpha_1$ for scenarios of reduction rates 1 ("No effect").}
    \label{fig:ScenariosWorst}
\end{figure}

\begin{figure}[!p]
    \centering     \includegraphics[width=1\linewidth,page=7]{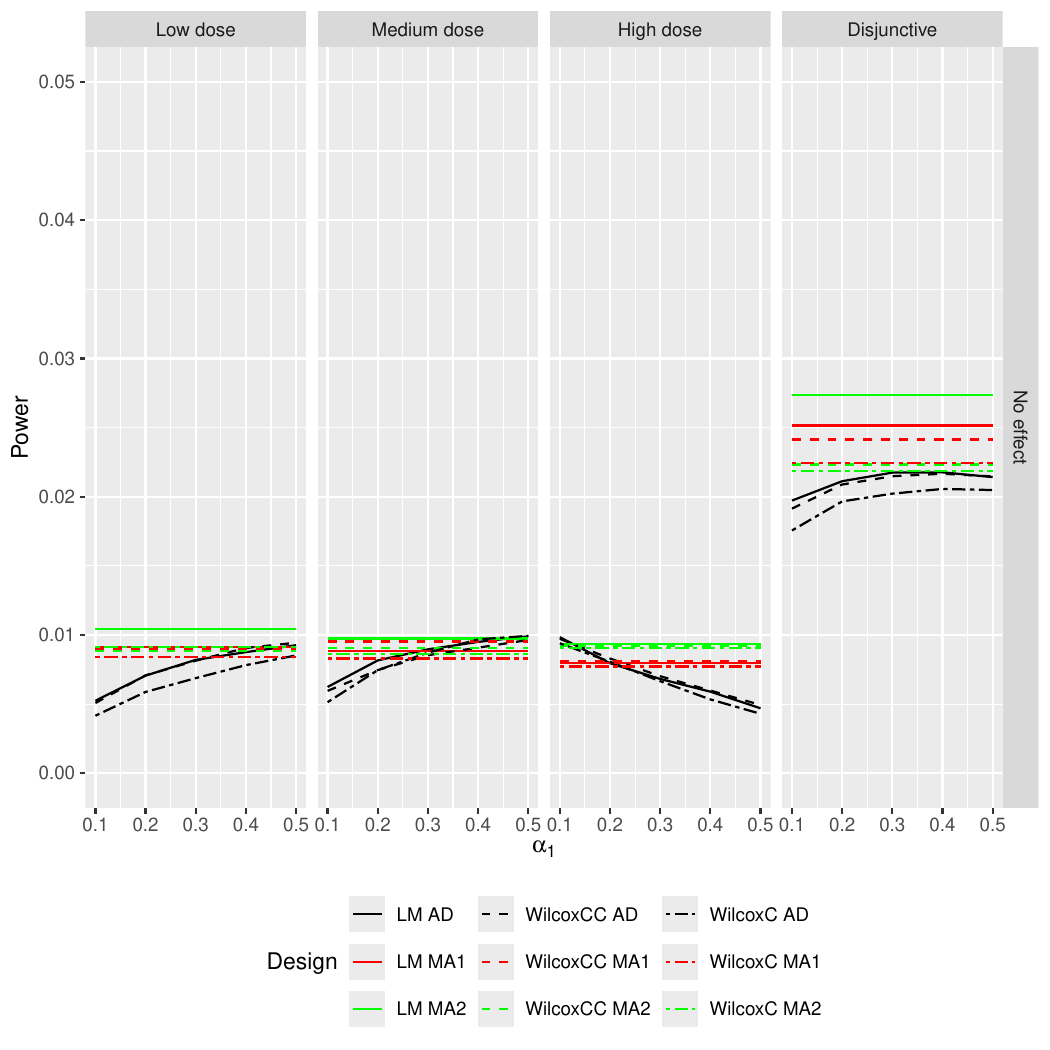}
    \caption[Onchocerciasis S2 - Error rates]{Onchocerciasis S2: Type I error rates for the low, medium, and high doses and the overall type I error rate are shown (in the columns) for simulation S2 (modified reduction rates) as a function of $\alpha_1$ for scenarios of reduction rates 1 ("No effect").}
    \label{fig:ScenariosWorst}
\end{figure}

\begin{figure}[!p]
    \centering     \includegraphics[width=1\linewidth,page=8]{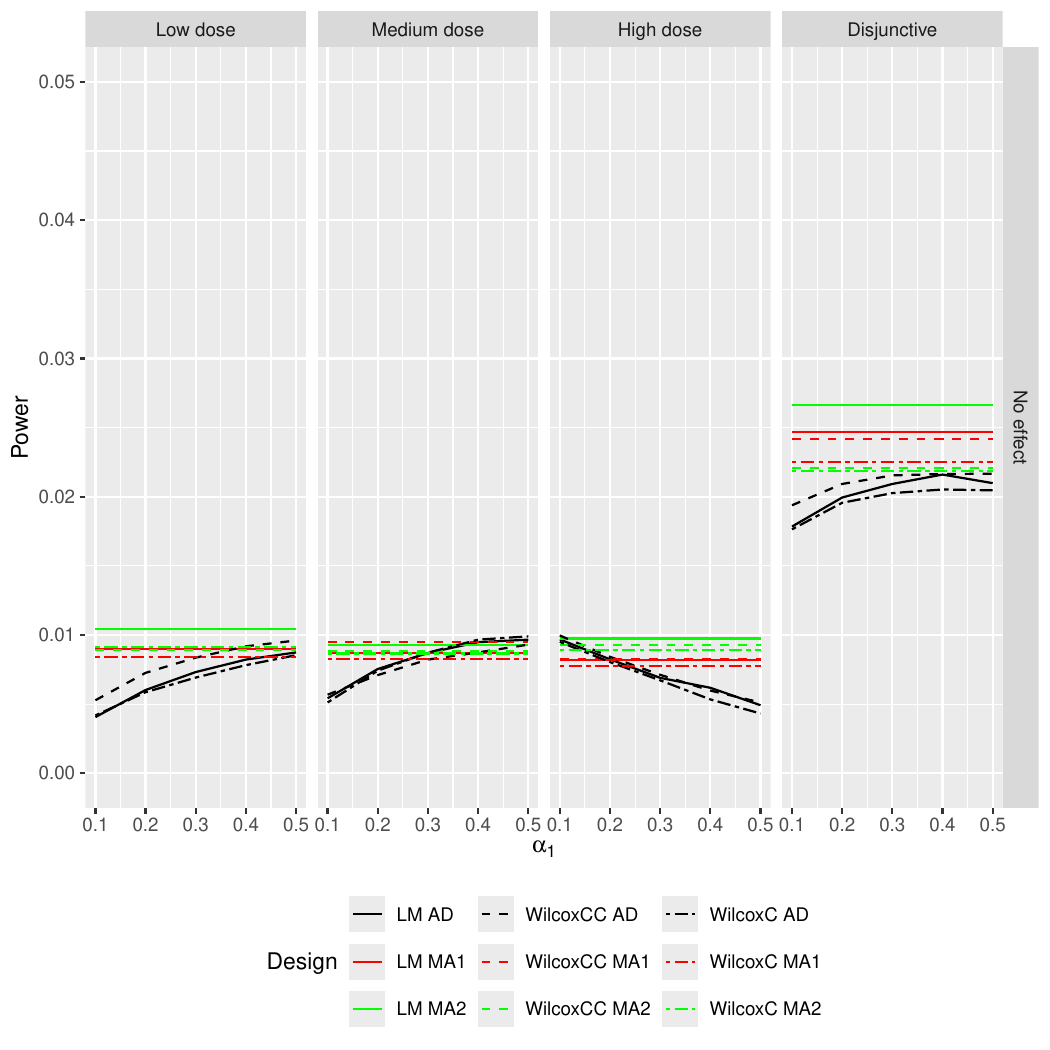}
    \caption[Onchocerciasis S3 - Error rates]{Onchocerciasis S3: Type I error rates for the low, medium, and high doses and the overall type I error rate are shown (in the columns) for simulation S3 (modified baseline parameters) as a function of $\alpha_1$ for scenarios of reduction rates 1 ("No effect")..}
    \label{fig:ScenariosWorst}
\end{figure}

\begin{figure}[!p]
    \centering     \includegraphics[width=1\linewidth,page=9]{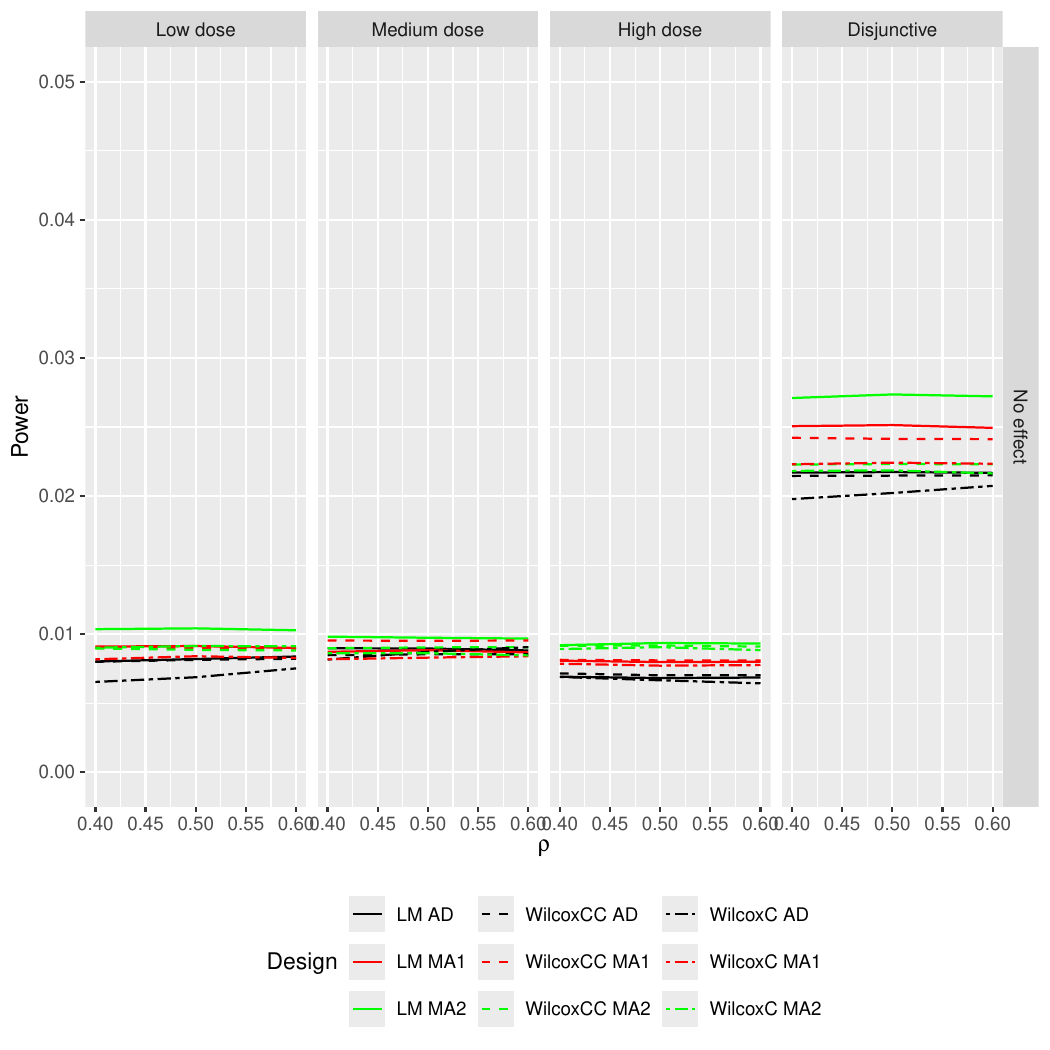}
    \caption[Onchocerciasis S4 - Error rates]{Onchocerciasis S4: Type I error rates for the low, medium, and high doses and the overall type I error rate are shown (in the columns) for simulation S4 as a function of $\rho$ for scenarios of reduction rates 1 ("No effect").}
    \label{fig:ScenariosWorst}
\end{figure}

\begin{figure}[!p]
    \centering     \includegraphics[width=1\linewidth,page=10]{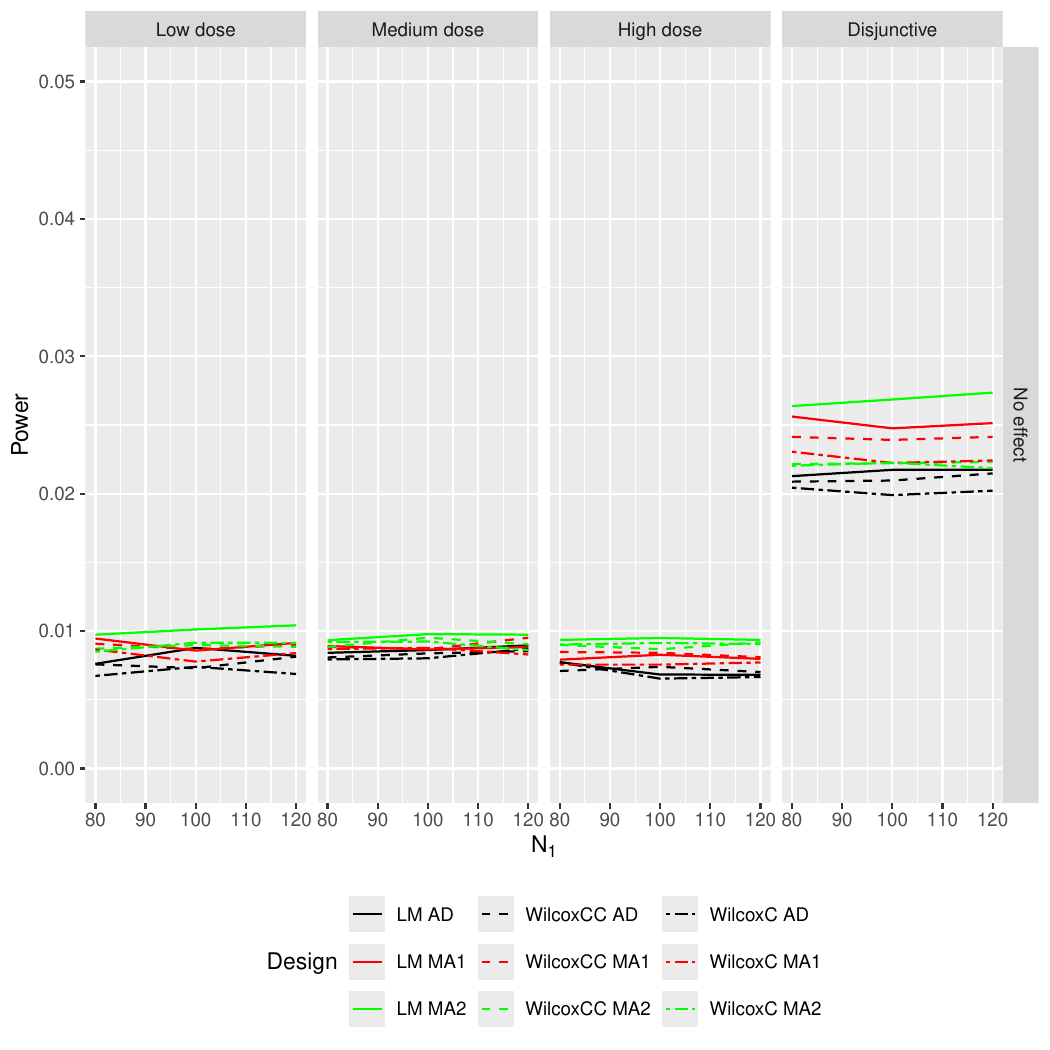}
    \caption[Onchocerciasis S5 - Error rates]{Onchocerciasis S5: Type I error rates for the low, medium, and high doses and the overall type I error rate are shown (in the columns) for simulation S5 as a function of $N_1$ for scenarios of reduction rates 1 ("No effect").}
    \label{fig:ScenariosWorst}
\end{figure}

\begin{figure}[!p]
    \centering     \includegraphics[width=1\linewidth,page=11]{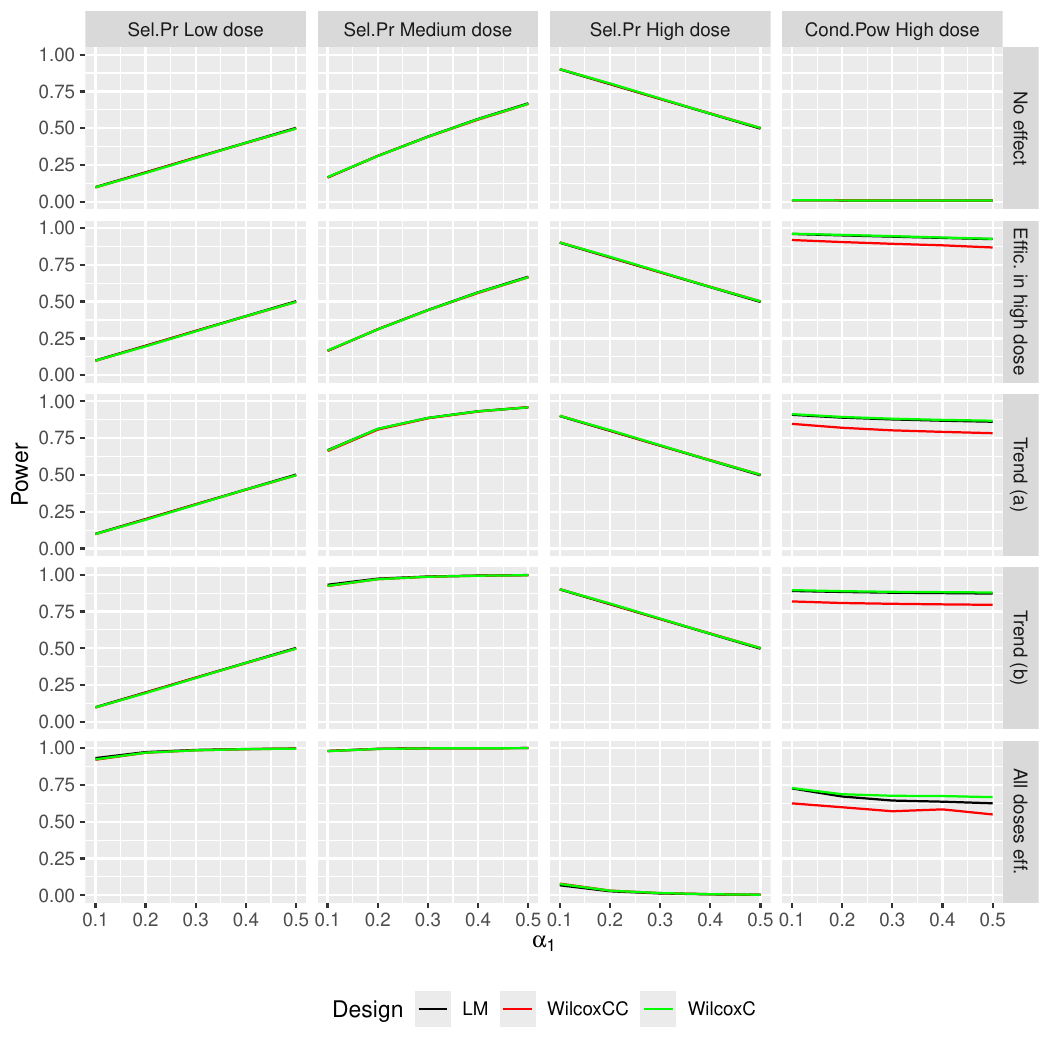}
    \caption[Onchocerciasis S1 - Selection probabilities and conditional probability]{Onchocerciasis S1: Selection probabilities (Sel.Prob) for low and medium dose and conditional probability (Cond.Prob) for high dose are depicted (in the columns) for simulation S1 (standard reduction rates) as a function of $\alpha_1 = \{0.1, 0.2, 0.3, 0.4, 0.5\}$ for scenarios of reduction rates 1 (in the rows)}
    \label{fig:ScenariosWorst}
\end{figure}

\begin{figure}[!p]
    \centering     \includegraphics[width=1\linewidth,page=12]{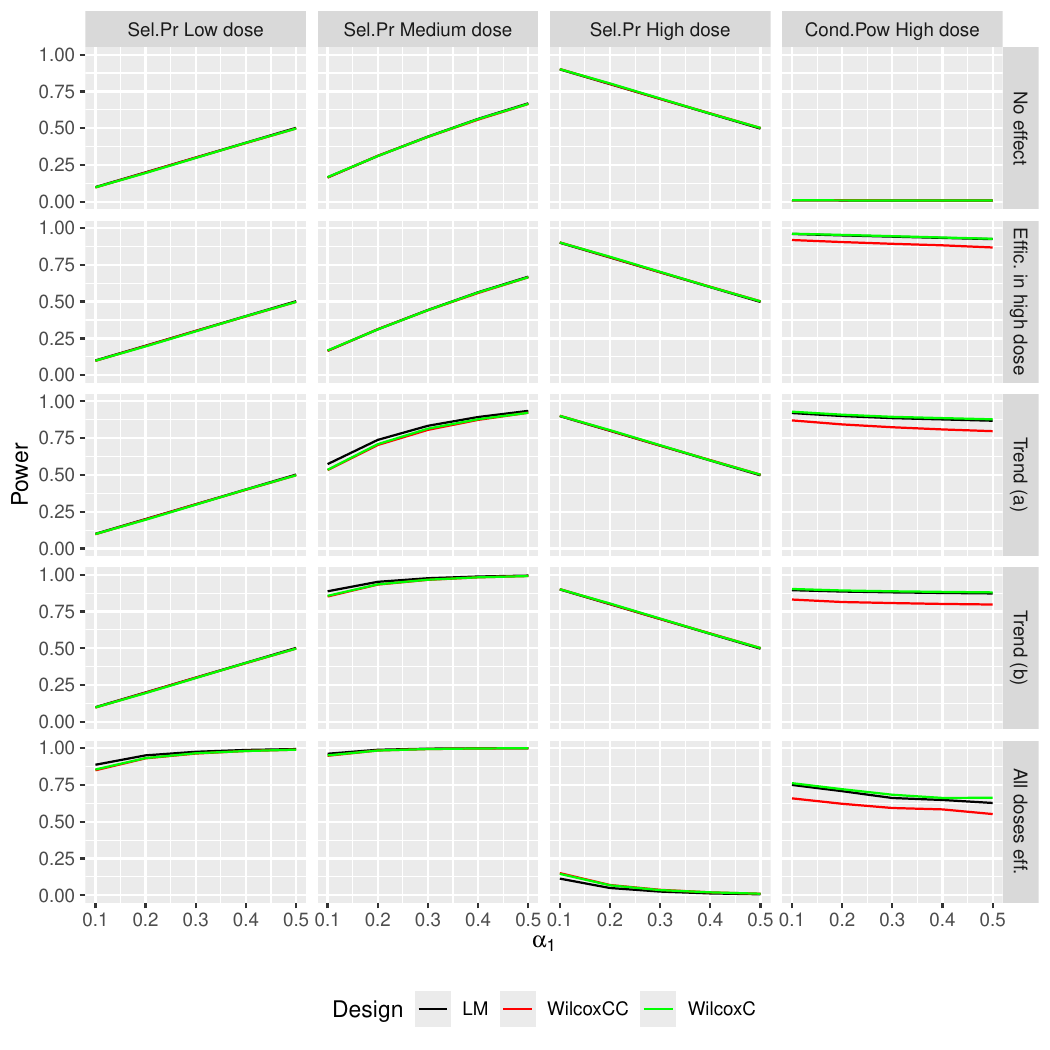}
    \caption[Onchocerciasis S2 - Selection probabilities and conditional probability]{Onchocerciasis S2: Selection probabilities (Sel.Pr) for all doses and conditional probability (Cond.Pow) for high dose are depicted (in the columns) for simulation S2 (modified reduction rates) as a function of $\alpha_1$ for several scenarios of reduction rates (in the rows).}
    \label{fig:ScenariosWorst}
\end{figure}

\begin{figure}[!p]
    \centering     \includegraphics[width=1\linewidth,page=13]{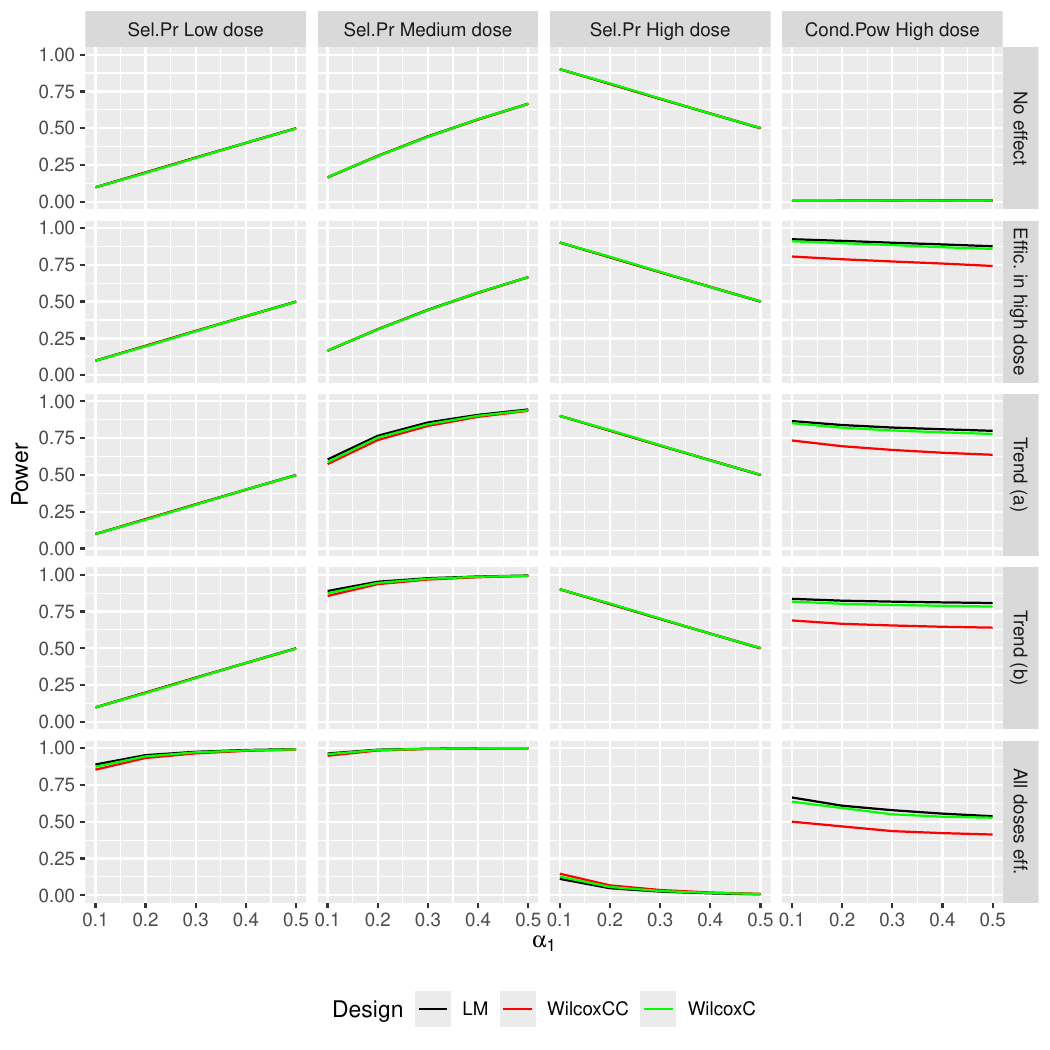}
    \caption[Onchocerciasis S3 - Selection probabilities and conditional probability]{Onchocerciasis S3: Selection probabilities (Sel.Pr) for all doses and conditional probability (Cond.Pow) for high dose are are depicted (in the columns) for simulation S3 (modified baseline parameters) as a function of $\alpha_1$ for several scenarios of reduction rates (in the rows).}
    \label{fig:ScenariosWorst}
\end{figure}

\begin{figure}[!p]
    \centering     \includegraphics[width=1\linewidth,page=14]{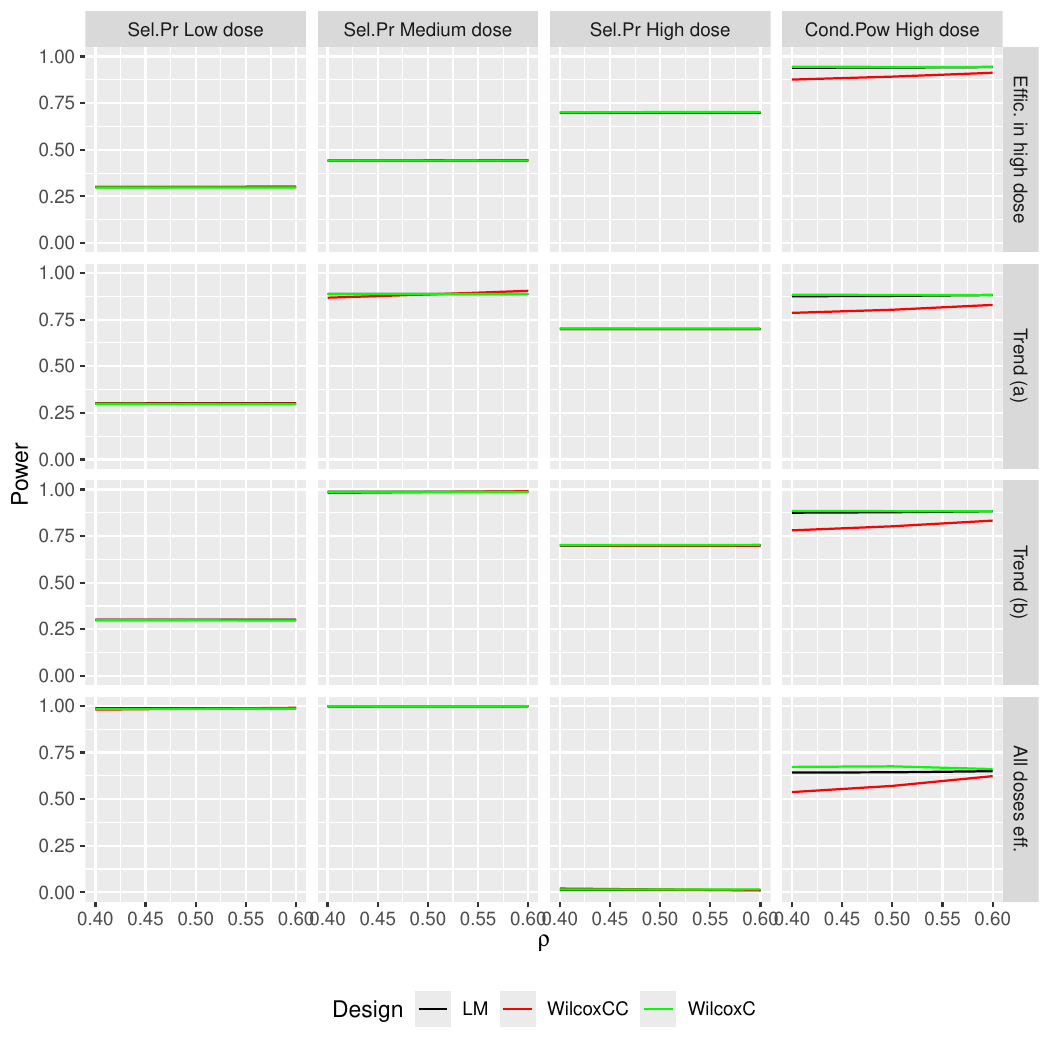}
    \caption[Onchocerciasis S4 - Selection probabilities and conditional probability]{Onchocerciasis S4: Selection probabilities (Sel.Pr) for all doses and conditional probability (Cond.Pow) for high dose are are depicted (in the columns) for simulation S4  as a function of $\rho$ for several scenarios of reduction rates (in the rows).}
    \label{fig:ScenariosWorst}
\end{figure}

\begin{figure}[!p]
    \centering     \includegraphics[width=1\linewidth,page=15]{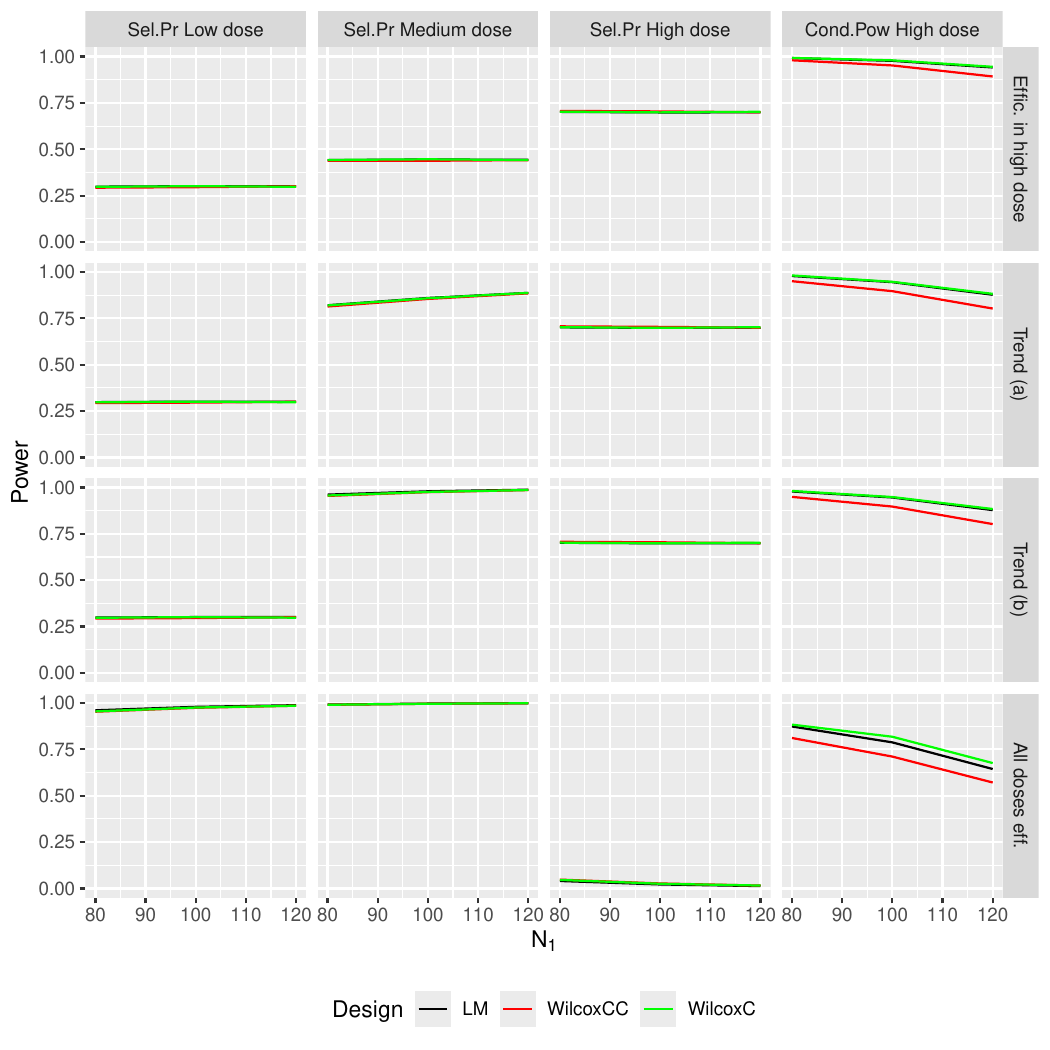}
    \caption[Onchocerciasis S5 - Selection probabilities and conditional probability]{Onchocerciasis S5: Selection probabilities (Sel.Pr) for all doses and conditional probability (Cond.Pow) for high dose are are depicted (in the columns) for simulation S5  as a function of $N_1$ for several scenarios of reduction rates (in the rows).}
    \label{fig:ScenariosWorst}
\end{figure}

\begin{figure}[!p]
    \centering     \includegraphics[width=1\linewidth,page=1]{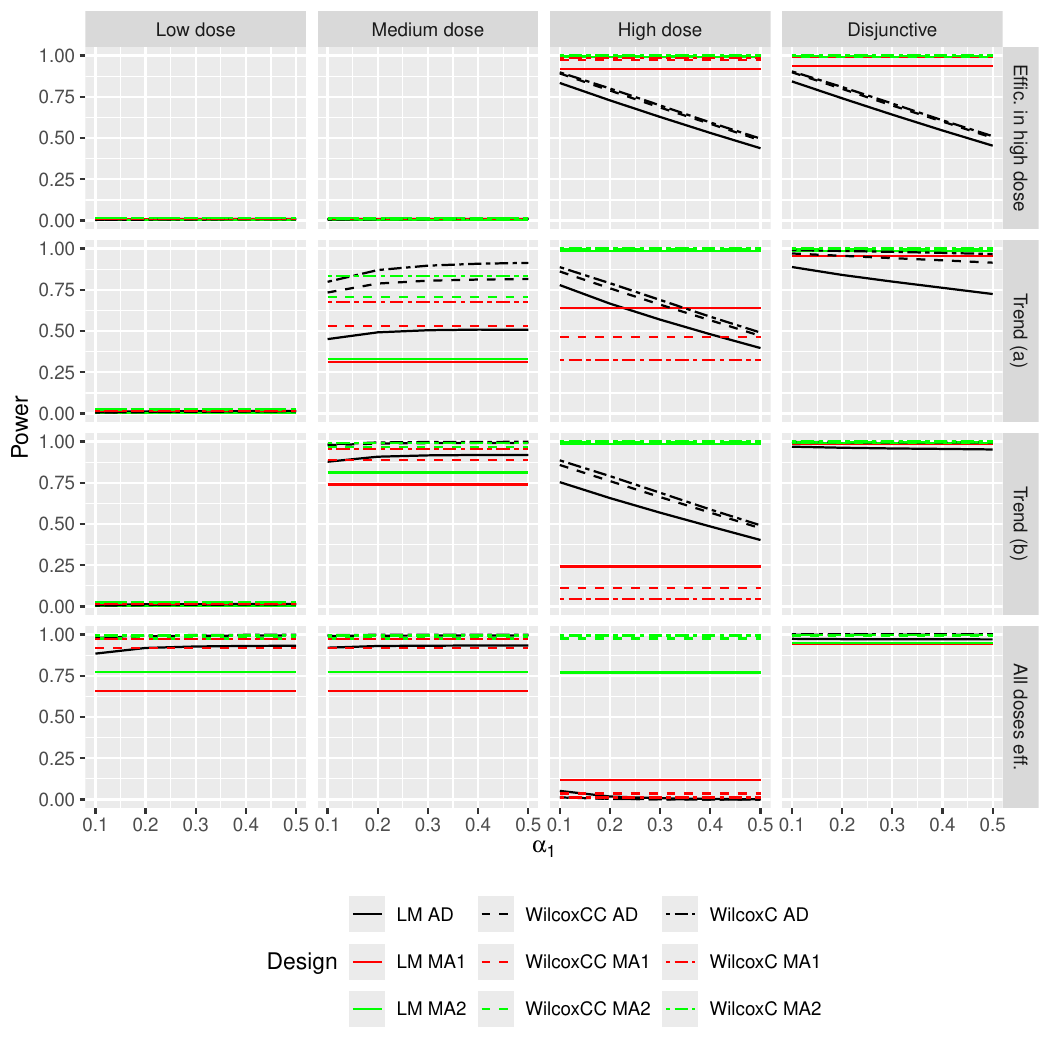}
    \caption[Loiasis S1]{Loiasis S1: Simulated power values for the low, medium, and high doses and the disjunctive power are depicted (in the columns) for simulation S1 (standard reduction rates) as a function of $\alpha_1 = \{0.1, 0.2, 0.3, 0.4, 0.5\}$ for several scenarios of reduction rates (in the rows)}
    \label{fig:ScenariosWorst}
\end{figure}

\begin{figure}[!p]
    \centering     \includegraphics[width=1\linewidth,page=2]{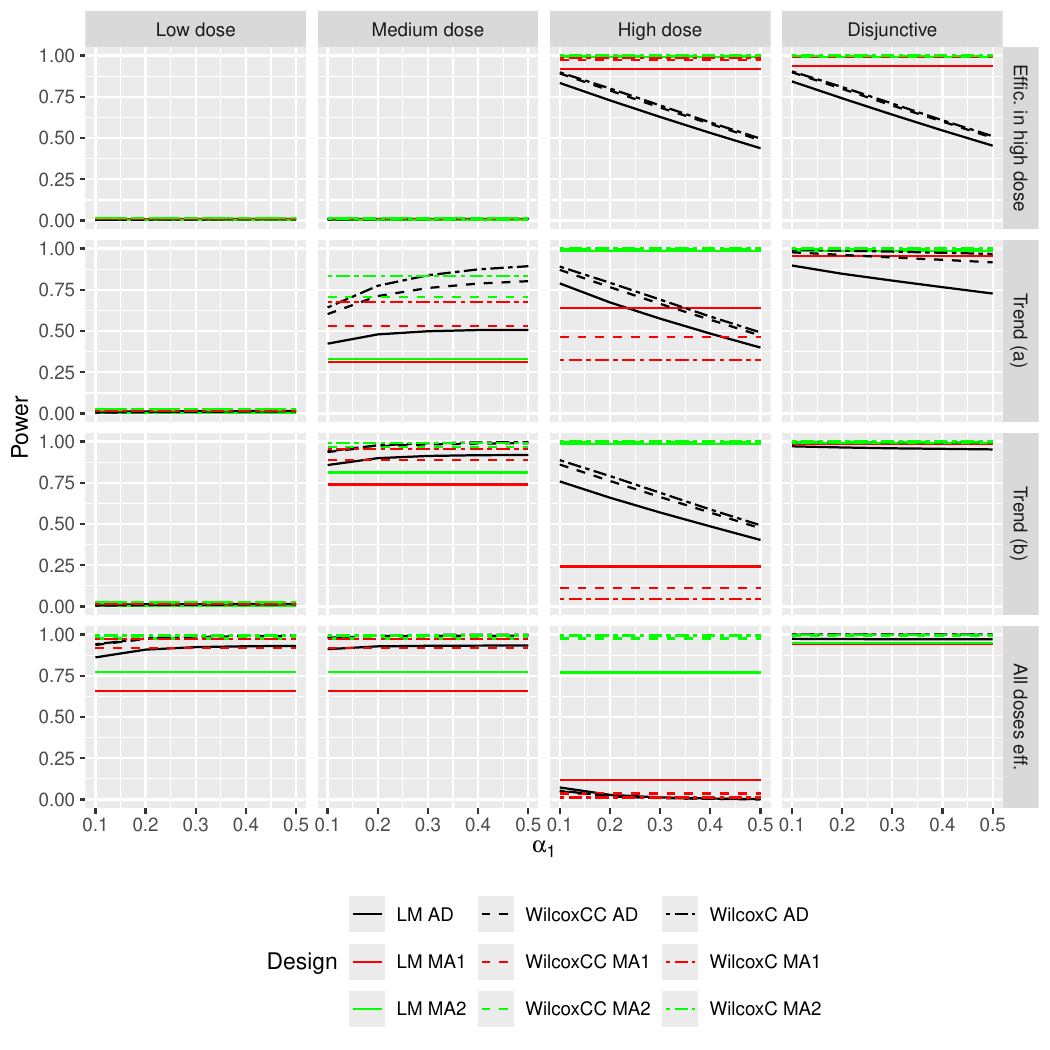}
    \caption[Loiasis S2]{Loiasis S2: Simulated power values for the low, medium, and high doses and the disjunctive power are depicted (in the columns) for simulation S2 (modified reduction rates) as a function of $\alpha_1$ for several scenarios of reduction rates (in the rows).}
    \label{fig:ScenariosWorst}
\end{figure}

\begin{figure}[!p]
    \centering     \includegraphics[width=1\linewidth,page=3]{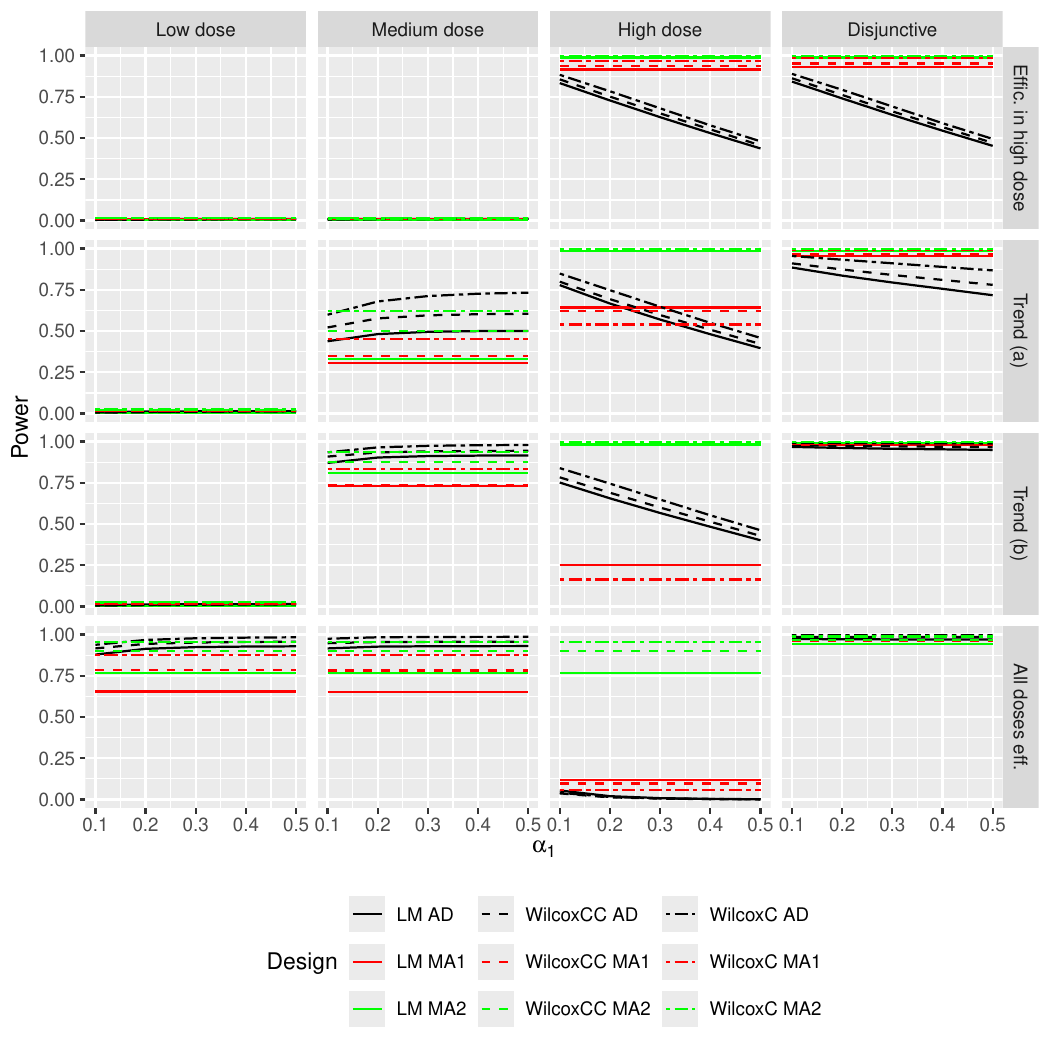}
    \caption[Loiasis S3]{Loiasis S3: Simulated power values for the low, medium, and high doses and the disjunctive power are depicted (in the columns) for simulation S3 (modified baseline parameters) as a function of $\alpha_1$ for several scenarios of reduction rates (in the rows).}
    \label{fig:ScenariosWorst}
\end{figure}

\begin{figure}[!p]
    \centering     \includegraphics[width=1\linewidth,page=4]{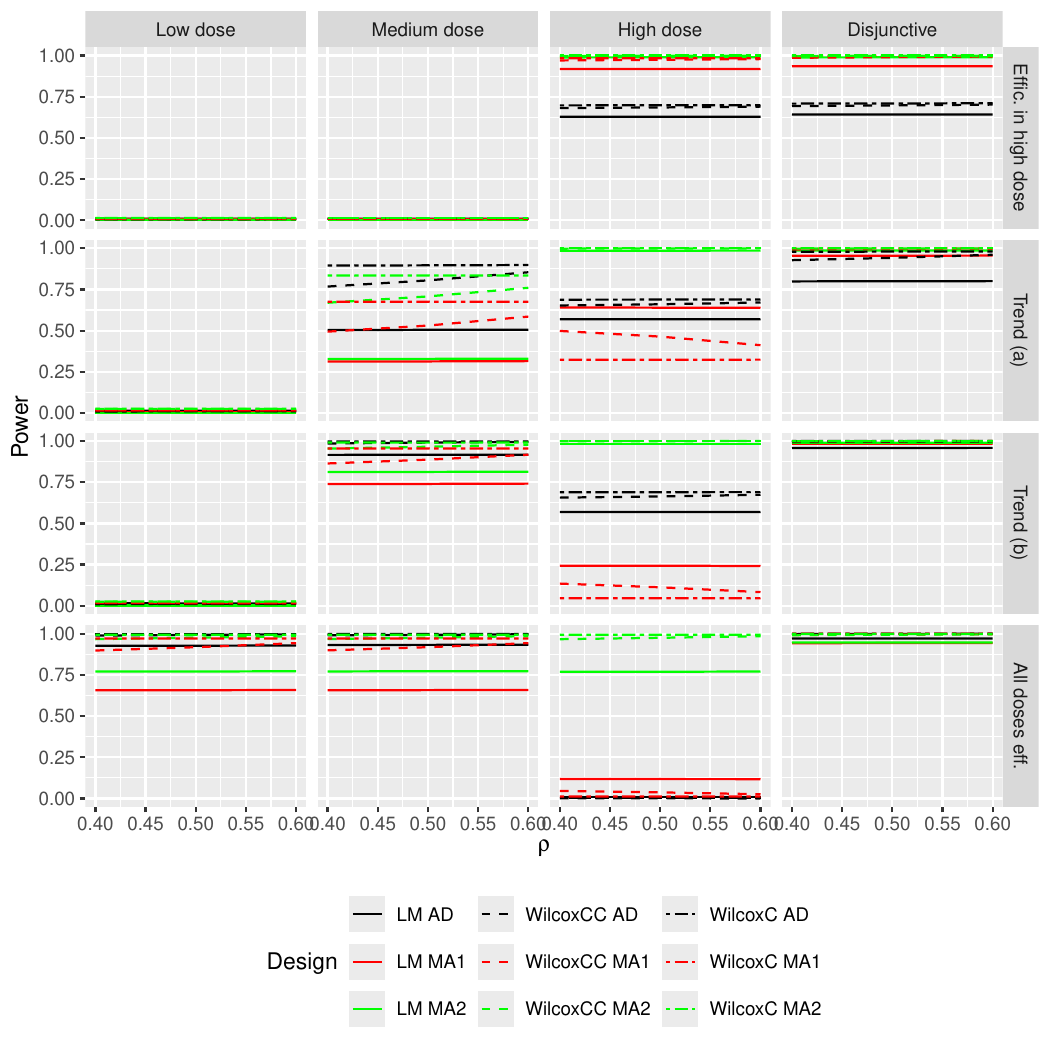}
    \caption[Loiasis S4]{Loiasis S4: Simulated power values for the low, medium, and high doses and the disjunctive power are depicted (in the columns) for simulation S4  as a function of $\rho$ for several scenarios of reduction rates (in the rows).}
    \label{fig:ScenariosWorst}
\end{figure}

\begin{figure}[!p]
    \centering     \includegraphics[width=1\linewidth,page=5]{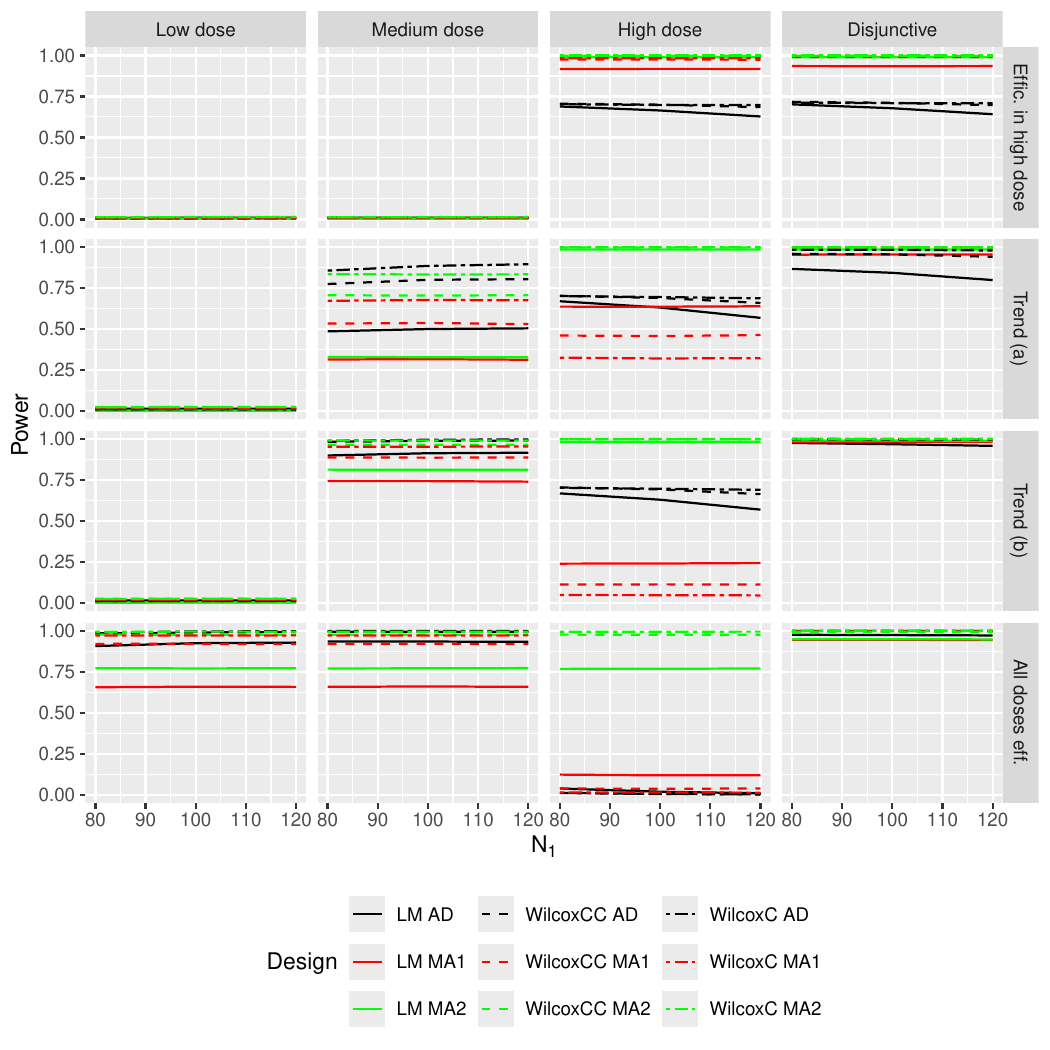}
    \caption[Loiasis S5]{Loiasis S5: Simulated power values for the low, medium, and high doses and the disjunctive power are depicted (in the columns) for simulation S5 (modified baseline parameters) as a function of $N_1$ for several scenarios of reduction rates (in the rows).}
    \label{fig:ScenariosWorst}
\end{figure}

\begin{figure}[!p]
    \centering     \includegraphics[width=1\linewidth,page=6]{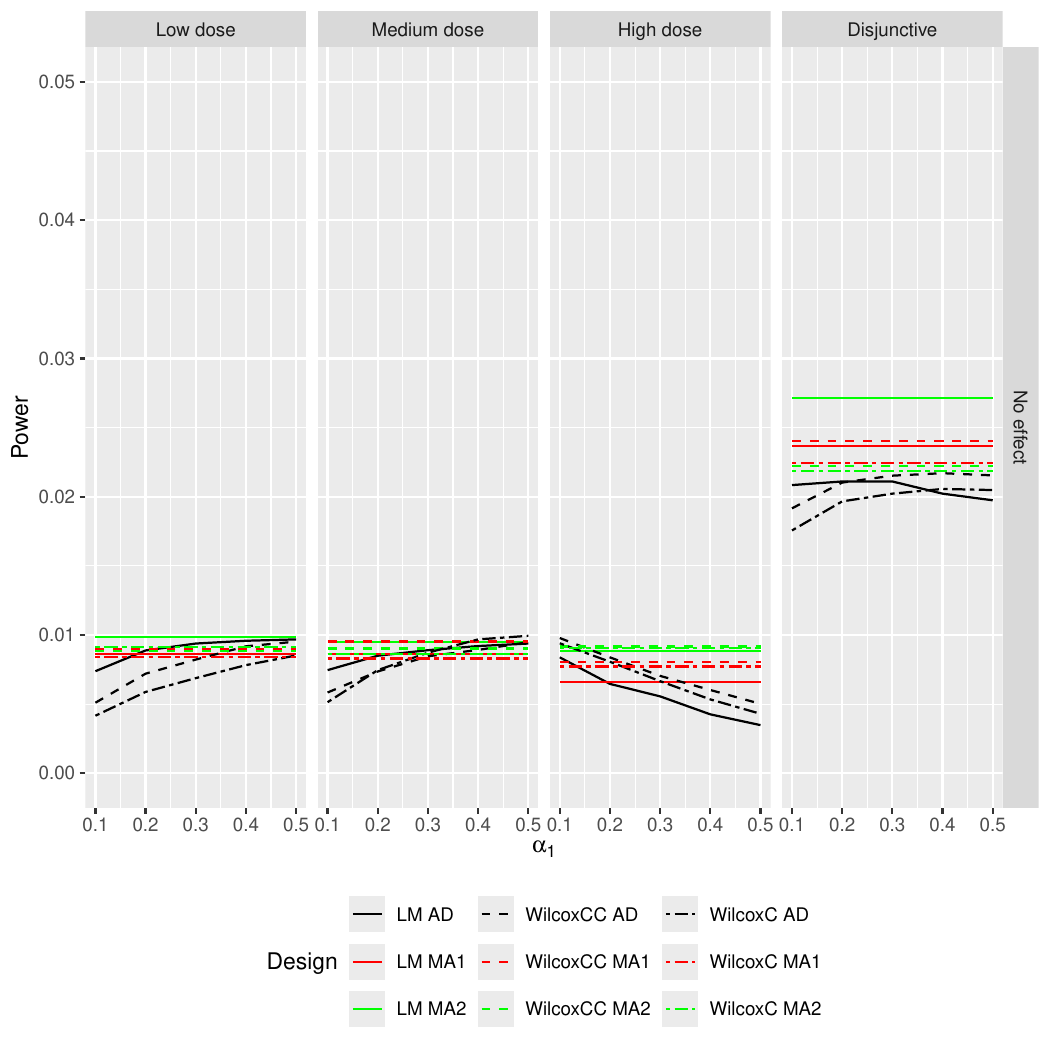}
    \caption[Loiasis S1 - Error rates]{Loiasis S1: Type I error rates for the low, medium, and high doses and the overall type I error rate are shown (in the columns) for simulation S1 as a function of $\alpha_1$ for scenarios of reduction rates 1 ("No effect").}
    \label{fig:ScenariosWorst}
\end{figure}

\begin{figure}[!p]
    \centering     \includegraphics[width=1\linewidth,page=7]{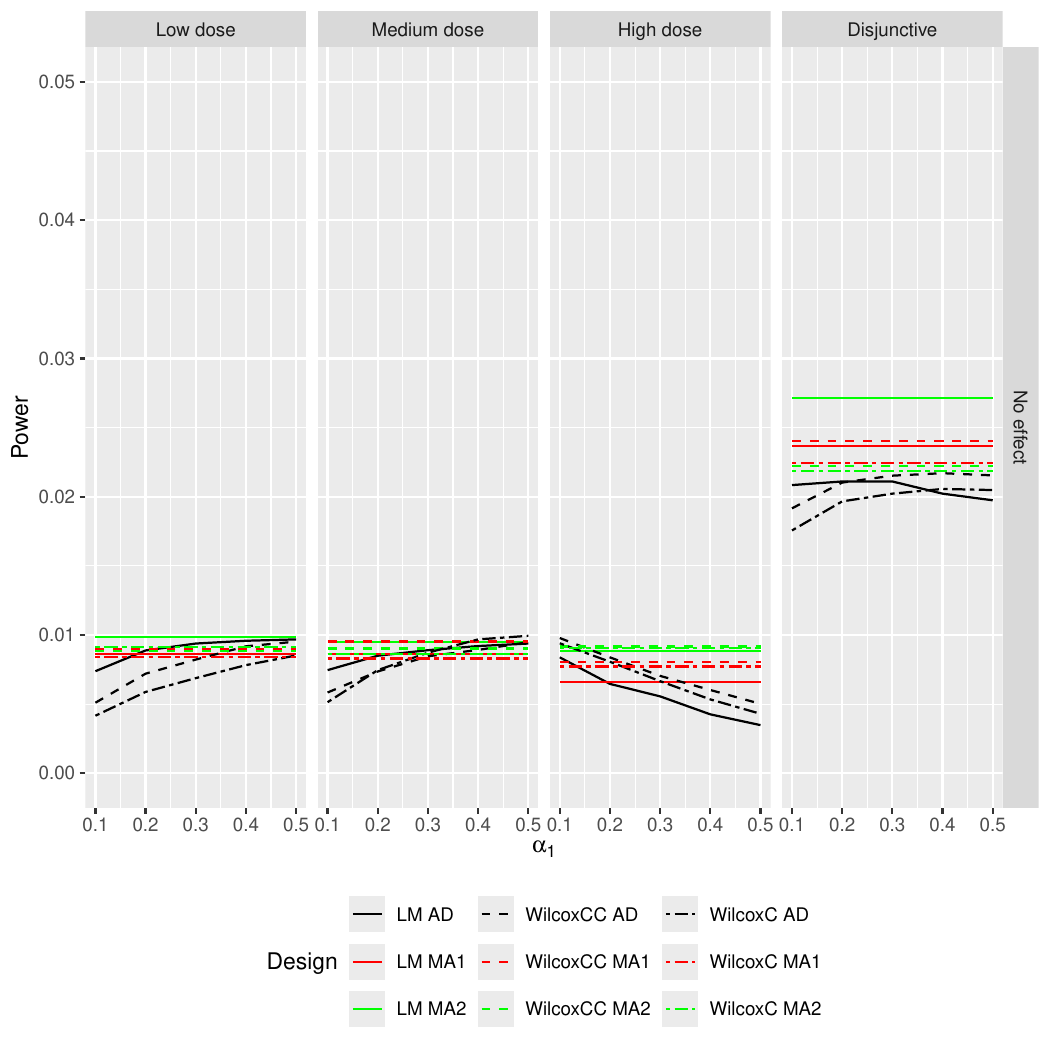}
    \caption[Loiasis S2 - Error rates]{Loiasis S2: Type I error rates for the low, medium, and high doses and the overall type I error rate are shown (in the columns) for simulation S2 (modified reduction rates) as a function of $\alpha_1$ for scenarios of reduction rates 1 ("No effect").}
    \label{fig:ScenariosWorst}
\end{figure}

\begin{figure}[!p]
    \centering     \includegraphics[width=1\linewidth,page=8]{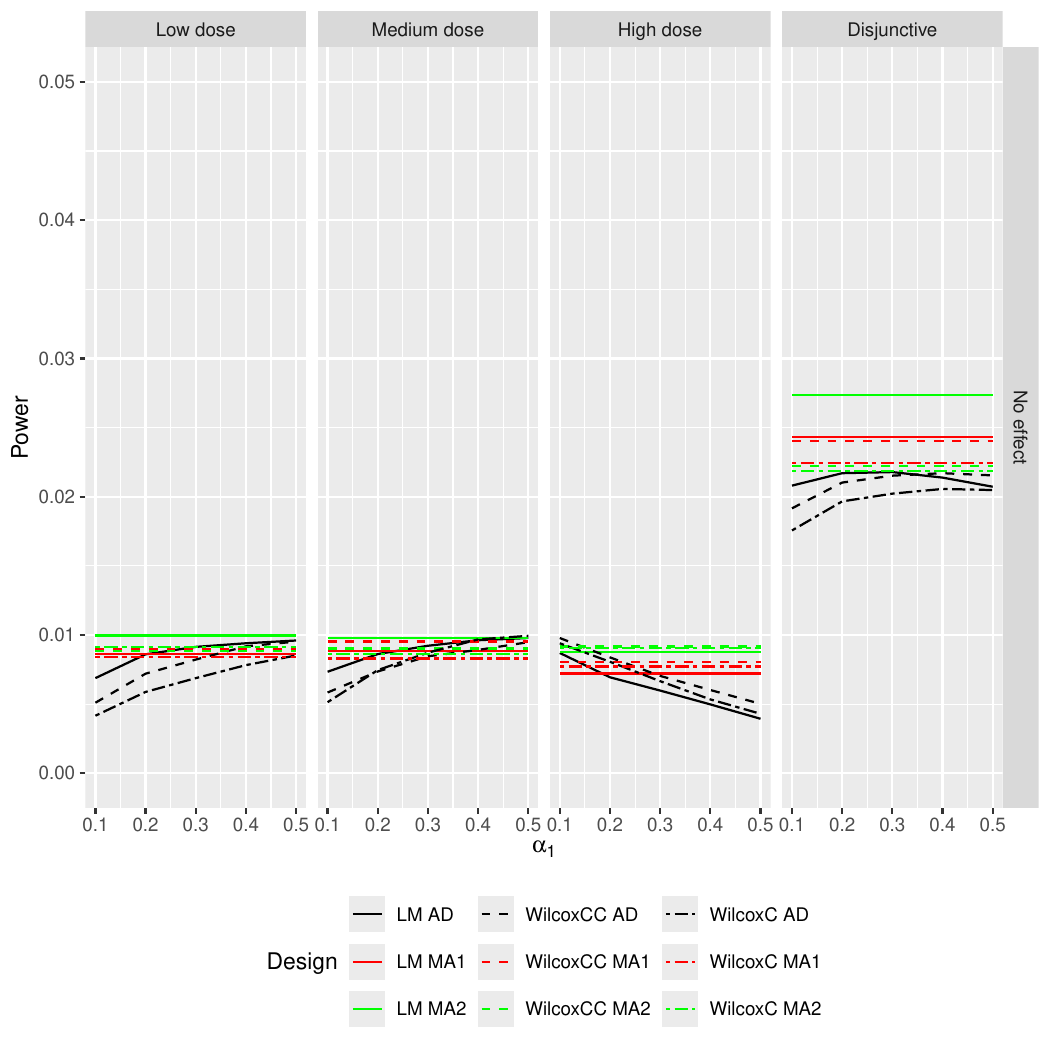}
    \caption[Loiasis S3 - Error rates]{Loiasis S3: Type I error rates for the low, medium, and high doses and the overall type I error rate are shown (in the columns) for simulation S3 (modified baseline parameters) as a function of $\alpha_1$ for scenarios of reduction rates 1 ("No effect").}
    \label{fig:ScenariosWorst}
\end{figure}

\begin{figure}[!p]
    \centering     \includegraphics[width=1\linewidth,page=9]{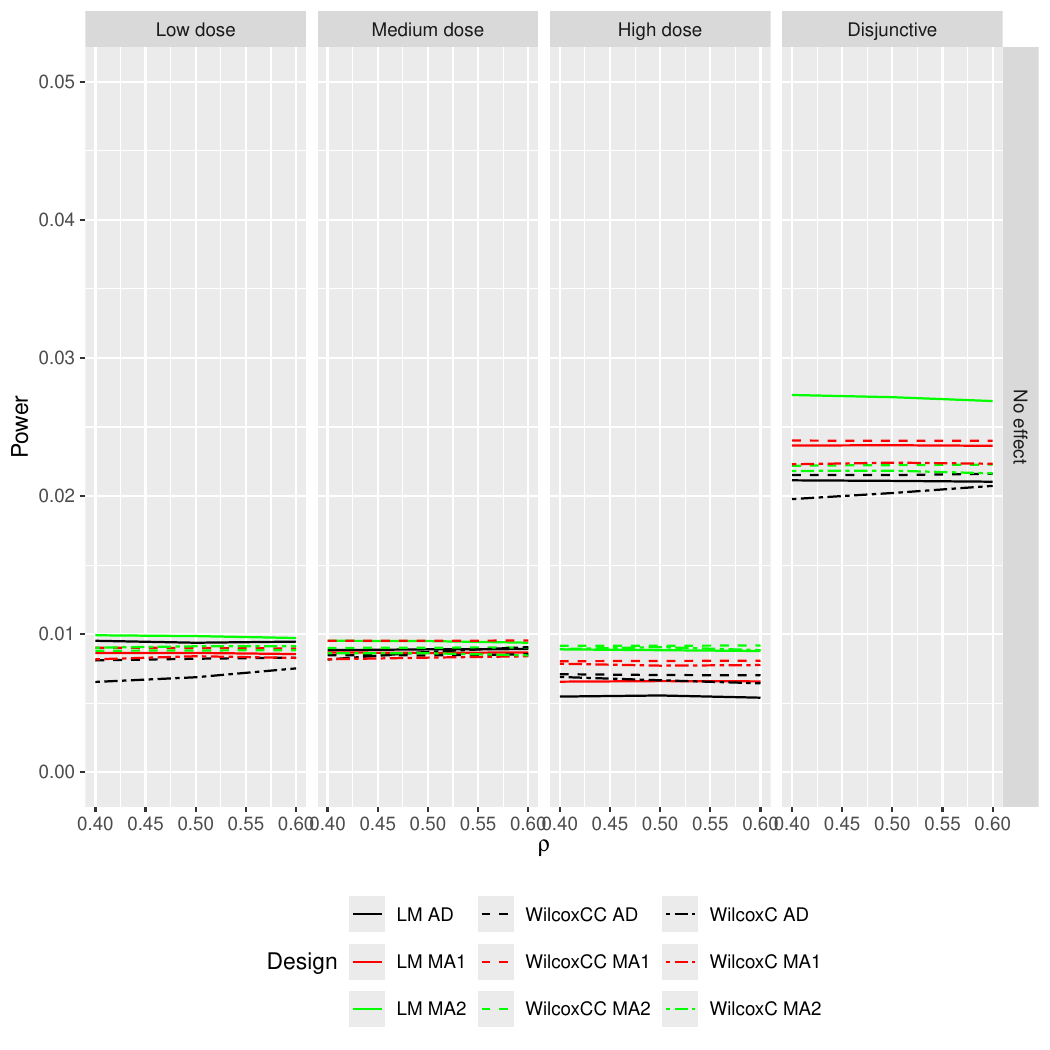}
    \caption[Loiasis S4 - Error rates]{Loiasis S4: Type I error rates for the low, medium, and high doses and the overall type I error rate are shown (in the columns) for simulation S4 as a function of $\rho$ for scenarios of reduction rates 1 ("No effect").}
    \label{fig:ScenariosWorst}
\end{figure}

\begin{figure}[!p]
    \centering     \includegraphics[width=1\linewidth,page=10]{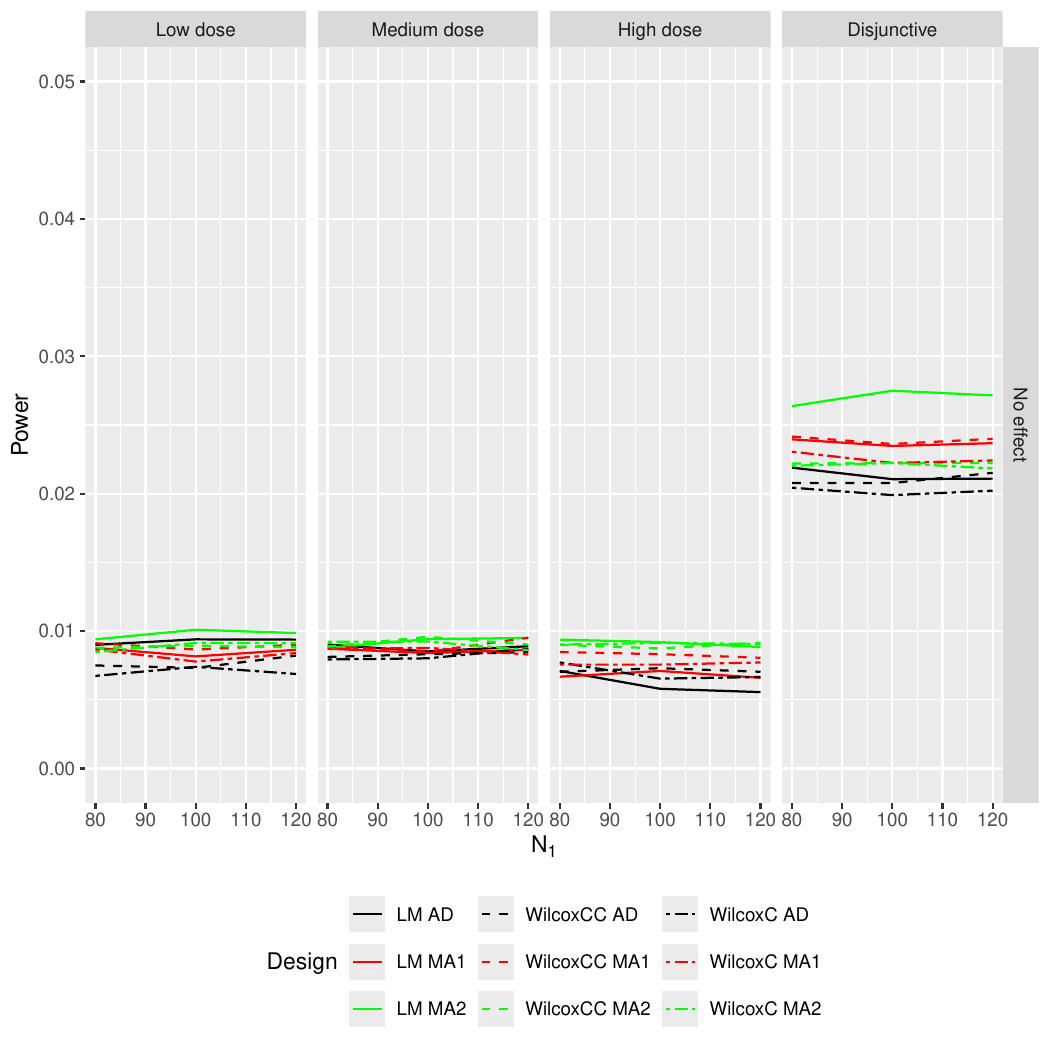}
    \caption[Loiasis S5 - Error rates]{Loiasis S5: Type I error rates for the low, medium, and high doses and the overall type I error rate are shown (in the columns) for simulation S5 as a function of $N_1$ for scenarios of reduction rates 1 ("No effect").}
    \label{fig:ScenariosWorst}
\end{figure}

\begin{figure}[!p]
    \centering     \includegraphics[width=1\linewidth,page=11]{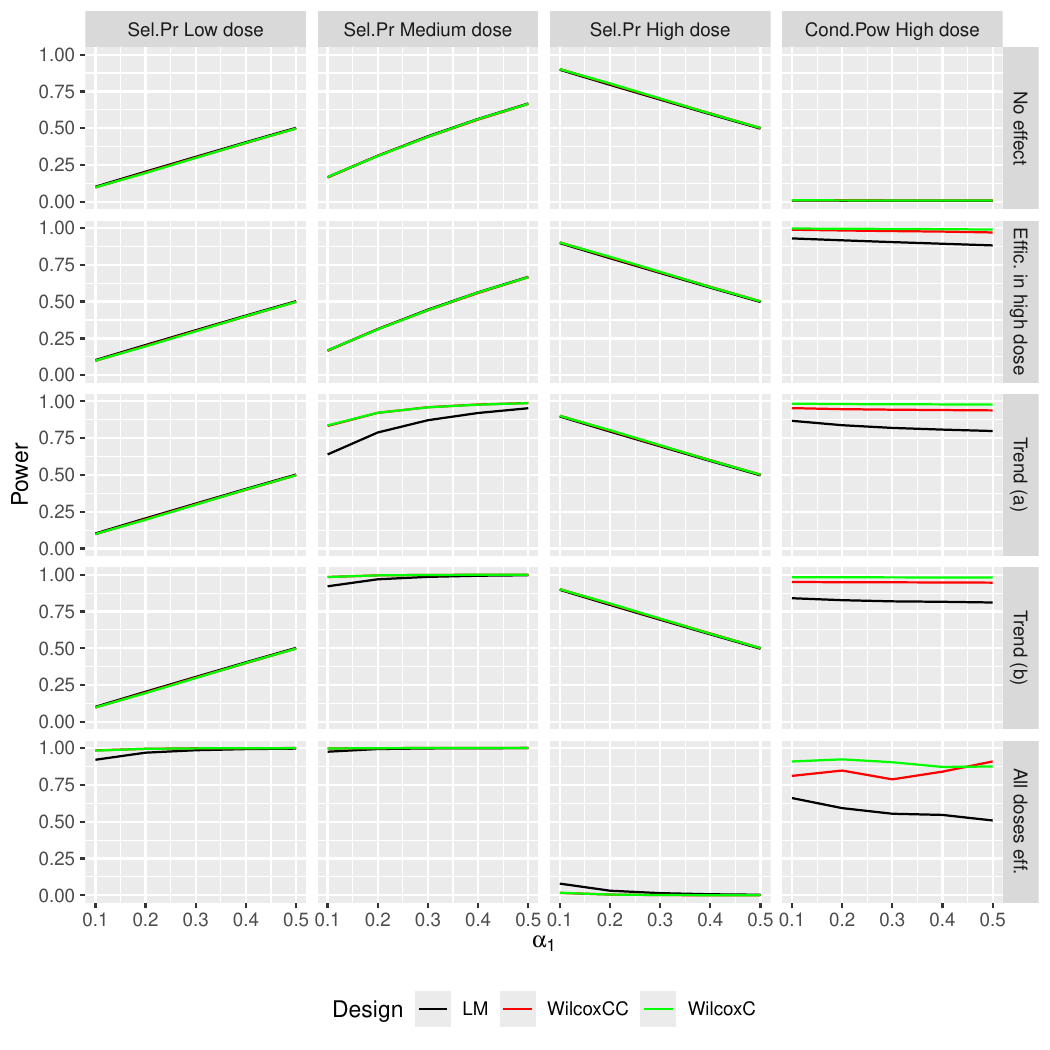}
    \caption[Loiasis S1 - Selection probabilities and conditional probability]{Loiasis S1: Selection probabilities (Sel.Prob) for the low and medium dose and conditional probability (Cond.Prob) for high dose are depicted (in the columns) for simulation S1 (standard reduction rates) as a function of $\alpha_1 = \{0.1, 0.2, 0.3, 0.4, 0.5\}$ for scenarios of reduction rates 1 (in the rows)}
    \label{fig:ScenariosWorst}
\end{figure}

\begin{figure}[!p]
    \centering     \includegraphics[width=1\linewidth,page=12]{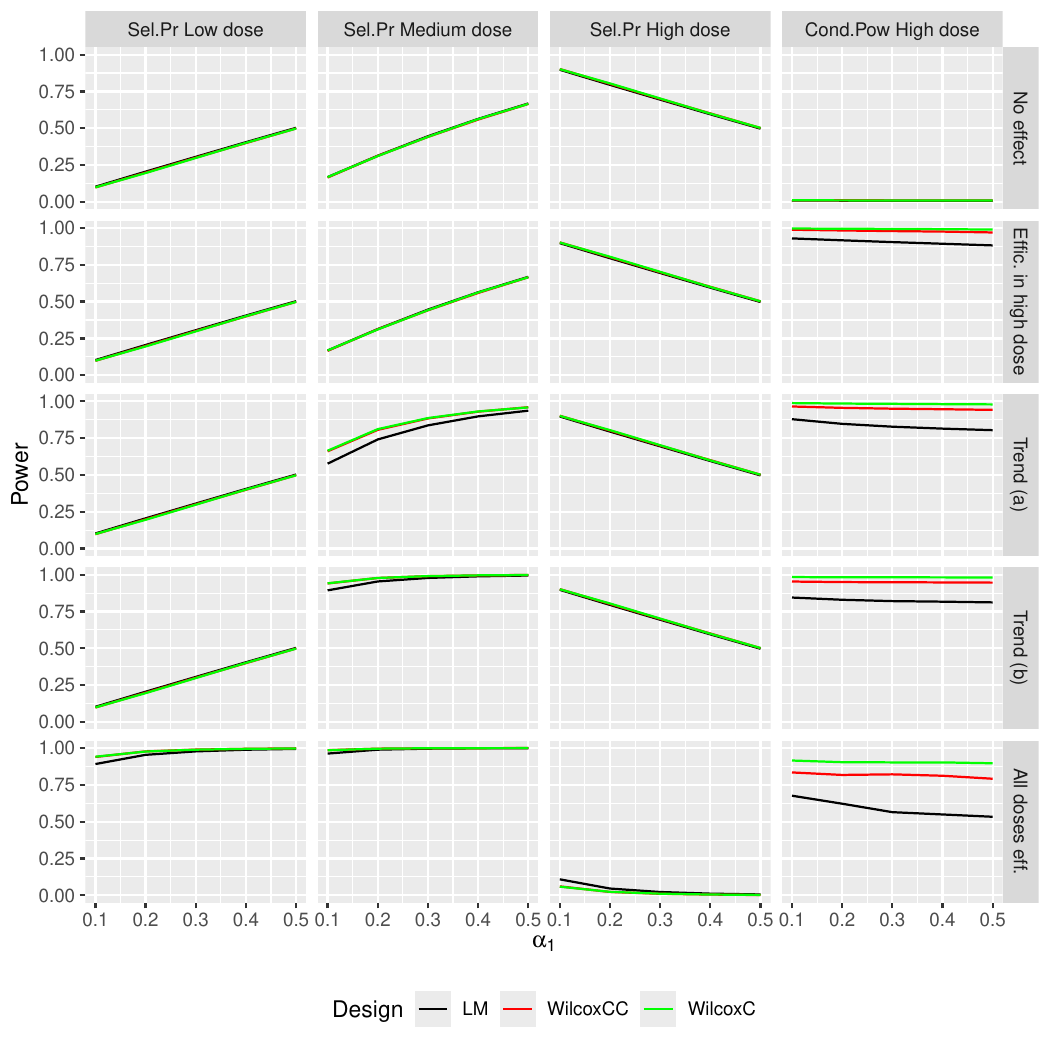}
    \caption[Loiasis S2 - Selection probabilities and conditional probability]{Loiasis S2: Selection probabilities (Sel.Pr) for all doses and conditional probability (Cond.Pow) for high dose are depicted (in the columns) for simulation S2 (modified reduction rates) as a function of $\alpha_1$ for several scenarios of reduction rates (in the rows).}
    \label{fig:ScenariosWorst}
\end{figure}

\begin{figure}[!p]
    \centering     \includegraphics[width=1\linewidth,page=13]{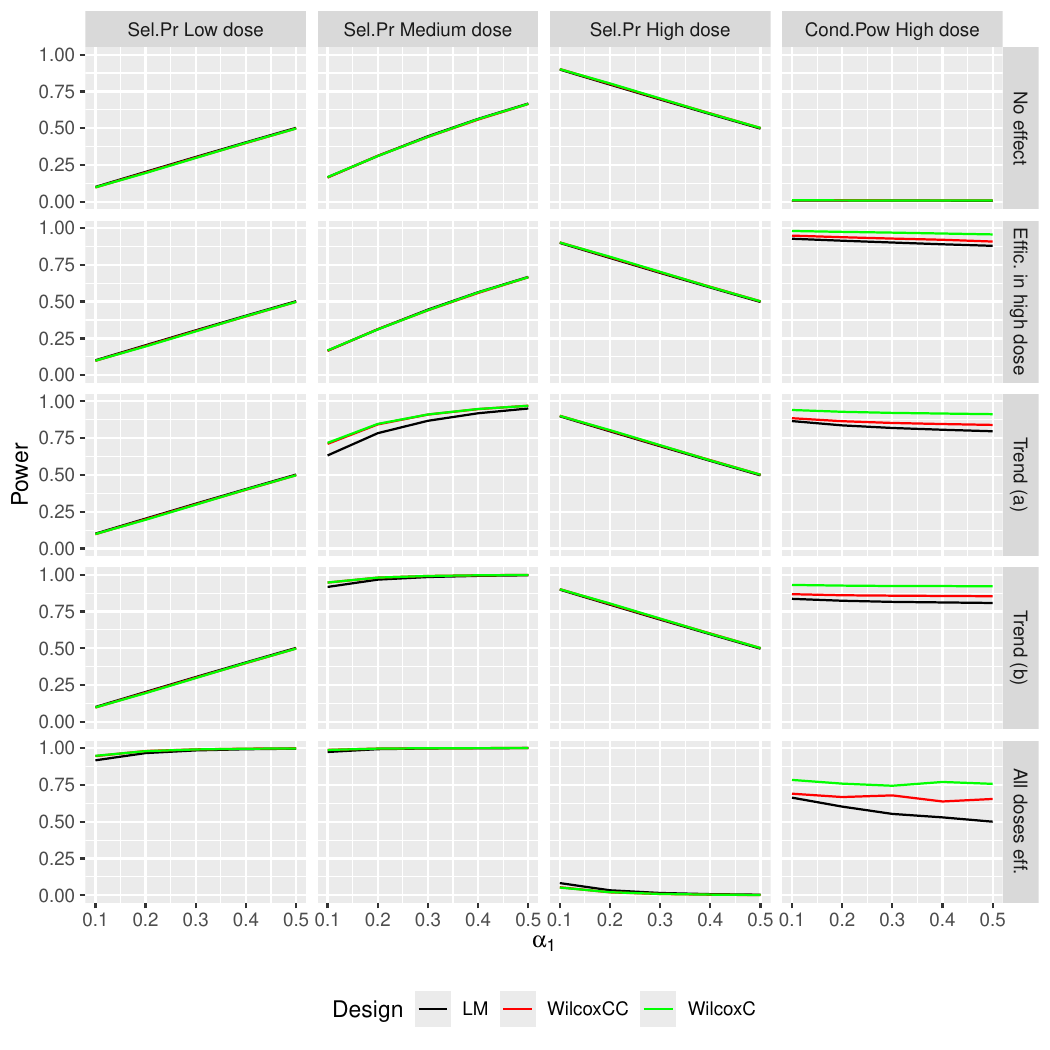}
    \caption[Loiasis S3 - Selection probabilities and conditional probability]{Loiasis S3: Selection probabilities (Sel.Pr) for all doses and conditional probability (Cond.Pow) for high dose are depicted (in the columns) for simulation S2 (modified baseline parameters) as a function of $\alpha_1$ for several scenarios of reduction rates (in the rows).}
    \label{fig:ScenariosWorst}
\end{figure}

\begin{figure}[!p]
    \centering     \includegraphics[width=1\linewidth,page=14]{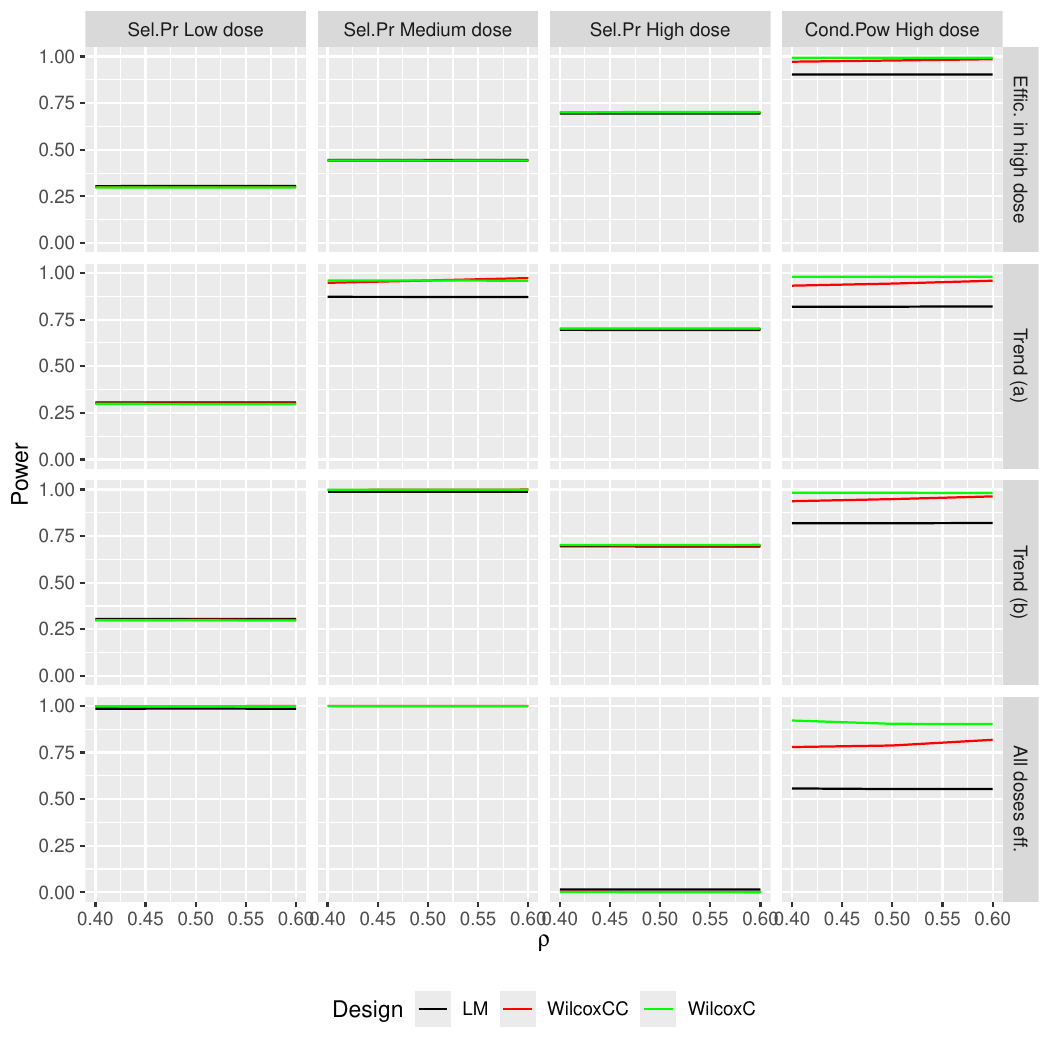}
    \caption[Loiasis S4 - Selection probabilities and conditional probability]{Loiasis S4: Selection probabilities (Sel.Pr) for all doses and conditional probability (Cond.Pow) for high dose are depicted (in the columns) for simulation S4 as a function of $\rho$ for several scenarios of reduction rates (in the rows).}
    \label{fig:ScenariosWorst}
\end{figure}

\begin{figure}[!p]
    \centering     \includegraphics[width=1\linewidth,page=15]{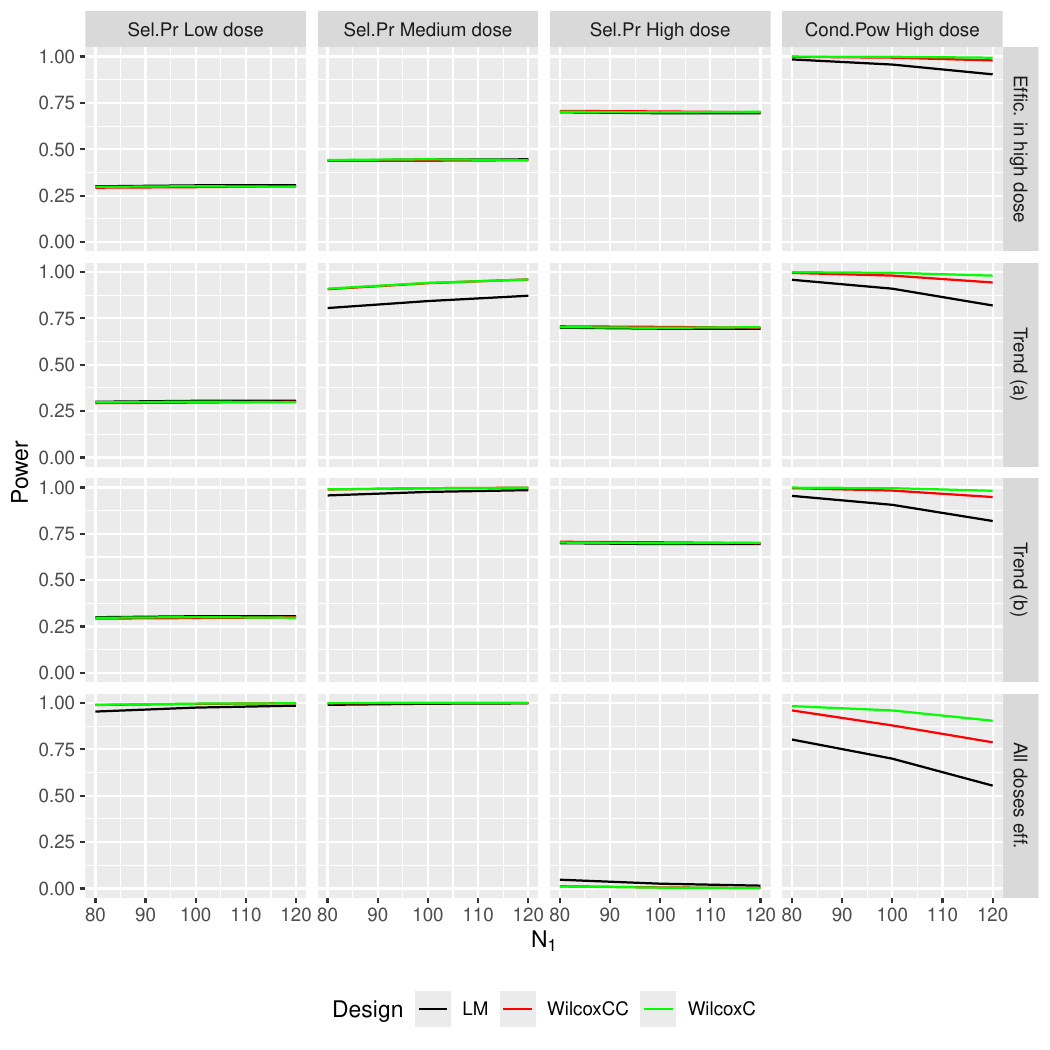}
    \caption[Loiasis S5 - Selection probabilities and conditional probability]{Loiasis S5: Selection probabilities (Sel.Pr) for all doses and conditional probability (Cond.Pow) for high dose are depicted (in the columns) for simulation S5  as a function of $N_1$ for several scenarios of reduction rates (in the rows).}
    \label{fig:ScenariosWorst}
\end{figure}

\end{document}